\documentclass[conference, twocolumn, 10pt]{IEEEtran}
\IEEEoverridecommandlockouts
\usepackage[english]{babel}
\addto\captionsenglish{\renewcommand{\figurename}{Fig.}}

\PassOptionsToPackage{utf8}{inputenc}
\usepackage{inputenc}

\PassOptionsToPackage{T1}{fontenc}
\usepackage{fontenc}
\usepackage{stfloats}
\usepackage{csquotes}
\usepackage{cite}
\usepackage{amssymb,amsfonts,amsthm}
\usepackage{textcomp}
\usepackage{xcolor}
\usepackage{xspace}
\usepackage[size=tiny]{todonotes}
\usepackage{balance}
\usepackage[all]{nowidow}
\PassOptionsToPackage{fleqn}{amsmath}       %
\usepackage{amsmath}

\usepackage[inline]{enumitem}

\newcommand{\qcscale}{0.55}

\newif\ifqclabels
\qclabelsfalse

\tikzset{
  qubit/.style        = {circle, draw=blue!80, fill=blue!10, thick,
                         minimum size=6mm, inner sep=0pt,
                         font=\sffamily\bfseries\tiny},
  coupling line/.style= {thick, draw=gray!70},
  coupler/.style      = {rectangle, draw=orange!90, fill=orange!25, thick,
                         minimum size=3.2mm, inner sep=0pt},
  qc layout/.style    = {scale=\qcscale, every node/.append style={transform shape}},
}

\newcommand{\qclink}[4]{%
  \draw[coupling line] (#1,#2) -- (#3,#4) node[midway,coupler]{};}

\newcommand{\qcgridlines}[4]{%
  \pgfmathtruncatemacro{\qcCmax}{#1-1}%
  \pgfmathtruncatemacro{\qcRmax}{#2-1}%
  \pgfmathtruncatemacro{\qcCmaxm}{#1-2}%
  \pgfmathtruncatemacro{\qcRmaxm}{#2-2}%
  \foreach \r in {0,...,\qcRmax}{%
    \foreach \c in {0,...,\qcCmaxm}{%
      \qclink{#3+\c}{#4+\r}{#3+\c+1}{#4+\r}}}%
  \foreach \c in {0,...,\qcCmax}{%
    \foreach \r in {0,...,\qcRmaxm}{%
      \qclink{#3+\c}{#4+\r}{#3+\c}{#4+\r+1}}}%
}
\newcommand{\qcgridnodes}[4]{%
  \pgfmathtruncatemacro{\qcCmax}{#1-1}%
  \pgfmathtruncatemacro{\qcRmax}{#2-1}%
  \foreach \r in {0,...,\qcRmax}{%
    \foreach \c in {0,...,\qcCmax}{%
      \pgfmathtruncatemacro{\qcIdx}{(#2-1-\r)*#1+\c+1}%
      \ifqclabels
        \node[qubit] at ({#3+\c},{#4+\r}) {\qcIdx};%
      \else
        \node[qubit] at ({#3+\c},{#4+\r}) {};%
      \fi}}%
}

\newcommand{\qcxarmlines}[5]{%
  \foreach \j in {1,...,#5}{%
    \pgfmathsetmacro{\qcAx}{#1+(#3)*(\j-1)}%
    \pgfmathsetmacro{\qcAy}{#2+(#4)*(\j-1)}%
    \pgfmathsetmacro{\qcBx}{\qcAx+(#3)}%
    \pgfmathsetmacro{\qcBy}{\qcAy+(#4)}%
    \qclink{\qcAx}{\qcAy}{\qcAx}{\qcBy}%
    \qclink{\qcAx}{\qcAy}{\qcBx}{\qcAy}%
    \qclink{\qcAx}{\qcBy}{\qcBx}{\qcBy}%
    \qclink{\qcBx}{\qcAy}{\qcBx}{\qcBy}%
  }%
}
\newcommand{\qcxarmnodes}[5]{%
  \foreach \j in {1,...,#5}{%
    \pgfmathsetmacro{\qcAx}{#1+(#3)*(\j-1)}%
    \pgfmathsetmacro{\qcAy}{#2+(#4)*(\j-1)}%
    \pgfmathsetmacro{\qcBx}{\qcAx+(#3)}%
    \pgfmathsetmacro{\qcBy}{\qcAy+(#4)}%
    \node[qubit] at (\qcAx,\qcBy) {};%
    \node[qubit] at (\qcBx,\qcAy) {};%
    \node[qubit] at (\qcBx,\qcBy) {};%
  }%
}

\newcommand{\qcxlayout}[1]{%
  \qcgridlines{3}{3}{1}{1}%
  \qcxarmlines{3}{3}{1}{1}{#1}%
  \qcxarmlines{1}{3}{-1}{1}{#1}%
  \qcxarmlines{1}{1}{-1}{-1}{#1}%
  \qcxarmlines{3}{1}{1}{-1}{#1}%
  \qcgridnodes{3}{3}{1}{1}%
  \qcxarmnodes{3}{3}{1}{1}{#1}%
  \qcxarmnodes{1}{3}{-1}{1}{#1}%
  \qcxarmnodes{1}{1}{-1}{-1}{#1}%
  \qcxarmnodes{3}{1}{1}{-1}{#1}%
}

\usepackage{booktabs}
\usepackage{siunitx}
\usepackage{tabularx}
\usepackage{adjustbox}
\usepackage{multirow}

\DeclareSIUnit{\eur}{\text{€}}

\usepackage{svg}

\usepackage{tikz}
\usepackage{subfigure}
\usepackage{graphicx}
\usepackage{circuitikz}
\usetikzlibrary{positioning}
\usetikzlibrary{arrows.meta}
\usetikzlibrary{fit,backgrounds}
\usetikzlibrary{calc}
\usetikzlibrary{patterns}
\usetikzlibrary{shapes.geometric}
\usetikzlibrary{decorations.markings,decorations.pathreplacing}

\usepackage[hidelinks]{hyperref}
\usepackage{cleveref}
\crefname{figure}{Fig.}{Figs.}
\crefname{section}{Section}{Sections}

\usepackage[vlined,ruled,linesnumbered,longend]{algorithm2e}

\usepackage{amsthm}
\newtheoremstyle{journalstyle}%
  {6pt}{6pt}%
  {\itshape}%
  {}%
  {\bfseries\sffamily}%
  {.}%
  {.5em}%
  {}%
\theoremstyle{journalstyle}

\newtheorem{abstraction}{Abstraction}
\newtheoremstyle{abstraction}
{1pt}   %
{1pt}   %
{\itshape} %
{}      %
{\bfseries} %
{.}     %
{ }     %
{}      %

\newsavebox{\measurebox}

\newcommand{\etal}{\textit{et~al}.\xspace}
\def\quad{\hskip1em\relax}
\usepackage{bm}

\newtheorem{example}{Example}
\newtheoremstyle{example}
{1pt}   %
{1pt}   %
{\itshape} %
{}      %
{\bfseries} %
{.}     %
{ }     %
{}      %

\renewcommand{\baselinestretch}{0.945}

\usepackage{titlesec}
\titlespacing*{\subsection}{0pt}{6pt}{4pt}     %
\titlespacing*{\subsubsection}{0pt}{4pt}{2pt}

\title{Physical Design Automation for\\ Planar Superconducting Quantum Chips}

\author{
    \IEEEauthorblockN{Michael Feldmeier, 
                      Marcel Walter,
                      Gerhard B. P. Huber,
                      Anirban Bhattacharjee,
                      Stefan Filipp,
                      and Robert Wille}

}

\begin{document}

\urlstyle{tt}

\maketitle

\begin{abstract}
Superconducting circuits have emerged as one of the most promising and scalable 
platforms for quantum computing, with substantial industrial adoption driving qubit counts steadily upward. Yet the layout of the corresponding chips is still primarily performed manually, consuming multiple days or weeks of expert effort, even for moderately sized designs. Existing automation approaches sidestep this bottleneck by simplifying the underlying problem and relaxing physical constraints. Therefore, they fall short of fabrication-ready results. To overcome this scalability wall without such compromises, we first introduce a formal abstraction that translates physical device characteristics into a rigorous geometric problem formulation. This abstraction bridges physical realization and design automation. Building upon this abstraction, we then propose a three-stage design automation flow comprising geometry-aware global partitioning, ILP-based port assignment, and a hierarchical routing pipeline. Evaluations conducted by our interdisciplinary team of design automation researchers and superconducting hardware experts confirm that the resulting flow produces complete, manufacturing-ready layouts that comply with the physical design rules at the push of a button. That is, it operates fully automatically within seconds to a few minutes, and solves complex instances up to 25× faster than the state of the art. All methods are released as the fully open-source tool \texttt{mqt-scpd} as part of the \emph{Munich Quantum Toolkit}, providing the first readily usable, end-to-end design flow for planar superconducting quantum chips that explicitly addresses the underlying physical characteristics and requirements.
\end{abstract}

\vspace{-4mm}
\section{Introduction}\label{sec:intro}
\vspace{-0.5mm}
\IEEEPARstart{C}{onventional} semiconductor technology is approaching fundamental physical boundaries. The continued shrinking of transistors has largely stagnated, marking the practical end of Moore's law and, with it, the era in which performance gains followed almost automatically from process scaling. At the same time, a broad class of computationally hard problems---ranging from quantum chemistry and materials science to combinatorial optimization---remains intractable on classical machines, motivating fundamentally new computing methodologies. \emph{Quantum Computing}~(QC) is among the most prominent of them, exploiting quantum mechanical effects such as superposition and entanglement to overcome these classical limitations~\cite{nielsen2000quantum}.

Several hardware technologies are competing to realize such machines, including trapped ions \cite{bruzewicz2019trapped,schoenberger2024shuttling}, neutral atoms \cite{bluvstein2024logical,schmid2024computational}, photonic systems, and superconducting circuits~\cite{ladd2010quantum,kunasaikaran2024framework}. Among these, \emph{Superconducting Quantum Computing}~(SCQC) has emerged as one of the most promising and scalable platforms, propelled by substantial industrial adoption by companies such as IBM, Google, IQM, and Rigetti~\cite{arute2019quantum,kim2023evidence,google2023suppressing}. Because superconducting chips are lithographically patterned from thin films, their design follows a process and flow closely resembling those established for conventional CMOS circuits: an abstract architecture must be translated into a physically fabricable layout that satisfies a strict set of design rules. The design automation flows developed for conventional computing, however, cannot be adopted one-to-one.

Several design tools for superconducting quantum chips exist, including the open-source CAD framework \emph{Qiskit Metal}~\cite{qiskitmetal}, \emph{EDA-Q}~\cite{edaQ} and its extension by Yu~\emph{et~al.}~\cite{consaware}, and the framework by Yang~\emph{et~al.}~\cite{yang2025scalable}. In practice, however, they provide little relief. Qiskit Metal, although widely used, lacks fully automatic design methodologies. The remaining approaches do offer automatic port assignment and routing, but suffer from other critical limitations. They lack support for essential physical features such as flux lines for tunable couplers or fixed-length resonator routing, and they rely on a simplified port-assignment scheme that applies only to regularly oriented qubit architectures. Bound to these rigid assumptions, these frameworks fail when confronted with the irregular topologies and complex geometries of modern quantum processors.

As a result, most SCQC chip design teams still perform the majority of layout 
steps manually. This is not only time-consuming and expensive but also 
error-prone: even for comparatively small designs with \SIrange{20}{30}{} qubits, the complete design cycle of layout, validation, and simulation takes teams of multiple people days to weeks, and every architectural change requires the manual layout effort to be repeated. As the qubit count increases, complexity increases accordingly. Manual layout drafting, therefore, constitutes a severe scalability wall for the next generation of quantum processors.

In this work, we close this gap---the absence of a design automation flow that handles irregular qubit topologies and the full set of physical features required by modern quantum processors---by introducing a comprehensive flow for automatic design of superconducting quantum chips. Our first contribution is a formal abstraction that translates physical device characteristics into a rigorous geometric problem formulation, thereby bridging the gap between the physical realization of superconducting quantum chips and the abstraction level required for design automation. To this end, each relevant aspect of superconducting quantum chip design (i.e., general physical background, layout, routing, and port assignment) is formalized. On that basis, we then propose a design automation flow that respects physical characteristics as well as requirements and comprises the following three steps:
\begin{enumerate}
    \item \emph{Global Partitioning}: analyzes the layout geometry to identify structural bottlenecks and open spaces, and divides the chip into regions with controlled routing capacities to prevent future congestion.
    \item \emph{Port Assignment Computation}: computes an optimized port assignment using \emph{Integer Linear Programming}~(ILP), thereby minimizing bridge usage and expected wire length.
    \item \emph{Routing}: realizes the physical wire routes through a hierarchical, four-stage routing pipeline that incrementally enforces all design rules.
\end{enumerate}
Evaluations  conducted by our interdisciplinary team of design automation researchers and superconducting hardware experts confirm that the resulting flow yields complete, manufacturing-ready chip layouts compliant with the physical design rules at the push of a button, i.e., fully automatically within seconds to a few minutes. Moreover, the proposed flow solves complex layouts up to 25× faster than the state of the art~\cite{yang2025scalable, consaware}.

Overall, this provides the first design flow for SCQC that is aware of the 
underlying physical characteristics and requirements. Moreover, in the spirit of open science and to allow researchers and hardware developers to adopt and extend the framework, an open-source implementation of the resulting methods has been made available as part of the \emph{Munich Quantum Toolkit}~(MQT,~\cite{Burgholzer2026}) and is released as the fully open-source tool \texttt{mqt-scpd}.\footnote{\href{https://github.com/munich-quantum-toolkit/scpd}{github.com/munich-quantum-toolkit/scpd}} Thereby, the work not only brings SCQC developers a level of design automation support comparable to what has been available in conventional circuit design for decades, but also provides an open-source foundation for future work. 

The remainder of this manuscript is structured as follows: \cref{sec:prelim} 
establishes the foundation for design automation by reviewing the relevant device physics and distilling it into a formal abstraction of components, constraints, and objectives. Building on this abstraction, \cref{sec:desau} identifies shortcomings in related work and proposes a physically rigid design-automation flow. \cref{sec:experiments} evaluates the resulting flow on planar SCQC layouts, confirming that the generated designs meet the requirements of established fabrication processes. Finally \cref{sec:conclusion} concludes the paper and outlines directions for future work.

\vspace{-1.5mm}
\section{From Physics to Abstraction}\label{sec:prelim}
\vspace{-1.0mm}
    While conventional physical design automation has matured over decades and routinely handles circuits with millions of components, the layout of superconducting quantum chips is still performed manually, requiring days or weeks even for small designs. This work seeks to transfer this established expertise to the design of superconducting quantum chips. Since the corresponding flows cannot be adopted one-to-one, however, this requires an abstraction that accurately captures the underlying physical principles while providing a formal framework for developing design automation techniques. This section introduces such an abstraction. To this end, for each relevant aspect of superconducting quantum chip design (i.e., general physical background, layout, routing, and port assignment), we first briefly review the corresponding physical principles and subsequently derive a formal abstraction that captures the relevant components $\mathcal{C}$, constraints $\mathcal{R}$, and optimization objectives $\mathcal{O}$. The component set $\mathcal{C}$ is further partitioned into layout components $\mathcal{C}_{\text{layout}}$ and routing components $\mathcal{C}_{\text{route}}$. Likewise, the constraint set $\mathcal{R}$ is partitioned into layout, routing, and assignment constraints, i.e., $\mathcal{R} = \mathcal{R}_{\text{layout}} \cup \mathcal{R}_{\text{route}} \cup \mathcal{R}_{\text{assign}}$, while the objective set $\mathcal{O}$ remains unified. 
    In this way, we establish a bridge between the physical realization of superconducting quantum chips and the abstraction level required for algorithm and tool development.
    
    The discussion of superconducting device physics is intentionally kept brief and primarily serves to motivate and contextualize the proposed abstractions. Readers interested in a more detailed treatment are referred to the cited literature. The resulting abstractions are summarized in dedicated text environments. Together, they provide a reusable foundation for the design automation method later proposed in \cref{sec:desau}.
    \vspace{-1.0mm}
    \subsection{Physical Background}
    \vspace{-1.0mm}
          Superconducting quantum circuits are typically patterned lithographically from thin films (commonly Nb, Al, or NbN) on low-loss substrates such as high-resistivity silicon or sapphire~\cite{krantz2019quantum}. Their \mbox{non-linearity} originates from \emph{Josephson Junctions}~(JJs), such that the lowest levels of the anharmonic circuit serve as a qubit, typically at \SIrange{4}{8}{\giga\hertz}, with thermal occupation suppressed inside a dilution refrigerator at \SIrange{10}{20}{\milli\kelvin}~\cite{koch2007charge, blais2021circuit}. Control and readout use microwave pulses routed on-chip via \emph{Coplanar Waveguides}~(CPWs), which are usually designed for a characteristic impedance of \SI{50}{\ohm} to match the external wiring and avoid reflections. Each qubit is commonly addressed by a capacitively coupled drive line, while readout uses resonators coupled to a common feedline. Frequency-tunable qubits and couplers additionally require a flux line, inductively coupled to a \emph{Superconducting Quantum Interference Device}~(SQUID), a superconducting loop interrupted by two Josephson junctions whose enclosed magnetic flux sets the critical current and therefore the transition frequency. Routing these lines splits the ground plane into isolated regions supporting spurious slotline resonances~\cite{ponchak2005excitation}, which are typically tied to a common ground by airbridges or bond wires~\cite{chen2014fabrication}. Airbridges are small superconducting arches that connect the separated ground regions over an intervening line, while bond wires are thin wires connecting the chip to its package. Bends are rounded to limit reflections, and neighboring lines separated to limit crosstalk~\cite{song2026microwave}. These constraints tighten as the qubit count grows and the number of control lines increases accordingly.
 \vspace{-1.0mm}       
\subsection{SCQC Layouts}
\vspace{-1.0mm}
\begin{figure}[!t]
    \centering

    \subfigure[\scriptsize Qubit Geom. A.]{%
        \makebox[0.11\textwidth][c]{%
            \includegraphics[width=0.082\textwidth]
            {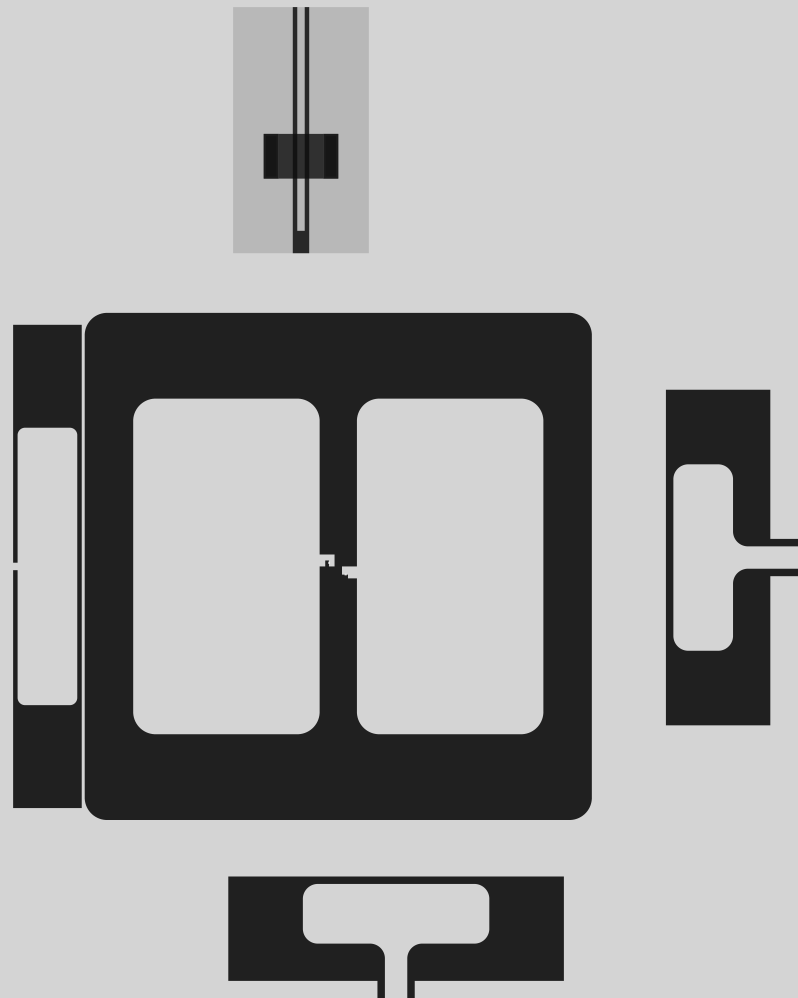}
        }%
        \label{fig:qubit_a}
    }
    \subfigure[\scriptsize Qubit Geom. B.]{%
        \makebox[0.105\textwidth][c]{%
            \includegraphics[width=0.095\textwidth]
            {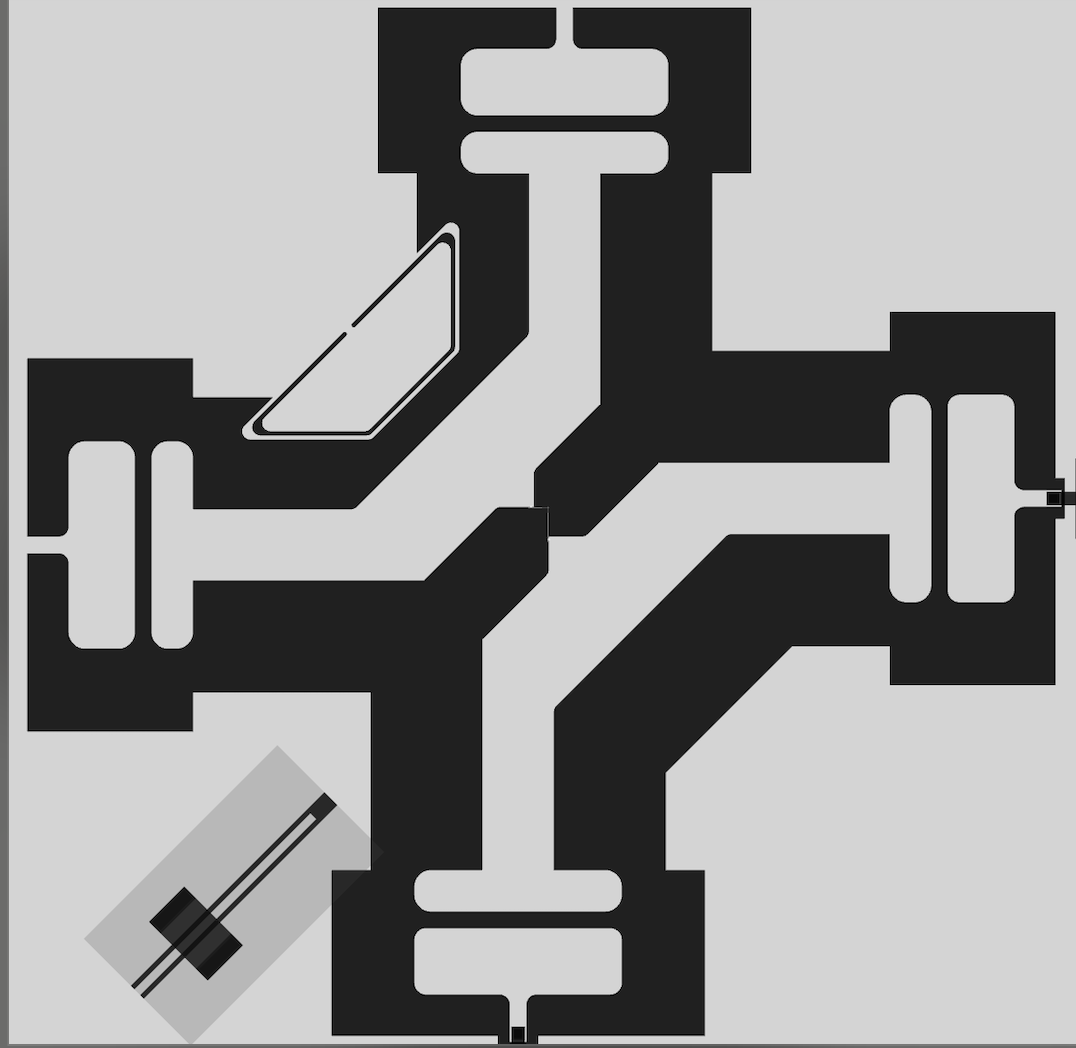}
        }%
        \label{fig:qubit_b}
    }
    \subfigure[\scriptsize Coupler.]{%
        \makebox[0.14\textwidth][c]{%
            \resizebox{.15\textwidth}{!}{\hbox{\input{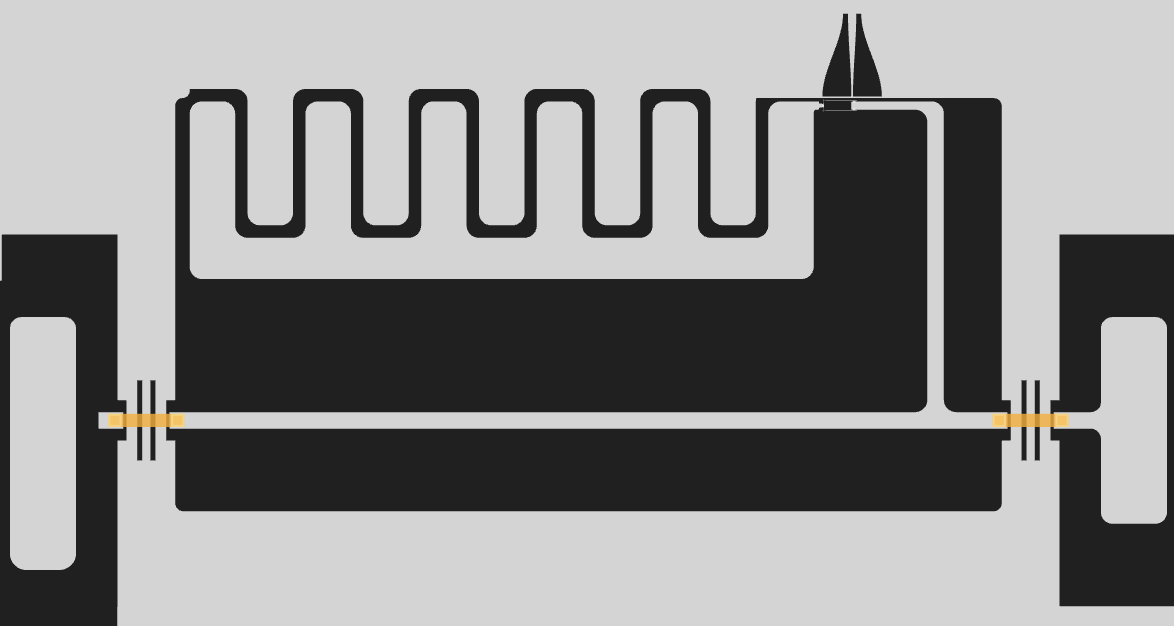}}}%
        }%
        \label{fig:coupler}
    }
    \hspace{-3mm}
    \subfigure[\scriptsize Launcher.]{%
        \makebox[0.08\textwidth][c]{%
            \includegraphics[width=0.06\textwidth]
            {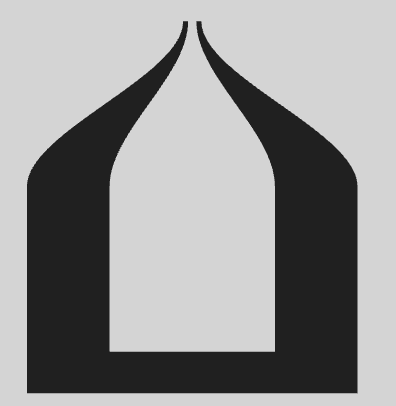}
        }%
        \label{fig:launcher}
    }

    \vspace{-1.7mm}
    \setcounter{subfigure}{4}
    \subfigure[Complete SCQC Layout Overview.]{%
        \scalebox{0.8}{
        \input{figures/fig01_layout_4q}}\ \ \label{fig:layout_overview}%
    }
    \vspace{-2.5mm}
    \caption{SCQC layout architecture: Detailed view of routing components showing (a), (b) two floating, fixed-frequency qubits of different geometry, where (a) supports coupling to two neighboring qubits and (b) to four, (c) a tunable coupler structure, and (d) peripheral launcher interfaces. These components are integrated into the (e) complete exemplar 4-qubit layout overview.}
    \label{fig:layout_architecture}
\end{figure}
\vspace{-1.0mm}
  While a \emph{Superconducting Quantum Chip} (SCQC) comprises numerous hardware structures, this work focuses on the three primary ones, namely \emph{qubits}, \emph{couplers}, and \emph{launchers}. Additional routing components are covered in the following \cref{subsec:abstraction_routing}. We introduce physically grounded abstractions for all three main components, first outlining their physical background and subsequently establishing their formal abstraction. In doing so, we define the layout component set $\mathcal{C}_{\text{layout}} \subseteq \mathcal{C}$ and the corresponding layout constraints $\mathcal{R}_{\text{layout}} \subseteq \mathcal{R}$.
    We begin our formalization using the running example of the layout shown in \cref{fig:layout_architecture}, starting with the superconducting qubits, followed by the couplers and the launchers.
    
    Physically, superconducting qubits are fragile artificial atoms (e.g., transmons~\cite{koch2007charge}) operating at millikelvin temperatures, whose delicate coherence requires strict shielding from thermal noise and electromagnetic crosstalk from neighboring signal lines~\cite{krantz2019quantum, song2026microwave}. Two independent design properties characterize them: A qubit is either \emph{fixed-frequency} or \emph{tunable}, where tunable qubits additionally require a flux line; also, the qubit is either \emph{grounded}, with a single island connected to ground through the JJ, or \emph{floating}, with two islands connected via the junction~\cite{blais2021circuit}. Independent of these properties, qubits can exhibit different physical footprints and layout structures, as illustrated by the two qubits in \cref{fig:qubit_a} and \cref{fig:qubit_b}, which are both fixed-frequency and floating, each comprising a main qubit structure and an additional open CPW stub. The two footprints offer different geometries, each better suited to a different architecture: \cref{fig:qubit_a} for 2-way and \cref{fig:qubit_b} for 4-way connectivity, with ports arranged to ensure reachability in their respective coupling schemes. Based on these physical requirements, we establish the following geometric abstraction:
    \vspace{-2.0mm}    
    \begin{abstraction}[Superconducting Qubits]
        A qubit is modeled as a layout component $C_q = (\mathcal{P}_q, \mathcal{T}_q) \in \mathcal{C}_{\text{layout}}$. 
        The geometry is defined by a set of obstacle polygons 
        $\mathcal{P}_q = \{P_{q_1}, P_{q_2}, \dots, P_{q_n}\}$, where each polygon 
        $P_{q_j} = (\mathbf{v}_{q_0}, \mathbf{v}_{q_1}, \dots, \mathbf{v}_{q_{m-1}})$ 
        with $\mathbf{v}_{q_i} \in \mathbb{R}^2$ is an ordered sequence of points.     
        All polygons in $\mathcal{P}_q$ are classified as routing obstacles according to the constraint 
        $R_o \in \mathcal{R}_{\text{layout}}$, defining strict keep-out regions that must not be crossed by signal wires.
        The interfaces are defined by the terminal set 
        $\mathcal{T}_q = \{t_{\text{read}}, t_{\text{ctrl}}\}$, where each terminal 
        $t = (\mathbf{p}, \phi) \in \mathbb{R}^2 \times [0, 2\pi)$ represents the precise coordinate location 
        $\mathbf{p}$ and the orientation angle $\phi$ of the respective interface pin.
        Spatially, the placement of qubits is governed by the array constraint 
        $R_{\text{arr}} \in \mathcal{R}_{\text{layout}}$, enforcing a regular grid-based arrangement.
    \end{abstraction}

    \vspace{-0.5mm}
    Following the computational cores, we define the coupling mechanisms that 
    facilitate inter-qubit communication and entanglement. From a hardware perspective, couplers are themselves superconducting qubits, again realized as floating designs with two islands connected via a JJ, but placed between neighboring qubits to mediate their interaction~\cite{krantz2019quantum, blais2021circuit}. As shown in \cref{fig:layout_overview}, adjacent qubits are connected through such elements, enabling two-qubit operations. Just as for qubits, couplers may be fixed-frequency or tunable: their frequency is either fixed or, as for the tunable couplers considered in this work, dynamically adjusted through an externally applied magnetic flux, which in turn tunes the effective qubit-qubit coupling relevant for the gates. This tunability allows the qubit-qubit interaction to be switched on and off on demand, suppressing residual coupling in the idle state and thereby reducing gate errors. As shown in \cref{fig:coupler}, the tunable coupler consists of a central island interfacing with the neighboring qubits and providing the connectivity for flux-based modulation of the coupling strength. In densely populated layouts, additional airbridges route across the coupler while preserving connectivity~\cite{chen2014fabrication,smlq-r7x6}, highlighted in red in \cref{fig:coupler} as the elevated structures on the coupler's left and right sides. To capture these components within a layout synthesis  framework, we translate the coupler physics into the following geometric formulation:
     \vspace{-1.5mm}
    \begin{abstraction}[Couplers]
    A coupler is modeled as a layout component $C_c = (\mathcal{P}_c, \mathcal{T}_c) \in \mathcal{C}_{\text{layout}}$.
    The geometry is defined by a set of polygons $\mathcal{P}_c = \{P_1, P_2, \dots, P_n\}$, which are classified as routing obstacles under the constraint 
    $R_o \in \mathcal{R}_{\text{layout}}$, forming a strict barrier that signal wires must not cross.
    The interfaces are defined by the terminal set $\mathcal{T}_c = \{t_{\text{flux}}\}$, where the terminal 
    $t_{\text{flux}} = (\mathbf{p}, \phi) \in \mathbb{R}^2 \times [0, 2\pi)$ represents the precise coordinate location $\mathbf{p}$ and the orientation angle $\phi$ of the interface pin. 
    Spatially, the placement of the component is governed by the adjacency constraint $R_{\text{adj}} \in \mathcal{R}_{\text{layout}}$, enforcing that each coupler is located between a pair of qubits.
    In addition, depending on the chip topology, couplers may provide auxiliary bridge terminals that serve as connection points for routing elements crossing the component, thereby extending the terminal structure.
    \end{abstraction}
    \vspace{-1.0mm}
    
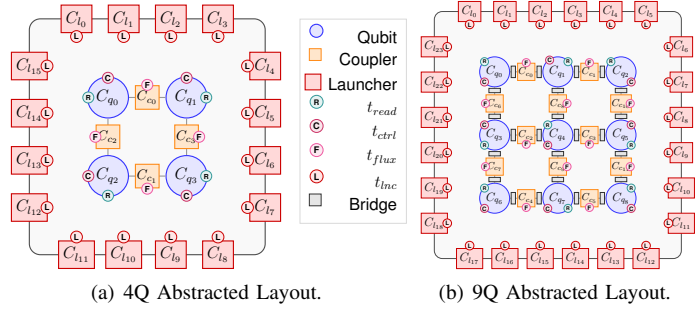
\begin{figure}[!t]
    \centering
        
    \subfigure[4Q Abstracted Layout.]{%
        \label{fig:abstraction_4q}
        \scalebox{0.52}{\hbox{\begin{tikzpicture}[
        qubit/.style={circle, draw=blue!80, fill=blue!10, thick, minimum size=9mm, font=\sffamily\large\bfseries},
        coupling line/.style={thick, draw=gray!80},
       coupler/.style={
            rectangle, 
            draw=orange!90, 
            fill=orange!20, 
            minimum width=6mm, 
            minimum height=6mm, 
            inner sep=1pt,
            font=\sffamily\normalsize\selectfont
        },
        launcher/.style={rectangle, draw=red!80!black, fill=red!15, thick, minimum size=7mm, font=\sffamily\large\bfseries},
        readout port/.style={circle, draw=teal!80, fill=teal!10, minimum size=2.8mm, font=\sffamily\bfseries\fontsize{5}{5}\selectfont, inner sep=0pt},
        control port/.style={circle, draw=purple!80, fill=purple!10, minimum size=2.8mm, font=\sffamily\bfseries\fontsize{6}{6}\selectfont, inner sep=0pt},
        flux port/.style={
            circle,
            draw=magenta!80,
            fill=magenta!10,
            minimum size=2.8mm,
            font=\sffamily\bfseries\fontsize{6}{6}\selectfont,
            inner sep=0pt
        },
        launcher port/.style={
            circle,
            draw=red!80!black,
            fill=red!10,
            minimum size=3mm,
            font=\sffamily\bfseries\fontsize{6}{6}\selectfont,
            inner sep=0pt
        },
        launcher qubit line/.style={thick, draw=cyan!80, dashed,font=\normalsize},
        launcher coupler line/.style={thick, draw=magenta!80, dotted},
        resonator line/.style={very thick, draw=green!60!black, line cap=round},
        bridge terminal/.style={
            rectangle,
            draw=black!80,
            fill=gray!20,
            minimum width=4mm,
            minimum height=1.5mm,
            inner sep=0pt
        },
    ]

    \def\rows{2}     %
    \def\cols{2}     %
    \def\spacing{2.0} %
    \def\k{4}        %
    \def\pad{1.0}    %
    \def\extdist{0.5} %
    \def\fportdist{0.32} %
    \def\lportdist{0.45}

    \pgfmathtruncatemacro{\maxc}{\cols-1}
    \pgfmathtruncatemacro{\maxr}{\rows-1}

    \draw[draw=black!70, fill=gray!5, thick, rounded corners=8pt] 
        ({0.5*\spacing - \pad}, {-0.5*\spacing + \pad}) rectangle ({(\cols+0.5)*\spacing + \pad}, {-(\rows+0.5)*\spacing - \pad});

    \foreach \r in {1,...,\rows} {
        \foreach \c in {1,...,\cols} {
            \pgfmathtruncatemacro{\qidx}{(\r-1)*\cols + \c - 1}
            
            \node[qubit] (Q-\qidx) at (\c*\spacing, -\r*\spacing) {$C_{q_{\qidx}}$};
            
            \ifcase\qidx
                \def\rAngle{180} \def\cAngle{90}
            \or
                \def\rAngle{0}  \def\cAngle{90}
            \or
                \def\rAngle{270} \def\cAngle{180}
            \or
                \def\rAngle{0} \def\cAngle{270}
            \else
                \def\rAngle{90}  \def\cAngle{0}
            \fi
            
            \node[readout port] (R-\qidx) at ([shift=({\rAngle}:0.5)] Q-\qidx) {R};
            \node[control port] (C-\qidx) at ([shift=({\cAngle}:0.5)] Q-\qidx) {C};
        }
    }

\foreach \r in {1,...,\rows} {
    \foreach \c in {1,...,\maxc} { 
        \pgfmathtruncatemacro{\qidxA}{(\r-1)*\cols + \c - 1}
        \pgfmathtruncatemacro{\qidxB}{(\r-1)*\cols + \c}
        \pgfmathtruncatemacro{\cidx}{(\r-1)*\maxc + \c - 1}

        \ifcase\cidx
            \def\fAngle{90}
        \or
            \def\fAngle{270}
        \else
            \def\fAngle{90}
        \fi

        \draw[coupling line] (Q-\qidxA) -- (Q-\qidxB) 
            node[midway, coupler] (C-H-\r-\c) {$C_{c_{\cidx}}$};

        \node[flux port] (F-\cidx) at ([shift=({\fAngle}:\fportdist)] C-H-\r-\c) {F};
    }
}

\foreach \c in {1,...,\cols} {
    \foreach \r in {1,...,\maxr} {
        \pgfmathtruncatemacro{\qidxA}{(\r-1)*\cols + \c - 1}
        \pgfmathtruncatemacro{\qidxB}{\r*\cols + \c - 1}
        \pgfmathtruncatemacro{\cidx}{\rows*(\cols-1) + (\c-1)*\maxr + \r - 1}

        \ifcase\cidx
            \def\fAngle{90}
        \or
            \def\fAngle{270}
        \or
            \def\fAngle{180}
        \or
            \def\fAngle{0}
        \else
            \def\fAngle{90}
        \fi

        \draw[coupling line] (Q-\qidxA) -- (Q-\qidxB) 
            node[midway, coupler] (C-V-\r-\c) {$C_{c_{\cidx}}$};

        \node[flux port] (F-\cidx) at ([shift=({\fAngle}:\fportdist)] C-V-\r-\c) {F};
    }
}

\foreach \i in {1,...,\k} {
    \pgfmathtruncatemacro{\idxTop}{\i - 1}
    \pgfmathtruncatemacro{\idxRight}{\k + \i - 1}
    \pgfmathtruncatemacro{\idxBottom}{2*\k + \i - 1}
    \pgfmathtruncatemacro{\idxLeft}{3*\k + \i - 1}

    \pgfmathtruncatemacro{\iRev}{\k - \i + 1}

    \node[launcher] (L-\idxTop) at ({(0.5*\spacing - \pad) + \i*(\cols*\spacing + 2*\pad)/(\k+1)}, {-0.5*\spacing + \pad}) {$C_{l_{\idxTop}}$};
    \coordinate (L-\idxTop_ext) at ([yshift=-\extdist cm] L-\idxTop);
    \node[launcher port] (LP-\idxTop) at ([shift=(270:\lportdist)] L-\idxTop) {L};

    \node[launcher] (L-\idxRight) at ({(\cols+0.5)*\spacing + \pad}, {(-0.5*\spacing + \pad) - \i*(\rows*\spacing + 2*\pad)/(\k+1)}) {$C_{l_{\idxRight}}$};
    \coordinate (L-\idxRight_ext) at ([xshift=-\extdist cm] L-\idxRight);
    \node[launcher port] (LP-\idxRight) at ([shift=(180:\lportdist)] L-\idxRight) {L};

    \node[launcher] (L-\idxBottom) at ({(0.5*\spacing - \pad) + \iRev*(\cols*\spacing + 2*\pad)/(\k+1)}, {-(\rows+0.5)*\spacing - \pad}) {$C_{l_{\idxBottom}}$};
    \coordinate (L-\idxBottom_ext) at ([yshift=\extdist cm] L-\idxBottom);
    \node[launcher port] (LP-\idxBottom) at ([shift=(90:\lportdist)] L-\idxBottom) {L};

    \node[launcher] (L-\idxLeft) at ({0.5*\spacing - \pad}, {(-0.5*\spacing + \pad) - \iRev*(\rows*\spacing + 2*\pad)/(\k+1)}) {$C_{l_{\idxLeft}}$};
    \coordinate (L-\idxLeft_ext) at ([xshift=\extdist cm] L-\idxLeft);
    \node[launcher port] (LP-\idxLeft) at ([shift=(0:\lportdist)] L-\idxLeft) {L};
}

        \matrix [draw=gray!30, fill=white,  anchor=north east, xshift=0.1cm] 
            at ({0.5*\spacing + \pad+7.5}, {-0.5*\spacing + \pad}) {
            \node[qubit, minimum size=2mm, scale=1.0] {}; & \node[font=\sffamily\large\selectfont] {Qubit}; \\
            \node[coupler, minimum size=3mm] {}; & \node[font=\sffamily\large\selectfont] {Coupler}; \\
            \node[launcher, minimum size=4mm] {}; & \node[font=\sffamily\large\selectfont] {Launcher}; \\
            \node[readout port, minimum size=3mm, scale=1.1] {R}; & \node[font=\sffamily\large\selectfont] {$t_{read}$}; \\
            \node[control port, minimum size=3mm, scale=1.1] {C}; & \node[font=\sffamily\large\selectfont] {$t_{ctrl}$}; \\
            \node[flux port, minimum size=3mm, scale=1.1] {F}; & \node[font=\sffamily\large\selectfont] {$t_{flux}$}; \\
            \node[launcher port, minimum size=3mm, scale=1.1] {L}; & \node[font=\sffamily\large\selectfont] {$t_{lnc}$}; \\
            \node[bridge terminal, minimum size=3mm, scale=1.0] {}; & \node[font=\sffamily\large\selectfont] {Bridge}; \\
        };

\end{tikzpicture}}}%
    }%
    \subfigure[9Q Abstracted Layout.]{%
        \label{fig:abstraction_9q}
        \scalebox{0.38}{\hbox{\begin{tikzpicture}[
        qubit/.style={circle, draw=blue!80, fill=blue!10, thick, minimum size=9mm, font=\sffamily\large\bfseries},
        coupling line/.style={thick, draw=gray!80},
       coupler/.style={
            rectangle, 
            draw=orange!90, 
            fill=orange!20, 
            minimum width=6mm, 
            minimum height=6mm, 
            inner sep=1pt,
            font=\sffamily\normalsize\selectfont
        },
        launcher/.style={rectangle, draw=red!80!black, fill=red!15, thick, minimum size=7mm, font=\sffamily\large\bfseries},
        readout port/.style={circle, draw=teal!80, fill=teal!10, minimum size=2.8mm, font=\sffamily\bfseries\fontsize{5}{5}\selectfont, inner sep=0pt},
        control port/.style={circle, draw=purple!80, fill=purple!10, minimum size=2.8mm, font=\sffamily\bfseries\fontsize{6}{6}\selectfont, inner sep=0pt},
        flux port/.style={
            circle,
            draw=magenta!80,
            fill=magenta!10,
            minimum size=2.8mm,
            font=\sffamily\bfseries\fontsize{6}{6}\selectfont,
            inner sep=0pt
        },
        launcher port/.style={
            circle,
            draw=red!80!black,
            fill=red!10,
            minimum size=3mm,
            font=\sffamily\bfseries\fontsize{6}{6}\selectfont,
            inner sep=0pt
        },
        launcher qubit line/.style={thick, draw=cyan!80, dashed,font=\normalsize},
        launcher coupler line/.style={thick, draw=magenta!80, dotted},
        resonator line/.style={very thick, draw=green!60!black, line cap=round},
        bridge terminal/.style={
            rectangle,
            draw=black!80,
            fill=gray!20,
            minimum width=4mm,
            minimum height=1.5mm,
            inner sep=0pt
        },
    bridge terminal/.style={
    rectangle,
    draw=black!80,
    fill=gray!20,
    minimum width=4mm,
    minimum height=1.5mm,
    inner sep=0pt
},
]

\def\rows{3}
\def\cols{3}
\def\spacing{2.2}
\def\pad{1.0}
\def\k{6}
\def\fportdist{0.32}
\def\lportdist{0.45}
\def\extdist{0.0}

\pgfmathtruncatemacro{\maxc}{\cols-1}
\pgfmathtruncatemacro{\maxr}{\rows-1}

\draw[draw=black!70, fill=gray!5, thick, rounded corners=8pt]
({0.5*\spacing - \pad}, {-0.5*\spacing + \pad})
rectangle ({(\cols+0.5)*\spacing + \pad}, {-(\rows+0.5)*\spacing - \pad});

\foreach \r in {1,...,\rows} {
    \foreach \c in {1,...,\cols} {
        \pgfmathtruncatemacro{\qidx}{(\r-1)*\cols + \c - 1}
        \node[qubit] (Q-\qidx) at (\c*\spacing, -\r*\spacing) {$C_{q_{\qidx}}$};

        \ifcase\qidx
            \def\rAngle{135}   %
            \def\cAngle{225}    %
        \or
            \def\rAngle{45}     %
            \def\cAngle{125}
        \or
            \def\rAngle{135}
            \def\cAngle{315}
        \or
            \def\rAngle{225}
            \def\cAngle{135}
        \or
            \def\rAngle{135}
            \def\cAngle{225}
        \or
            \def\rAngle{315}
            \def\cAngle{45}

        \or
            \def\rAngle{135}
            \def\cAngle{225}
        \or
            \def\rAngle{315}
            \def\cAngle{225}
         \or
            \def\rAngle{45}
            \def\cAngle{315}
        \fi
        \node[readout port] at ([shift=({\rAngle}:0.5)] Q-\qidx) {R};
        \node[control port] at ([shift=({\cAngle}:0.5)] Q-\qidx) {C};
    }
}

\foreach \r in {1,...,\rows} {
    \foreach \c in {1,...,\maxc} {
        \pgfmathtruncatemacro{\qidxA}{(\r-1)*\cols + \c - 1}
        \pgfmathtruncatemacro{\qidxB}{(\r-1)*\cols + \c}
        \pgfmathtruncatemacro{\cidx}{(\r-1)*\maxc + \c - 1}

        \draw[coupling line] (Q-\qidxA) -- (Q-\qidxB)
            node[midway, coupler] (C-\cidx) {$C_{c_{\cidx}}$};

        \ifcase\cidx
        \def\fAngle{90}
        \or
            \def\fAngle{90}
        \or
            \def\fAngle{270}
        \or
            \def\fAngle{270}
        \or
            \def\fAngle{270}
        \or
            \def\fAngle{270}
        \fi
        
        \node[flux port]
            at ([shift=({\fAngle}:\fportdist)] C-\cidx)
            {F};

        \node[bridge terminal, rotate=90] at ([shift=(180:0.45)] C-\cidx) {};
        \node[bridge terminal, rotate=90] at ([shift=(0:0.45)] C-\cidx) {};
    
    }
}

\foreach \c in {1,...,\cols} {
    \foreach \r in {1,...,\maxr} {
        \pgfmathtruncatemacro{\qidxA}{(\r-1)*\cols + \c - 1}
        \pgfmathtruncatemacro{\qidxB}{\r*\cols + \c - 1}
        \pgfmathtruncatemacro{\cidx}{\rows*(\cols-1) + (\c-1)*\maxr + \r - 1}

        \draw[coupling line] (Q-\qidxA) -- (Q-\qidxB)
            node[midway, coupler] (C-\cidx) {$C_{c_{\cidx}}$};

        \ifcase\cidx
        \or\or\or\or\or\or
            \def\fAngle{180}
        \or
            \def\fAngle{180}
        \or
            \def\fAngle{0}
        \or
            \def\fAngle{0}
        \or
            \def\fAngle{0}
        \or
            \def\fAngle{0}
        \fi
        
        \node[flux port]
            at ([shift=({\fAngle}:\fportdist)] C-\cidx)
            {F};

        \node[bridge terminal] at ([shift=(90:0.45)] C-\cidx) {};
        \node[bridge terminal] at ([shift=(270:0.45)] C-\cidx) {};
    }
}

\foreach \i in {1,...,\k} {
    \pgfmathtruncatemacro{\idxTop}{\i - 1}
    \pgfmathtruncatemacro{\idxRight}{\k + \i - 1}
    \pgfmathtruncatemacro{\idxBottom}{2*\k + \i - 1}
    \pgfmathtruncatemacro{\idxLeft}{3*\k + \i - 1}

    \pgfmathtruncatemacro{\iRev}{\k - \i + 1}

    \node[launcher] (L-\idxTop) at ({(0.5*\spacing - \pad) + \i*(\cols*\spacing + 2*\pad)/(\k+1)}, {-0.5*\spacing + \pad}) {$C_{l_{\idxTop}}$};
    \coordinate (L-\idxTop_ext) at ([yshift=-\extdist cm] L-\idxTop);
    \node[launcher port] (LP-\idxTop) at ([shift=(270:\lportdist)] L-\idxTop) {L};

    \node[launcher] (L-\idxRight) at ({(\cols+0.5)*\spacing + \pad}, {(-0.5*\spacing + \pad) - \i*(\rows*\spacing + 2*\pad)/(\k+1)}) {$C_{l_{\idxRight}}$};
    \coordinate (L-\idxRight_ext) at ([xshift=-\extdist cm] L-\idxRight);
    \node[launcher port] (LP-\idxRight) at ([shift=(180:\lportdist)] L-\idxRight) {L};

    \node[launcher] (L-\idxBottom) at ({(0.5*\spacing - \pad) + \iRev*(\cols*\spacing + 2*\pad)/(\k+1)}, {-(\rows+0.5)*\spacing - \pad}) {$C_{l_{\idxBottom}}$};
    \coordinate (L-\idxBottom_ext) at ([yshift=\extdist cm] L-\idxBottom);
    \node[launcher port] (LP-\idxBottom) at ([shift=(90:\lportdist)] L-\idxBottom) {L};

    \node[launcher] (L-\idxLeft) at ({0.5*\spacing - \pad}, {(-0.5*\spacing + \pad) - \iRev*(\rows*\spacing + 2*\pad)/(\k+1)}) {$C_{l_{\idxLeft}}$};
    \coordinate (L-\idxLeft_ext) at ([xshift=\extdist cm] L-\idxLeft);
    \node[launcher port] (LP-\idxLeft) at ([shift=(0:\lportdist)] L-\idxLeft) {L};
}

\end{tikzpicture}}}%
    }%
    \vspace{-2mm}
    \caption{Abstracted representations of the 4-qubit and 9-qubit layouts, derived from their physical chip designs.}
    \label{fig:abstraction}
\end{figure}

    Finally, we review the external interface components that connect the internal 
    quantum circuitry to the macroscopic environment, i.e., the launchers. Physically, launchers are macroscopic wire-bond pads or CPW transitions located at the chip periphery to provide the interface between external control electronics and the on-chip superconducting circuitry~\cite{krantz2019quantum}. As shown in \cref{fig:launcher}, a launcher forms a representative edge-interface structure connecting off-chip and on-chip signal networks, providing access points for microwave control and readout signals and facilitating wire bonding to the package. As illustrated in the overall layout in \cref{fig:layout_overview}, launchers are positioned exclusively along the perimeter of the chip. Due to their comparatively large footprint, they form fixed boundary structures within the layout and must be carefully integrated with the surrounding routing fabric. To reflect this structural constraint in the routing engine, we mathematically 
    formalize these physical interfaces as static components:
    \vspace{-1.5mm}
   \begin{abstraction}[Launchers]
    A launcher is modeled as a layout component $C_l = (\mathcal{P}_l, \mathcal{T}_l) \in \mathcal{C}_{\text{layout}}$.     
    The geometry is defined by a set of polygons $\mathcal{P}_l = \{P_1, \dots, P_n\}$, classified as routing obstacles under 
    $R_o \in \mathcal{R}_{\text{layout}}$. Spatially, the placement of the component is governed by the boundary constraint 
    $R_{\text{bnd}} \in \mathcal{R}_{\text{layout}}$, enforcing that launchers are restricted to the outer boundary of the chip area.  The interfaces are defined by the terminal set $\mathcal{T}_l = \{t_{\text{lnc}}\}$, where 
    $t_{\text{lnc}} = (\mathbf{p}, \phi) \in \mathbb{R}^2 \times [0, 2\pi)$ denotes the position $\mathbf{p}$ and orientation $\phi$ of the interface pin, serving as a fixed source terminal for routing.
    \end{abstraction}
    \vspace{-3.0mm}
    \begin{example}[Chip Layout Abstraction]
    Throughout this work, two representative superconducting quantum chip layouts serve as running examples: the four-qubit (4Q) design shown in \cref{fig:layout_architecture} and a nine-qubit (9Q) design. \Cref{fig:abstraction} illustrates how both are translated into their corresponding design abstractions. For the 4Q design, this yields the component set $\mathcal{C}_{\text{layout}} = \{C_{l_0}, \ldots, C_{l_{15}}, C_{q_0}, \ldots, C_{q_3}, C_{c_0}, \ldots, C_{c_3}\}$, governed by the layout constraint set $\mathcal{R}_{\text{layout}}$; analogous structures apply to the 9Q case. For clarity, all components are depicted as simplified geometric shapes in \cref{fig:abstraction_4q,fig:abstraction_9q}.
\end{example}
    \vspace{-2.0mm}

In summary, the entire layout of an SCQC chip is formalized as a structured composition of three distinct component types, namely qubits, couplers, and launchers, forming the component set $\mathcal{C}_{\text{layout}}$. By unifying these under the shared formal abstraction $C = (\mathcal{P}, \mathcal{T})$, physical layout generation is reduced to a geometric abstraction. In this context, we introduced the layout component set $\mathcal{C}_{\text{layout}}$ and the corresponding constraint set $\mathcal{R}_{\text{layout}}$. As the placement is assumed to be valid and fixed, all layout constraints in $\mathcal{R}_{\text{layout}}$, except for the obstacle constraint $R_o$, are already satisfied. The constraint $R_o$ represents a special case, as it must be explicitly enforced during the routing phase to ensure that signal wires do not intersect with obstacle regions. The behavior of the resulting system is further governed by additional routing constraints $\mathcal{R}_{\text{route}}$ and corresponding optimization objectives $\mathcal{O}$, extending the existing constraints in $\mathcal{R}$.

\begin{figure}[!t]
    \centering

    \refstepcounter{subfigure}\label{fig:min_bending_radius}
\refstepcounter{subfigure}\label{fig:min_straight_length}
\refstepcounter{subfigure}\label{fig:min_clearance}

\refstepcounter{subfigure}\label{fig:wire_launcher_coupler}
\refstepcounter{subfigure}\label{fig:wire_feedline}
\refstepcounter{subfigure}\label{fig:wire_resonator}\label{fig:wire_launcher_qubit}
\refstepcounter{subfigure}\label{fig:airbridge}
\refstepcounter{subfigure}\label{fig:cpw_coupler}
    \vspace{-1mm}
    \subfigure[\scriptsize Routed 4Q Layout.]{%
        \label{fig:layout_4q_routed}
        
        \input{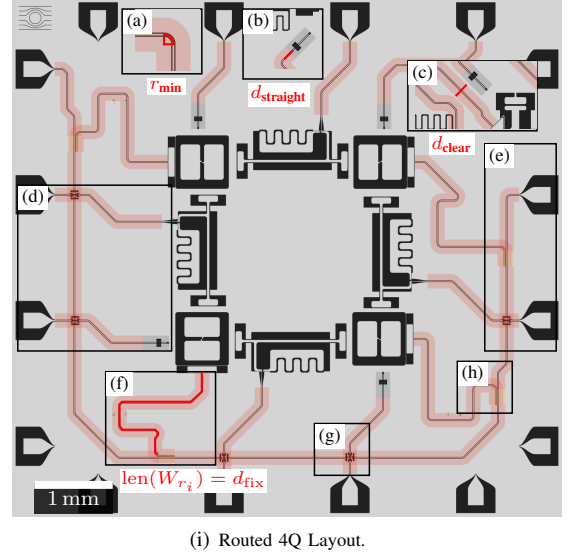}%
    }%

    \vspace{-2.5mm}
    \caption{Overview of routing constraints, wire constructs, and auxiliary routing 
    components used in superconducting quantum chip layouts: (a) minimum bending radius 
    $R_r$, (b) minimum straight length $R_s$, (c) minimum clearance $R_c$, (d) launcher-port 
    wires $W_p$, (e) feedline wires $W_f$, (f) resonator wires $W_r$ , (g) bridges, and 
    (h) CPW couplers. Subfigure (i) a fully routed layout.}
    \label{fig:routing_overview}
\end{figure}
    \vspace{-1.0mm}
   \subsection{Routing}
    \vspace{-1.0mm}
   \label{subsec:abstraction_routing}
    Building upon the geometric abstractions, this section introduces routing constraints $\mathcal{R}_{\text{route}} \subseteq \mathcal{R}$ and objectives $\mathcal{O}$. Additionally, routing constructs $\mathcal{W}$ and routing components $\mathcal{C}_{\text{route}} \subseteq \mathcal{C}$ are defined. Finally, we highlight inner layout regions as routing bottlenecks. To justify the geometric routing rules, we must first examine the underlying physical vulnerabilities:
    
    The qubits are controlled using microwave signals and, for flux-tunable couplers, by DC currents on the flux lines~\cite{krantz2019quantum}. Since these signals propagate as electromagnetic waves along the coplanar waveguides, their transmission is degraded by impedance discontinuities: At a sharp $90^\circ$ corner or a tight kink, the local impedance changes abruptly, so part of the wave is reflected rather than transmitted~\cite{276748}. This is especially detrimental at the source and target terminals, where a bend directly at the port reflects the signal before it has settled into the CPW's guided mode. In addition to avoiding impedance mismatches at bends, two additional effects need to be considered while routing. First, the resonant frequency of a coplanar-waveguide resonator is set by its physical length, which must therefore be controlled precisely to place each resonator at its target frequency~\cite{goppl2008coplanar}. Second, and most critically, the electromagnetic field of a coplanar waveguide is not fully confined to the conductor but extends into the surrounding substrate and ground plane; when two lines run too close, their overlapping fields couple capacitively and inductively into neighboring lines, causing microwave crosstalk~\cite{song2026microwave}. Beyond these effects, longer lines expose a larger chip area to potential interference and occupy valuable routing space. The following abstraction maps these physical statements into geometric boundaries and algorithmic objectives.
    \vspace{-1.5mm}
    \begin{abstraction}[Routing Constraints and Objectives]
    As illustrated in \cref{fig:routing_overview}, superconducting quantum chip routing is subject to several geometry-dependent constraints. To capture these requirements, all pathways are modeled as geometric curves with a uniform coplanar waveguide (CPW) base width $d_{\text{cpw}}$. The routing constraint set $\mathcal{R}_{\text{route}} \subseteq \mathcal{R}$ comprises:
    \begin{enumerate*}[label=\roman*)]
        \item a minimum bending radius constraint $R_r$, enforced through smooth bends with radius $r_{\min}$ rather than hard corners, highlighted in \cref{fig:min_bending_radius},
        \item straight-segment constraints $R_s$ at connection interfaces, defined by $d_{\text{straight}}$ and visible at the terminal regions in the layout shown in \cref{fig:min_straight_length},
        \item an exact length constraint $R_l$ for resonators $\text{length}(W_{r_i})=d_{\text{fix}_i}$, realized through meander structures to achieve a prescribed electrical length as shown in \cref{fig:wire_resonator}, and
        \item a minimum clearance constraint $R_c$ between neighboring conductors, defined by $d_{\text{clear}}$ and visualized as additional keep-out boundaries throughout \cref{fig:min_clearance}.
    \end{enumerate*}
    Together, these constraints form $\mathcal{R}_{\text{route}}$.
    In addition, layout quality is optimized through the objective set $\mathcal{O}$ consisting of
    \begin{enumerate*}[label=\alph*)]
        \item minimizing total routed wire length $O_l$, and
        \item maximizing available wire clearance $O_c$.
        \item minimizing bends in wires $O_a$.
    \end{enumerate*}
    \end{abstraction}
    \vspace{-1.5mm}
    While these geometric rules define the overall design space, their applicability depends on the operational role of each pathway. The following overview outlines these distinct physical tasks. Lines connecting external launchers to ports form isolated point-to-point channels, as shown in \cref{fig:wire_launcher_qubit}, delivering highly sensitive microwave pulses directly to qubit control ports ($t_{\text{ctr}}$) for gate operations or, as shown in \cref{fig:wire_launcher_coupler}, to flux ports ($t_{\text{flux}}$) for frequency tuning~\cite{krantz2019quantum}. These paths require strict isolation to avoid gate errors~\cite{song2026microwave}.
    Quantum-state measurement is performed using readout resonators, shown in \cref{fig:wire_resonator}, which couple to the corresponding qubit readout ports ($t_{\text{read}}$)~\cite{blais2021circuit}. The resonator response is modified by the qubit state, allowing the qubit state to be inferred by probing the resonator. A resonator supports a standing-wave mode whose frequency scales inversely with its physical length. Its length must therefore be controlled precisely to achieve the desired resonant frequency. The readout resonators are coupled to a common feedline, illustrated in \cref{fig:wire_feedline}, which serves as a transmission pathway across the chip. The feedline delivers the probe tones to the resonators, collects the resulting readout signals, and routes them to external amplification and measurement systems~\cite{goppl2008coplanar}. Based on these functional requirements, the physical pathways are categorized into distinct sets. The following abstraction formally defines these wire types and maps their specific geometric constraints and optimization goals:
    \vspace{-1.5mm}
    \begin{abstraction}[Wiring Constructs]
    As shown in \cref{fig:wire_launcher_port,fig:wire_resonator,fig:wire_feedline}, the layout is partitioned into three distinct sets of wire types, forming the set $\mathcal{W} = \{W_p, W_f, W_r\}$:
    \begin{enumerate*}[label=\roman*)]
        \item \emph{Launcher-to-Port} lines ($W_p$), which establish direct point-to-point connections from an external launcher to qubit control or coupler flux interfaces, as illustrated in \cref{fig:wire_launcher_port}, and are subject to $R_r, R_s, R_c, O_l, O_c,O_a$; and
        \item \emph{Resonators} ($W_r$), which are fixed-length physical structures routed from a feedline to each individual qubit readout interface, as depicted in \cref{fig:wire_resonator}, and are strictly bound by $R_r, R_l, R_c, O_c, O_a$.
        \item \emph{Feedlines} ($W_f$), which form continuous pathways from one launcher to another, as shown in \cref{fig:wire_feedline}, limited by a maximum resonator utilization capacity $r$ (maximum number of resonators connected to a single feedline) and governed by $R_r, R_s, O_c, O_a$;
    \end{enumerate*}
    \end{abstraction}
    \vspace{-1.5mm}
    
    Beyond these main pathways, the strictly two-dimensional nature of planar chip architectures makes overlap-free routing physically impossible: In a two-dimensional qubit-coupler grid, the inner components cannot be addressed without crossing elements on the edge of the grid. To resolve such intersections, airbridges are employed as small three-dimensional arches that lift one coplanar waveguide over another, as shown in \cref{fig:airbridge}~\cite{chen2014fabrication,smlq-r7x6}. As further illustrated in \cref{fig:routing_overview}, they are placed wherever they cross any transmission line structures. However, each airbridge locally alters the waveguide impedance, adding parasitic capacitance to ground and scattering part of the signal at the crossing~\cite{chen2014fabrication}. Additionally, the layout requires dedicated CPW couplers, illustrated in \cref{fig:cpw_coupler}, which transfer energy between two coplanar waveguide sections through their overlapping electromagnetic fields. These coupling regions are instantiated throughout the layout wherever resonators connect to feedlines. To capture these physically sensitive crossing and coupling regions, they are modeled as dedicated layout components. The following abstraction maps airbridges and CPW couplers into geometric objects while extending $\mathcal{O}$ with an additional optimization objective.
    \vspace{-1.5mm}
    \begin{abstraction}[Wiring Components]
    To handle planar intersections and signal transitions, the framework introduces the global bridge minimization objective $O_b \in \mathcal{O}$ to minimize the total number of instantiated airbridges ($\min N_{\text{bridges}}$). The physical elements are modeled as routing component tuples $C_i = (\mathcal{P}_i, \mathcal{T}_i) \in \mathcal{C}_{\text{route}}$, where $\mathcal{P}_i = \{P_1, P_2, \dots, P_n\}$ defines a set of obstacle polygons acting as routing barriers under $R_o \in \mathcal{R}_{\text{route}}$, and $\mathcal{T}_i$ specifies the interface terminals with $t = (\mathbf{p}, \phi) \in \mathbb{R}^2 \times [0, 2\pi)$. This framework introduces two routing components:
    \begin{enumerate*}[label=\roman*)]
        \item \emph{Airbridges} ($C_b \in \mathcal{C}_{\text{route}}$), with terminal set $\mathcal{T}_b$ of four pins enabling a localized 3D arch for planar wire crossings, directly evaluated by the objective $O_b$; and
        \item \emph{CPW-Couplers} ($C_{\text{cpw}} \in \mathcal{C}_{\text{route}}$), with terminal set $\mathcal{T}_{\text{cpw}} = \{t_{f0}, t_{f1}, t_{r0}, t_{r1}\}$, acting as static routing hubs connecting the feedline ($W_f$) and resonators ($W_r$).
    \end{enumerate*}
\end{abstraction}
 \vspace{-1.5mm}       

\begin{example}[From Abstraction to Routing]
    Consider the running example, the physical 4Q layout, from \cref{fig:layout_architecture}, consisting of four qubits, four couplers, and sixteen launchers, each represented as a layout component $C=(\mathcal{P},\mathcal{T})$. Applying the layout abstraction introduced above transforms this physical design into the abstract representation shown in \cref{fig:abstraction_4q}, yielding the component set $\mathcal{C}_{\text{layout}}$ together with the associated layout constraints $\mathcal{R}_{\text{layout}}$. Building upon this representation, the routing components $C_b$ and $C_{cpw}$, the wire constructs $\mathcal{W}=\{W_p,W_f,W_r\}$, the routing constraints $\mathcal{R}_{\text{route}}$, and the optimization objectives $\mathcal{O}$ jointly define the corresponding routing problem. This abstract routing formulation is illustrated in \cref{fig:assignment_primary}. Solving the resulting optimization problem yields the routed layout shown in \cref{fig:layout_4q_routed}.
\end{example}
\vspace{-1.0mm}

In planar chip architectures, inner layout regions may become physically inaccessible due to the rigid boundaries of pre-placed components. For example, the flux interface of coupler $C_{c_8}$ in the center of the 9Q layout shown in \cref{fig:abstraction_9q} is fully enclosed by surrounding couplers, preventing any conventional intersection-free routing path. To overcome this topological bottleneck, the airbridge structures visible at the sides of the coupler in \cref{fig:coupler} are instantiated above these blocking components, forming additional routing channels in the abstracted layout of \cref{fig:abstraction_9q}. By locally lifting wires into the third dimension, they enable access to otherwise unreachable interior regions. Nevertheless, for many architectures, not all inner ports can be reached even with these bridges, as the number of available bridges per component is limited.

In summary, the layout and routing of the SCQC can be described through the component sets $\mathcal{C}_{\text{layout}}$ and $\mathcal{C}_{\text{route}}$, together with the corresponding constraint sets $\mathcal{R}_{\text{layout}}$ and $\mathcal{R}_{\text{route}}$. The interconnect structure is captured by the wire constructs $\mathcal{W}$, while layout quality is governed by the objective set $\mathcal{O}$.
The remaining aspect is the introduction of assignment constraints $\mathcal{R}_{\text{assign}}$, which define the mapping of wire constructs to available component terminals.
\vspace{-1.0mm}
\subsection{Port Assignment}
\vspace{-1.0mm}
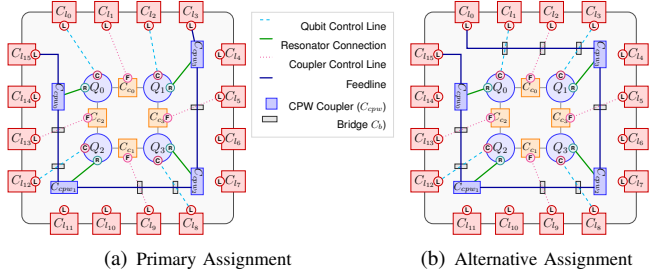
\begin{figure}[!t]
    \centering
    \subfigure[\scriptsize Primary Assignment]{%
        \label{fig:assignment_primary}
        \makebox[0.2\textwidth][c]{%
            \scalebox{0.4}{\hbox{\begin{tikzpicture}[
        qubit/.style={circle, draw=blue!80, fill=blue!10, thick, minimum size=9mm, font=\sffamily\large\bfseries},
        coupling line/.style={thick, draw=gray!80},
       coupler/.style={
            rectangle, 
            draw=orange!90, 
            fill=orange!20, 
            minimum width=6mm, 
            minimum height=6mm, 
            inner sep=1pt,
            font=\sffamily\normalsize\selectfont
        },
        launcher/.style={rectangle, draw=red!80!black, fill=red!15, thick, minimum size=7mm, font=\sffamily\large\bfseries},
        readout port/.style={circle, draw=teal!80, fill=teal!10, minimum size=2.8mm, font=\sffamily\bfseries\fontsize{5}{5}\selectfont, inner sep=0pt},
        control port/.style={circle, draw=purple!80, fill=purple!10, minimum size=2.8mm, font=\sffamily\bfseries\fontsize{6}{6}\selectfont, inner sep=0pt},
        flux port/.style={
            circle,
            draw=magenta!80,
            fill=magenta!10,
            minimum size=2.8mm,
            font=\sffamily\bfseries\fontsize{6}{6}\selectfont,
            inner sep=0pt
        },
        launcher port/.style={
            circle,
            draw=red!80!black,
            fill=red!10,
            minimum size=3mm,
            font=\sffamily\bfseries\fontsize{6}{6}\selectfont,
            inner sep=0pt
        },
        launcher qubit line/.style={thick, draw=cyan!80, dashed,font=\normalsize},
        launcher coupler line/.style={thick, draw=magenta!80, dotted},
        feedline/.style={very thick, draw=blue!60!black, line cap=round},
        resonator line/.style={very thick, draw=green!60!black, line cap=round},
        bridge terminal/.style={
            rectangle,
            draw=black!80,
            fill=gray!20,
            minimum width=4mm,
            minimum height=1.5mm,
            inner sep=0pt
        },
        cpw coupler/.style={
            rectangle,
            draw=blue!80,
            fill=blue!20,
            minimum width=4mm,
            minimum height=4mm,
            inner sep=0pt,
            font=\normalsize\selectfont
        },
    ]

    \def\rows{2}     %
    \def\cols{2}     %
    \def\spacing{2.0} %
    \def\k{4}        %
    \def\pad{1.5}    %
    \def\extdist{1.2} %
    \def\lportdist{0.45}
    \def\fportdist{0.32} %

    \pgfmathtruncatemacro{\maxc}{\cols-1}
    \pgfmathtruncatemacro{\maxr}{\rows-1}

    \draw[draw=black!70, fill=gray!5, thick, rounded corners=8pt] 
        ({0.5*\spacing - \pad}, {-0.5*\spacing + \pad}) rectangle ({(\cols+0.5)*\spacing + \pad}, {-(\rows+0.5)*\spacing - \pad});

    \foreach \r in {1,...,\rows} {
        \foreach \c in {1,...,\cols} {
            \pgfmathtruncatemacro{\qidx}{(\r-1)*\cols + \c - 1}
            
            \node[qubit] (Q-\qidx) at (\c*\spacing, -\r*\spacing) {$Q_{\qidx}$};
            
            \ifcase\qidx
                \def\rAngle{180} \def\cAngle{90}
            \or
                \def\rAngle{0}  \def\cAngle{90}
            \or
                \def\rAngle{270} \def\cAngle{180}
            \or
                \def\rAngle{0} \def\cAngle{270}
            \else
                \def\rAngle{90}  \def\cAngle{0}
            \fi
            
            \node[readout port] (R-\qidx) at ([shift=({\rAngle}:0.41)] Q-\qidx) {R};
            \node[control port] (C-\qidx) at ([shift=({\cAngle}:0.41)] Q-\qidx) {C};
        }
    }

\foreach \r in {1,...,\rows} {
    \foreach \c in {1,...,\maxc} { 
        \pgfmathtruncatemacro{\qidxA}{(\r-1)*\cols + \c - 1}
        \pgfmathtruncatemacro{\qidxB}{(\r-1)*\cols + \c}
        \pgfmathtruncatemacro{\cidx}{(\r-1)*\maxc + \c - 1}

        \ifcase\cidx
            \def\fAngle{90}
        \or
            \def\fAngle{270}
        \else
            \def\fAngle{90}
        \fi

        \draw[coupling line] (Q-\qidxA) -- (Q-\qidxB) 
            node[midway, coupler] (C-H-\r-\c) {$C_{c_{\cidx}}$};

        \node[flux port] (F-\cidx) at ([shift=({\fAngle}:\fportdist)] C-H-\r-\c) {F};
    }
}

\foreach \c in {1,...,\cols} {
    \foreach \r in {1,...,\maxr} {
        \pgfmathtruncatemacro{\qidxA}{(\r-1)*\cols + \c - 1}
        \pgfmathtruncatemacro{\qidxB}{\r*\cols + \c - 1}
        \pgfmathtruncatemacro{\cidx}{\rows*(\cols-1) + (\c-1)*\maxr + \r - 1}

        \ifcase\cidx
            \def\fAngle{90}
        \or
            \def\fAngle{270}
        \or
            \def\fAngle{180}
        \or
            \def\fAngle{0}
        \else
            \def\fAngle{90}
        \fi

        \draw[coupling line] (Q-\qidxA) -- (Q-\qidxB) 
            node[midway, coupler] (C-V-\r-\c) {$C_{c_{\cidx}}$};

        \node[flux port] (F-\cidx) at ([shift=({\fAngle}:\fportdist)] C-V-\r-\c) {F};
    }
}

    \foreach \i in {1,...,\k} {
        \pgfmathtruncatemacro{\idxTop}{\i - 1}
        \pgfmathtruncatemacro{\idxBottom}{\k + \i - 1}
        \pgfmathtruncatemacro{\idxLeft}{2*\k + \i - 1}
        \pgfmathtruncatemacro{\idxRight}{3*\k + \i - 1}
    
        \pgfmathtruncatemacro{\lblTop}{\i - 1}                     %
        \pgfmathtruncatemacro{\lblRight}{\k + \i - 1}               %
        \pgfmathtruncatemacro{\lblBottom}{2*\k + (\k - \i + 1) - 1} %
        \pgfmathtruncatemacro{\lblLeft}{3*\k + (\k - \i + 1) - 1}   %
    
        \node[launcher] (L-\idxTop) at ({(0.5*\spacing - \pad) + \i*(\cols*\spacing + 2*\pad)/(\k+1)}, {-0.5*\spacing + \pad}) {$C_{l_{\lblTop}}$};
        \coordinate (L-\idxTop_ext) at ([yshift=-\extdist cm] L-\idxTop);
        \node[launcher port] (LP-\idxTop) at ([shift=(270:\lportdist)] L-\idxTop) {L};
    
        \node[launcher] (L-\idxBottom) at ({(0.5*\spacing - \pad) + \i*(\cols*\spacing + 2*\pad)/(\k+1)}, {-(\rows+0.5)*\spacing - \pad}) {$C_{l_{\lblBottom}}$};
        \coordinate (L-\idxBottom_ext) at ([yshift=\extdist cm] L-\idxBottom);
        \node[launcher port] (LP-\idxBottom) at ([shift=(90:\lportdist)] L-\idxBottom) {L};
    
        \node[launcher] (L-\idxLeft) at ({0.5*\spacing - \pad}, {(-0.5*\spacing + \pad) - \i*(\rows*\spacing + 2*\pad)/(\k+1)}) {$C_{l_{\lblLeft}}$};
        \coordinate (L-\idxLeft_ext) at ([xshift=\extdist cm] L-\idxLeft);
        \node[launcher port] (LP-\idxLeft) at ([shift=(0:\lportdist)] L-\idxLeft) {L};
    
        \node[launcher] (L-\idxRight) at ({(\cols+0.5)*\spacing + \pad}, {(-0.5*\spacing + \pad) - \i*(\rows*\spacing + 2*\pad)/(\k+1)}) {$C_{l_{\lblRight}}$};
        \coordinate (L-\idxRight_ext) at ([xshift=-\extdist cm] L-\idxRight);
        \node[launcher port] (LP-\idxRight) at ([shift=(180:\lportdist)] L-\idxRight) {L};
    }

    \node[bridge terminal] at (0.7,-3.4) {};
    \node[bridge terminal] at (0.7,-4.6) {};
    \node[bridge terminal,rotate=90] at (3.4,-5.3) {};
    \node[bridge terminal,rotate=90] at (4.58,-5.3) {};
    \node[bridge terminal] at (5.3,-2.65) {};
    
    \draw[launcher qubit line] (LP-0) -- (C-0);
    \draw[launcher qubit line] (LP-11) -- (C-2);
    \draw[launcher qubit line] (LP-7) -- (C-3);
    \draw[launcher qubit line] (LP-2) -- (C-1);

    \draw[ resonator line] (L-9_ext) -- (R-0);
    \draw[ resonator line] (L-4_ext) -- (R-2);
    \draw[ resonator line] (L-15_ext) -- (R-3);
    \draw[ resonator line] (L-12_ext) -- (R-1);

    \node[cpw coupler, rotate=270] (CPW-0) at (L-9_ext) {$C_{cpw_0}$};
    \node[cpw coupler] (CPW-1) at (L-4_ext) {$C_{cpw_1}$};
    \node[cpw coupler, rotate=270] (CPW-2) at (L-15_ext) {$C_{cpw_2}$};
    \node[cpw coupler, rotate=270] (CPW-3) at (L-12_ext) {$C_{cpw_3}$};

    \draw[feedline] (LP-8) -- (L-8_ext) -- (CPW-0.west) ;

    \draw[feedline]  (CPW-0.east) -- (L-10_ext) -- (L-11_ext) ;

    \draw[feedline]   (CPW-1.east) -- (L-5_ext) -- (L-6_ext) -- (L-7_ext) ;

    \draw[feedline]    (CPW-2.west)  -- (L-14_ext) -- (L-13_ext) -- (CPW-3.east) -- (CPW-3) -- (LP-3);

    \draw[launcher coupler line] (LP-1) -- (F-0);
    \draw[launcher coupler line] (LP-10) -- (F-2);
    \draw[launcher coupler line] (LP-13) -- (F-3);
    \draw[launcher coupler line] (LP-6) -- (F-1);

\matrix [draw=gray!30, fill=white, anchor=north east, xshift=0.8cm] 
    at ({0.5*\spacing + \pad+8.5}, {-0.5*\spacing + \pad}) {
    
    \node[inner sep=0pt, minimum width=6mm,  label={center:\tikz\draw[launcher qubit line] (0,0) -- (0.4,0);},scale=1.2] {}; 
    & \node[font=\sffamily\normalsize\selectfont] {Qubit Control Line}; \\
    
    \node[inner sep=0pt, minimum width=6mm,  label={center:\tikz\draw[resonator line] (0,0) -- (0.4,0);},scale=1.2] {}; 
    & \node[font=\sffamily\normalsize\selectfont] {Resonator Connection}; \\
    
    \node[inner sep=0pt, minimum width=6mm,  label={center:\tikz\draw[launcher coupler line] (0,0) -- (0.4,0);},scale=1.2] {}; 
    & \node[font=\sffamily\normalsize\selectfont] {Coupler Control Line}; \\

    \node[inner sep=0pt, minimum width=6mm,  label={center:\tikz\draw[feedline] (0,0) -- (0.4,0);},scale=1.2] {}; 
    & \node[font=\sffamily\normalsize\selectfont] {Feedline}; \\[2mm]
    \node[cpw coupler, minimum width=3mm, minimum height=3mm,scale=1.2] {}; 
    & \node[font=\sffamily\normalsize\selectfont] {CPW Coupler ($C_{cpw}$)}; \\
    \vspace{2mm}
    \node[bridge terminal,scale=1.2] {}; 
    & \node[font=\sffamily\normalsize\selectfont] {Bridge $C_{b}$)}; \\
};

\end{tikzpicture}}}%
        }%
    }%
    \hspace*{6mm}%
    \subfigure[\scriptsize Alternative Assignment]{%
    \label{fig:assignment_alternative}
        \makebox[0.2\textwidth][c]{%
            \scalebox{0.4}{\hbox{\begin{tikzpicture}[
        qubit/.style={circle, draw=blue!80, fill=blue!10, thick, minimum size=9mm, font=\sffamily\large\bfseries},
        coupling line/.style={thick, draw=gray!80},
       coupler/.style={
            rectangle, 
            draw=orange!90, 
            fill=orange!20, 
            minimum width=6mm, 
            minimum height=6mm, 
            inner sep=1pt,
            font=\sffamily\normalsize\selectfont
        },
        launcher/.style={rectangle, draw=red!80!black, fill=red!15, thick, minimum size=7mm, font=\sffamily\large\bfseries},
        readout port/.style={circle, draw=teal!80, fill=teal!10, minimum size=2.8mm, font=\sffamily\bfseries\fontsize{5}{5}\selectfont, inner sep=0pt},
        control port/.style={circle, draw=purple!80, fill=purple!10, minimum size=2.8mm, font=\sffamily\bfseries\fontsize{6}{6}\selectfont, inner sep=0pt},
        flux port/.style={
            circle,
            draw=magenta!80,
            fill=magenta!10,
            minimum size=2.8mm,
            font=\sffamily\bfseries\fontsize{6}{6}\selectfont,
            inner sep=0pt
        },
        launcher port/.style={
            circle,
            draw=red!80!black,
            fill=red!10,
            minimum size=3mm,
            font=\sffamily\bfseries\fontsize{6}{6}\selectfont,
            inner sep=0pt
        },
        launcher qubit line/.style={thick, draw=cyan!80, dashed,font=\normalsize},
        launcher coupler line/.style={thick, draw=magenta!80, dotted},
        feedline/.style={very thick, draw=blue!60!black, line cap=round},
        resonator line/.style={very thick, draw=green!60!black, line cap=round},
        bridge terminal/.style={
            rectangle,
            draw=black!80,
            fill=gray!20,
            minimum width=4mm,
            minimum height=1.5mm,
            inner sep=0pt
        },
        cpw coupler/.style={
            rectangle,
            draw=blue!80,
            fill=blue!20,
            minimum width=4mm,
            minimum height=4mm,
            inner sep=0pt,
            font=\normalsize\selectfont
        },
    ]

    \def\rows{2}     %
    \def\cols{2}     %
    \def\spacing{2.0} %
    \def\k{4}        %
    \def\pad{1.5}    %
    \def\extdist{1.2} %
    \def\lportdist{0.45}
    \def\fportdist{0.32} %

    \pgfmathtruncatemacro{\maxc}{\cols-1}
    \pgfmathtruncatemacro{\maxr}{\rows-1}

    \draw[draw=black!70, fill=gray!5, thick, rounded corners=8pt] 
        ({0.5*\spacing - \pad}, {-0.5*\spacing + \pad}) rectangle ({(\cols+0.5)*\spacing + \pad}, {-(\rows+0.5)*\spacing - \pad});

    \foreach \r in {1,...,\rows} {
        \foreach \c in {1,...,\cols} {
            \pgfmathtruncatemacro{\qidx}{(\r-1)*\cols + \c - 1}
            
            \node[qubit] (Q-\qidx) at (\c*\spacing, -\r*\spacing) {$Q_{\qidx}$};
            
            \ifcase\qidx
                \def\rAngle{180} \def\cAngle{90}
            \or
                \def\rAngle{0}  \def\cAngle{90}
            \or
                \def\rAngle{270} \def\cAngle{180}
            \or
                \def\rAngle{0} \def\cAngle{270}
            \else
                \def\rAngle{90}  \def\cAngle{0}
            \fi
            
            \node[readout port] (R-\qidx) at ([shift=({\rAngle}:0.41)] Q-\qidx) {R};
            \node[control port] (C-\qidx) at ([shift=({\cAngle}:0.41)] Q-\qidx) {C};
        }
    }

\foreach \r in {1,...,\rows} {
    \foreach \c in {1,...,\maxc} { 
        \pgfmathtruncatemacro{\qidxA}{(\r-1)*\cols + \c - 1}
        \pgfmathtruncatemacro{\qidxB}{(\r-1)*\cols + \c}
        \pgfmathtruncatemacro{\cidx}{(\r-1)*\maxc + \c - 1}

        \ifcase\cidx
            \def\fAngle{90}
        \or
            \def\fAngle{270}
        \else
            \def\fAngle{90}
        \fi

        \draw[coupling line] (Q-\qidxA) -- (Q-\qidxB) 
            node[midway, coupler] (C-H-\r-\c) {$C_{c_{\cidx}}$};

        \node[flux port] (F-\cidx) at ([shift=({\fAngle}:\fportdist)] C-H-\r-\c) {F};
    }
}

\foreach \c in {1,...,\cols} {
    \foreach \r in {1,...,\maxr} {
        \pgfmathtruncatemacro{\qidxA}{(\r-1)*\cols + \c - 1}
        \pgfmathtruncatemacro{\qidxB}{\r*\cols + \c - 1}
        \pgfmathtruncatemacro{\cidx}{\rows*(\cols-1) + (\c-1)*\maxr + \r - 1}

        \ifcase\cidx
            \def\fAngle{90}
        \or
            \def\fAngle{270}
        \or
            \def\fAngle{180}
        \or
            \def\fAngle{0}
        \else
            \def\fAngle{90}
        \fi

        \draw[coupling line] (Q-\qidxA) -- (Q-\qidxB) 
            node[midway, coupler] (C-V-\r-\c) {$C_{c_{\cidx}}$};

        \node[flux port] (F-\cidx) at ([shift=({\fAngle}:\fportdist)] C-V-\r-\c) {F};
    }
}

    \foreach \i in {1,...,\k} {
        \pgfmathtruncatemacro{\idxTop}{\i - 1}
        \pgfmathtruncatemacro{\idxBottom}{\k + \i - 1}
        \pgfmathtruncatemacro{\idxLeft}{2*\k + \i - 1}
        \pgfmathtruncatemacro{\idxRight}{3*\k + \i - 1}
    
        \pgfmathtruncatemacro{\lblTop}{\i - 1}                     %
        \pgfmathtruncatemacro{\lblRight}{\k + \i - 1}               %
        \pgfmathtruncatemacro{\lblBottom}{2*\k + (\k - \i + 1) - 1} %
        \pgfmathtruncatemacro{\lblLeft}{3*\k + (\k - \i + 1) - 1}   %
    
        \node[launcher] (L-\idxTop) at ({(0.5*\spacing - \pad) + \i*(\cols*\spacing + 2*\pad)/(\k+1)}, {-0.5*\spacing + \pad}) {$C_{l_{\lblTop}}$};
        \coordinate (L-\idxTop_ext) at ([yshift=-\extdist cm] L-\idxTop);
        \node[launcher port] (LP-\idxTop) at ([shift=(270:\lportdist)] L-\idxTop) {L};
    
        \node[launcher] (L-\idxBottom) at ({(0.5*\spacing - \pad) + \i*(\cols*\spacing + 2*\pad)/(\k+1)}, {-(\rows+0.5)*\spacing - \pad}) {$C_{l_{\lblBottom}}$};
        \coordinate (L-\idxBottom_ext) at ([yshift=\extdist cm] L-\idxBottom);
        \node[launcher port] (LP-\idxBottom) at ([shift=(90:\lportdist)] L-\idxBottom) {L};
    
        \node[launcher] (L-\idxLeft) at ({0.5*\spacing - \pad}, {(-0.5*\spacing + \pad) - \i*(\rows*\spacing + 2*\pad)/(\k+1)}) {$C_{l_{\lblLeft}}$};
        \coordinate (L-\idxLeft_ext) at ([xshift=\extdist cm] L-\idxLeft);
        \node[launcher port] (LP-\idxLeft) at ([shift=(0:\lportdist)] L-\idxLeft) {L};
    
        \node[launcher] (L-\idxRight) at ({(\cols+0.5)*\spacing + \pad}, {(-0.5*\spacing + \pad) - \i*(\rows*\spacing + 2*\pad)/(\k+1)}) {$C_{l_{\lblRight}}$};
        \coordinate (L-\idxRight_ext) at ([xshift=-\extdist cm] L-\idxRight);
        \node[launcher port] (LP-\idxRight) at ([shift=(180:\lportdist)] L-\idxRight) {L};
    }

    \node[bridge terminal] at (0.7,-3.4) {};
    \node[bridge terminal] at (0.7,-4.6) {};
    \node[bridge terminal,rotate=90] at (3.4,-5.3) {};
    \node[bridge terminal,rotate=90] at (4.58,-5.3) {};
    \node[bridge terminal] at (5.3,-2.65) {};

    \node[bridge terminal,rotate=90] at (4.6,-0.7) {};
    \node[bridge terminal,rotate=90] at (3.4,-0.7) {};
    \node[bridge terminal,rotate=90] at (2.15,-0.7) {};
    
    \draw[launcher qubit line] (LP-1) -- (C-0);
    \draw[launcher qubit line] (LP-11) -- (C-2);
    \draw[launcher qubit line] (LP-7) -- (C-3);
    \draw[launcher qubit line] (LP-3) -- (C-1);

    \draw[ resonator line] (L-9_ext) -- (R-0);
    \draw[ resonator line] (L-4_ext) -- (R-2);
    \draw[ resonator line] (L-15_ext) -- (R-3);
    \draw[ resonator line] (L-12_ext) -- (R-1);

    \node[cpw coupler, rotate=270] (CPW-0) at (L-9_ext) {$C_{cpw_0}$};
    \node[cpw coupler] (CPW-1) at (L-4_ext) {$C_{cpw_1}$};
    \node[cpw coupler, rotate=270] (CPW-2) at (L-15_ext) {$C_{cpw_2}$};
    \node[cpw coupler, rotate=270] (CPW-3) at (L-12_ext) {$C_{cpw_3}$};

    \draw[feedline] (LP-8) -- (L-8_ext) -- (CPW-0.west) ;

    \draw[feedline]  (CPW-0.east) -- (L-10_ext) -- (L-11_ext) ;

    \draw[feedline]   (CPW-1.east) -- (L-5_ext) -- (L-6_ext) -- (L-7_ext) ;

    \draw[feedline]    (CPW-2.west)  -- (L-14_ext) -- (L-13_ext) -- (CPW-3.east) ;

    \draw[feedline]  (L-3_ext) -- (L-2_ext) -- (L-1_ext) -- (L-0_ext)-- (LP-0);

    \draw[launcher coupler line] (LP-2) -- (F-0);
    \draw[launcher coupler line] (LP-10) -- (F-2);
    \draw[launcher coupler line] (LP-13) -- (F-3);
    \draw[launcher coupler line] (LP-6) -- (F-1);

\end{tikzpicture}}}%
        }%
    }%
    
    \caption{Comparison between (a) the primary port assignment and (b) an alternative assignment.}
    \label{fig:assignment_comparison}
\end{figure}

    Building upon the routing formulation, we now introduce assignment constraints $\mathcal{R}_{\text{assign}} \subseteq \mathcal{R}$, which govern the mapping of routing constructs $\mathcal{W}$. Specifically, port assignment associates each wire element (e.g., $W_{p_1}, W_{f_1}$) with a sequence of terminals, defining the order in which terminals are connected along the wire.
    Each qubit control port and coupler flux port must connect to exactly one launcher, resulting in $|W_p| = q + c$ launcher-to-port wires for $q$ qubits and $c$ couplers. The number of required feedlines is determined by the maximum resonator utilization capacity $r_\text{util}$ as the upper bound of assignable CPW couplers for a feedline, yielding $|W_f| = \left\lceil \frac{q}{r_\text{util}} \right\rceil$ feedlines, each connecting two launchers. Furthermore, each qubit requires exactly one resonator connected to a feedline, resulting in $|W_r| = q$ resonator wires.
    Furthermore, the structure of the wire constructs is governed by assignment constraints $\mathcal{R}_{\text{assign}}$. While launcher-to-port wires $W_p$ and resonators $W_r$ each connect exactly two terminals, forming source-to-target pairs $(t_{\text{src}}, t_{\text{tgt}})$, feedlines $W_f$ are defined as ordered sequences of terminals. Specifically, a feedline starts at a launcher, traverses a sequence of intermediate terminals equipped with CPW couplers and airbridges, and terminates at another launcher.  These structures are enforced by constraints 
    $\mathcal{R}_{\text{assign}} = \{R_p, R_r, R_f, R_{\text{lnc}}, R_{\text{port}}\}$. Here, $R_p \in \mathcal{R}_{\text{assign}}$ restricts each launcher-to-port wire $W_p=(t_{\text{lnc}}, t)$ to exactly two terminals, consisting of one launcher terminal $t_{\text{lnc}}$ and one control or flux terminal $t \in \{t_{\text{ctrl}}, t_{\text{flux}}\}$. Likewise, $R_r \in \mathcal{R}_{\text{assign}}$ restricts each resonator wire $W_r=(t_r,t_{\text{read}})$ to exactly two terminals, connecting a CPW-coupler resonator terminal $t_r$ to a qubit readout terminal $t_{\text{read}}$. Furthermore, $R_f \in \mathcal{R}_{\text{assign}}$ enforces that each feedline $W_f$ forms a valid ordered terminal sequence starting and ending at launcher terminals. It traverses at most $r_\text{util}$ intermediate CPW-coupler terminals, while the number of traversed bridge terminals is unconstrained. Constraint $R_{\text{lnc}} \in \mathcal{R}_{\text{assign}}$ ensures that each launcher terminal is used by at most one wire, and $R_{\text{port}} \in \mathcal{R}_{\text{assign}}$ guarantees a one-to-one assignment between launcher-to-port wires and control or flux terminals, such that every terminal $t_{\text{ctrl}}$ and $t_{\text{flux}}$ appears in exactly one wire $W_p$.
    \vspace{-1.5mm}
    \begin{example}[Valid Assignments and Their Quality]
    \Cref{fig:assignment_primary} illustrates the abstract routing assignment based on the wire set $\mathcal{W}=\{W_p,W_f,W_r\}$. For example, the launcher-to-port wire $W_{p_1}=(t_{lnr_0},t_{ctrl,0})$ connects launcher $C_{l_0}$ to the control terminal of qubit $C_{q_0}$. Likewise, the resonator wire $W_{r_0}=(t_{r_0},t_{read,0})$ connects a resonator terminal of the CPW coupler $C_{cpw_0}$ to the readout terminal of $C_{q_0}$. The feedline is represented by an ordered terminal sequence, e.g., $W_{f_0}=t_{lnr_8},t_{f_0},t_{f_1},t_{b_0},t_{b_1},\ldots,t_{lnr_3})$,
    starting at launcher $C_{l_8}$, traversing multiple CPW couplers and bridges, and finally terminating at launcher $C_{l_3}$. Overall, the assignment contains $|W_f|=1$ feedline, $|W_p|=8$ launcher-to-port wires, and $|W_r|=4$ resonators, while requiring a total of $|C_b|=5$ instantiated bridges. For comparison, \cref{fig:assignment_alternative} shows an alternative valid assignment with a different feedline mapping, e.g., $W_{f_0}=(t_{lnr_8},\ldots,t_{lnr_0})$. Although both assignments satisfy the constraints in $\mathcal{R}_{assign}$, the alternative solution results in a larger number of bridge crossings as well as wire length for $W_p$ and therefore a less favorable routing solution with respect to the optimization objectives.
    \end{example}
    \vspace{-1.5mm}
    The example further demonstrates that the assignment has a significant impact on the optimization objectives $\mathcal{O}$, in particular $O_b$, $O_c$, $O_a$, and $O_l$. Different assignment choices directly affect wire crossings, spacing, and total path length, leading to substantial variations in routing quality. Consequently, suboptimal assignments degrade routing solutions, highlighting the need for methods that can jointly optimize these competing objectives.

    In summary, the complete planar chip design automation task is modeled through the sets $\mathcal{C}$, $\mathcal{W}$, $\mathcal{R}$, and $\mathcal{O}$. The component set $\mathcal{C} = \mathcal{C}_{\text{layout}} \cup \mathcal{C}_{\text{route}} = \{C_q, C_l, C_c, C_b, C_{\text{cpw}}\}$ collects all components, while the interconnect network is given by $\mathcal{W} = \{W_p, W_f, W_r\}$.  Constraints are partitioned as $\mathcal{R} = \mathcal{R}_{\text{layout}} \cup \mathcal{R}_{\text{route}} \cup \mathcal{R}_{\text{assign}}$, and optimization is driven by $\mathcal{O}$. Based on this formalization, the subsequent design automation problem consists of determining an assignment and routing of $\mathcal{W}$ that satisfies $\mathcal{R}$ while optimizing $\mathcal{O}$.

\vspace{-0.5mm}
\section{Design Automation}\label{sec:desau}
\vspace{-1.0mm}
 In this section, we first briefly explain why previous approaches cannot address the design automation tasks reviewed in \cref{subsec:short}. We then provide an overview of the general idea of the proposed design automation flow in \cref{subsec:gen_idea}. Subsequently, we present the details of each step, starting with the global partitioning in \cref{subsec:global_part}, followed by the port assignment computation in \cref{subsec:assignm}, and finally the routing in \cref{subsec:routing}.
 \vspace{-0.5mm}
 \subsection{Shortcomings of Related Work}\label{subsec:short}
 \vspace{-0.5mm}
 Existing open-source and academic design automation frameworks for superconducting quantum circuits include Quantum Metal, EDA-Q \cite{edaQ}, the framework proposed by Yang~\etal~\cite{yang2025scalable}, and the extension introduced by Yu~\etal~\cite{consaware}. These are reviewed in the following.

Quantum Metal, developed by IBM, is an open-source framework for quantum circuit design and simulation that provides integrated point-to-point and fixed-length routing capabilities for simple scenarios. However, the tool exhibits several critical shortcomings: It lacks support for port assignment computation and cannot fulfill the assignment constraints $R_\text{assign}$. Additionally, it offers no native support for the flux-control ports $t_\text{flux}$ that tunable couplers require. Furthermore, its routing strategies neither incorporate mechanisms to resolve routing constraint violations ($R_\text{routing}$), nor are they capable of optimizing wire length ($O_l$), clearance ($O_c$), wire bends ($O_a$), or bridges ($O_b$). Finally, the framework relies on simplified layout representations and does not model complex polygonal obstacles or couplers as static routing obstacles.

EDA-Q~\cite{edaQ} and its extension by Yu~\etal~\cite{consaware}, both developed at the Laboratory for Advanced Computing and Intelligence Engineering at the Information Engineering University, Zhengzhou, as well as the approach by Yang~\etal~\cite{yang2025scalable}, offer assignment and routing solutions for planar and flip-chip superconducting quantum circuits. However, these methods rely on rigid, regular row-wise assignment schemes designed for highly uniform chip architectures. Consequently, they lack support for generalized port assignment constraints $R_\text{assign}$, including the specific requirements for resonators ($W_r$), feedlines ($W_f$), and flux-control ports ($t_\text{flux}$). Furthermore, none of these frameworks model tunable couplers as static geometric obstacles or account for dynamic fixed-length resonator routing ($R_l$). As a result, existing approaches are inherently limited to regular architectures and cannot address complex quantum chip layouts featuring irregular geometries, tunable couplers, and tight spatial clearance constraints.

\begin{figure}[!t]
        \centering
        \subfigure[Conventional Layout.]{
            \includegraphics[width=0.22\textwidth]{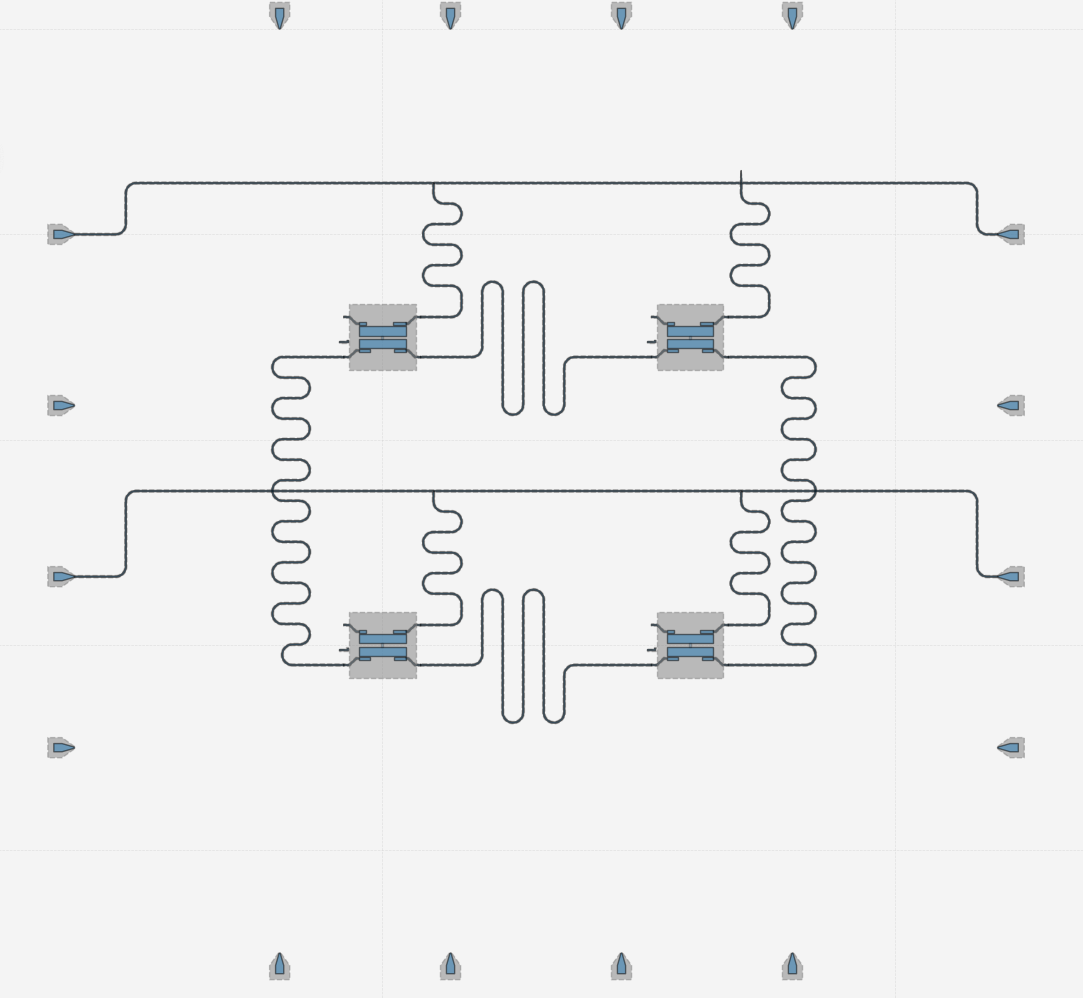}
        }
        \subfigure[Proposed Layout.]{
            \includegraphics[width=0.22\textwidth]{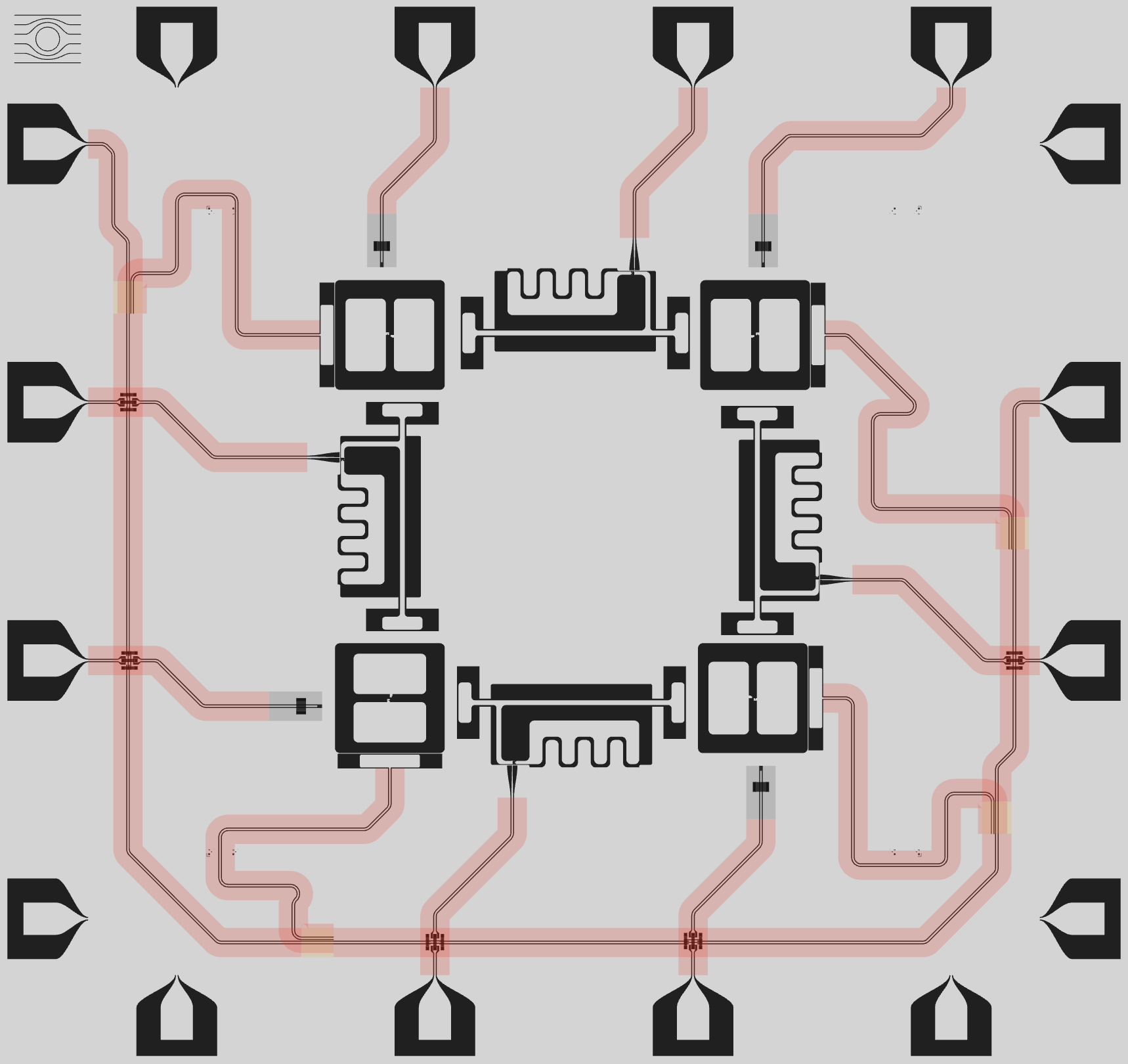}
        }
         \vspace{-2.0mm}
        \caption{Comparison of assignment and routing complexity between (a) conventional layout generation and (b) a layout generated by the proposed method. Existing approaches consider point-to-point assignments for $q+2r$ wires $W_p$ distributed across $r$ rows. The flow proposed in this work solves assignments for $q+c$ control wires $W_p$ as well as $q$ resonator wires $W_r$ and feedline connections for approximately $\left\lceil q/r_{\text{util}} \right\rceil$ qubits. As a result, the problem involves ordered port assignments rather than point-to-point mappings alone. Assignment optimization further incorporates wirelength and bridge minimization objectives. For this 4-qubit layout, the assignment complexity increases from a total of 8 in (a) to 14 targets in (b) and from 15 to 34 targets for a 9-qubit case. In addition, routing complexity is significantly higher due to fixed-length resonator constraints, increased wiring demand, coupler and bridge placement, and feedline routing.}
        \label{fig:related}
    \end{figure}
In conclusion, as visually evidenced by the layout comparison in \cref{fig:related}, existing approaches either tackle simplified, lower-complexity design problems or lack the mandatory features. Since the reviewed approaches do not satisfy the requirements of the targeted design problem, the following section outlines the general concept of the proposed flow.
 \vspace{-1.0mm}
\subsection{General Idea}
 \vspace{-0.5mm}
\label{subsec:gen_idea}
 The following presents a high-level view of the proposed design automation flow, along with a breakdown of each stage. The first step, the \emph{Global Partitioning}, maps the layout based on the component geometries of $C$ (\cref{fig:regions}) to routing regions $G$ with edge capacities between adjacent cells (\cref{fig:final_partitions}). This step is required because SCQC layouts contain narrow passages between components (bottlenecks) and enclosed free-space regions accessible only through these passages (cavities). As the wires routed through these confined regions must still maintain the minimum clearance $d_\text{clear}$, only a limited number of wires fit through each passage, and port assignment decisions made without this geometric awareness, therefore, frequently produce configurations that are physically unroutable in the later routing stage ($R_c$). To prevent this, the partitioning automatically identifies bottlenecks and cavities before routing, assigns realistic capacities to the edges between $G$-cells based on the local geometry, and enforces physical capacity limits on dedicated bridges to guarantee downstream routability.

The subsequent \emph{Port Assignment Computation} takes the edge capacities of the $G$-cells (\cref{fig:final_partitions}) and computes all net assignments $W$ ($W_p$, $W_r$, $W_f$) while satisfying the assignment constraints $R_\text{assign}$ (\cref{fig:assignment_primary}). These high-level assignments must respect the local geometric bottlenecks and cavities, which is achieved by explicitly reusing the cavity characteristics extracted during the global partitioning step. The assignment proceeds in two main steps: first, dedicated bridges are assigned to the inner target ports subject to capacity limits; then, the outer ports and connections are assigned. Both steps optimize primarily for bridge usage ($O_b$) and secondarily for wire length ($O_l$).

Finally, the \emph{Routing} stage takes the assigned nets $W$ (\cref{fig:assignment_primary}) as input, at this stage consisting only of source and target ports for $W_p$ and $W_r$, or an ordered terminal sequence for $W_f$ and transforms them into fully routed wires satisfying all design rules $R_\text{routing}$ (\cref{fig:layout_4q_routed}). Since enforcing all physical, geometric, and length constraints simultaneously is computationally intractable for large circuits, the routing task is decomposed into a four-stage hierarchical pipeline that incrementally incorporates fine-grained constraints, based on A$^*$ search and a cyclic rip-up-and-reroute strategy for the first three stages: Global routing first routes $W_p$ and $W_r$ across the extracted $G$-cells. Detailed routing then refines these wires on the partitioning grid while enforcing the clearance constraint $R_c$. Next, a Dubins A$^*$ router refines the wires on a finer grid: this 
curvature-aware A$^*$ variant expands arc-move primitives rather than straight grid steps and thus inherently satisfies the minimum bending radius $R_r$ and straight length $R_s$. It searches for the shortest path ($O_l$), penalizes angle changes ($O_a$) and proximity to other wires ($O_c$), and dynamically matches the fixed-length requirement $R_l$ for resonators. Finally, the feedline routing generates the feedline wires $W_f$ using the same Dubins A$^*$ approach.

With the general concept provided, the next section presents a detailed breakdown of the three steps in the design flow, starting with global partitioning.
 \vspace{-0.5mm}
\subsection{Global Partitioning}
\begin{figure}[!t]
        \centering
        \subfigure[Clearance Violation.]{%
        \makebox[0.15\textwidth][c]{%
            \resizebox{.3\linewidth}{!}{\hbox{\input{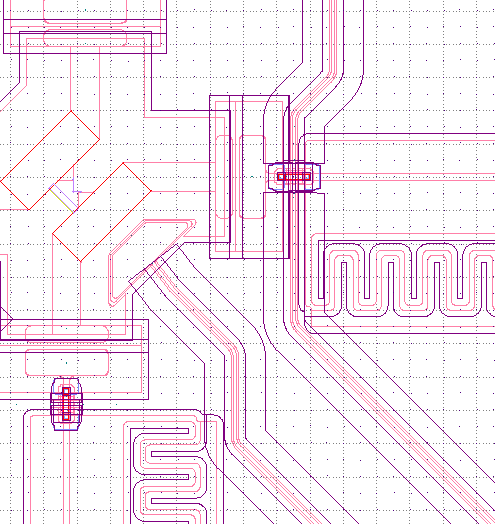}}}%
            \label{fig:clearance_violation}
            }
        }
        \hfil
        \subfigure[Spatial Routing Conflict.]{%
        \makebox[0.2\textwidth][c]{%
         \resizebox{.4\linewidth}{!}{\hbox{\input{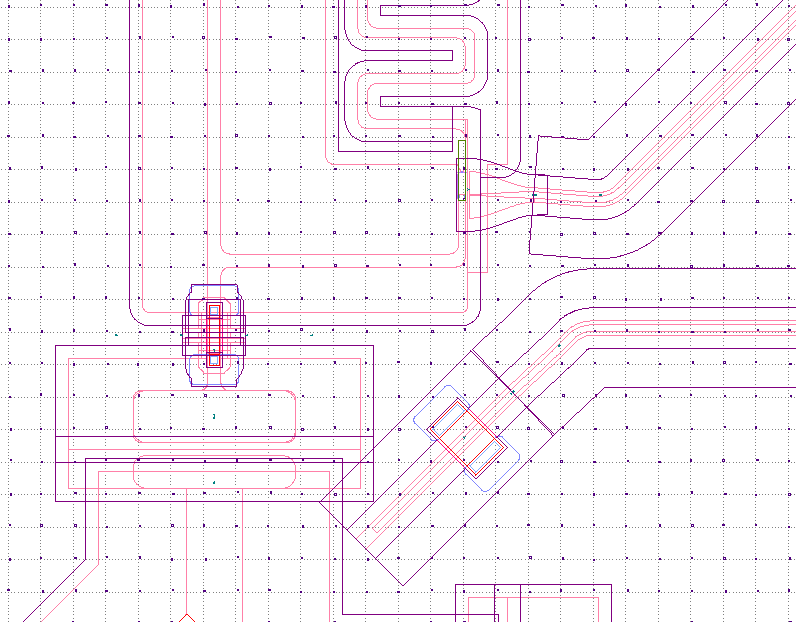}}}%
         }\ \ \label{fig:spatial_conflict}%
         }
         \vspace{-1.0mm}
        \caption{Illustrations of layout cavities and bottlenecks: (a) clearance violation within a cavity, and (b) spatially infeasible bridge assignment.}
        \label{fig:cavities_bottlenecks}
    \end{figure}

 \vspace{-0.5mm}
    \label{subsec:global_part}
    SCQC layouts exhibit irregular polygonal geometries with cavities and narrow bottlenecks, while a high minimum clearance $d_\text{clear}$ must be enforced between nets ($R_c$). Consequently, port assignments satisfying $R_\text{assign}$ may still be unroutable within such constrained regions.
     \vspace{-1.5mm}
    \begin{example}[Cavities and Bottlenecks]
    \Cref{fig:cavities_bottlenecks} illustrates layout cavities and bottlenecks that lead to routing violations if not handled properly prior to routing attempts. In \cref{fig:clearance_violation}, a cavity contains two bridges alongside a mandatory qubit port $t_\text{read}$. Assigning both bridges as routing channels violates the minimum wire clearance, as the available space is insufficient for three parallel wires. Similarly, in \cref{fig:spatial_conflict}, a dedicated bridge could theoretically be assigned; however, due to the spatial arrangement of the two neighboring ports, such an assignment would render the layout completely unroutable.
    \end{example}
    \vspace{-1.5mm}
    \vspace{-1.5mm}
    To resolve this problem, we propose forming geometry-aware partitions $G = \{G_0, G_1, \dots, G_k\}$, where adjacent cells are assigned edge capacities $g_{kj}$ that limit the total number of allowed nets, information that is later 
    leveraged during both port assignment and global routing. To this end, the layout is first rasterized onto a partitioning grid on which all relevant bottlenecks are detected (\cref{subsub:detect}). Their topological dependencies are then captured in a cavity tree, from which the final geometry-aware partitions are derived (\cref{subsub:cavity}).
    \subsubsection{Grid Construction and Bottleneck Detection}\label{subsub:detect}
    The basis for the geometric analysis is a discretized representation of the layout. To this end, the continuous geometry is rasterized onto a dedicated partitioning grid with dimensions $p_\text{width} \times p_\text{height}$, incorporating the polygonal obstacle shapes of qubits, couplers, and launchers. The layout is then separated into the outer region $G$ and the inner regions $I = \{I_0, I_1, \dots\}$, and the obstacle-free area of the outer region is uniformly partitioned into initial global cells (gcells) based on width $g_\text{width}$ and height $g_\text{height}$.
    \vspace{-1.5mm}
    \begin{example}[Inner and Outer Partitions]
        \Cref{fig:regions} illustrates the distinction between the four inner partitions ($I_0, I_1, I_2, I_3$), which are accessible only through dedicated bridges, and the initial rectangular outer partitions ($G_0, \dots, G_{63}$). These outer partitions serve as a baseline layout that is subsequently transformed into geometry-aware partitions. Furthermore, \cref{fig:rasterization} provides a detailed view of the rasterized component geometries $C$, where the minimum straight path $d_\text{straight}$ and clearance $d_\text{clear}$ are already marked as obstacles for each target node.
    \end{example}
    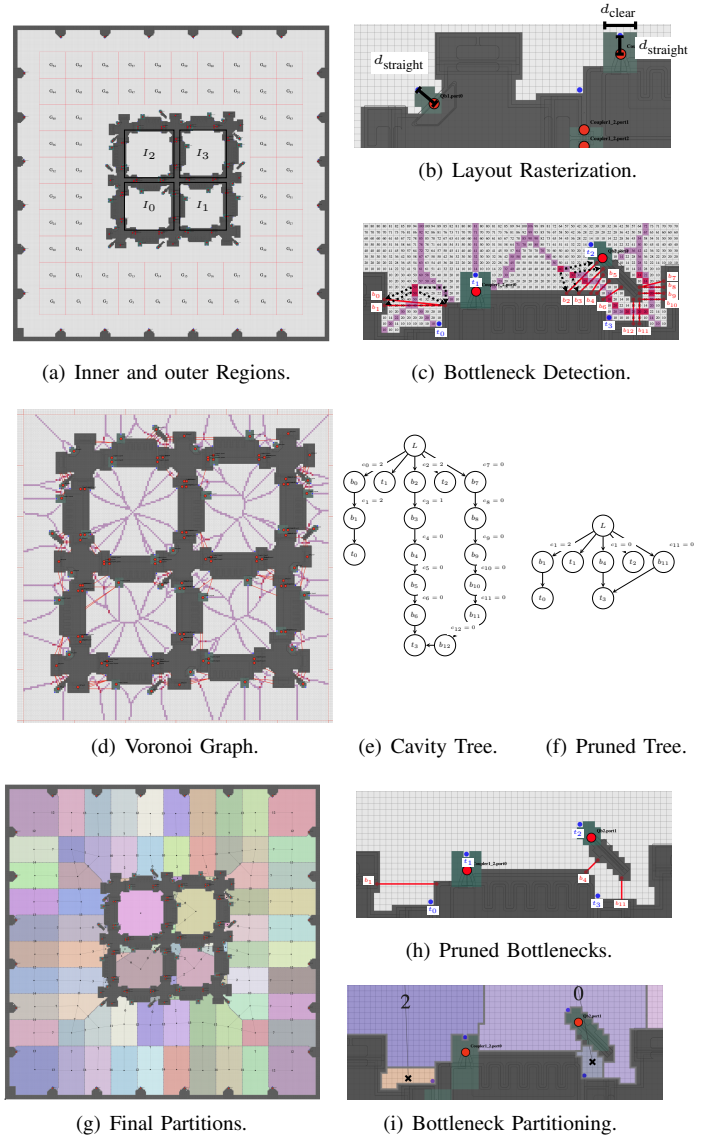
\begin{figure}[!t]  
        \centering
        \subfigure[Inner and outer Regions.]{%
        \label{fig:regions}
        \scalebox{0.77}{\hbox{\input{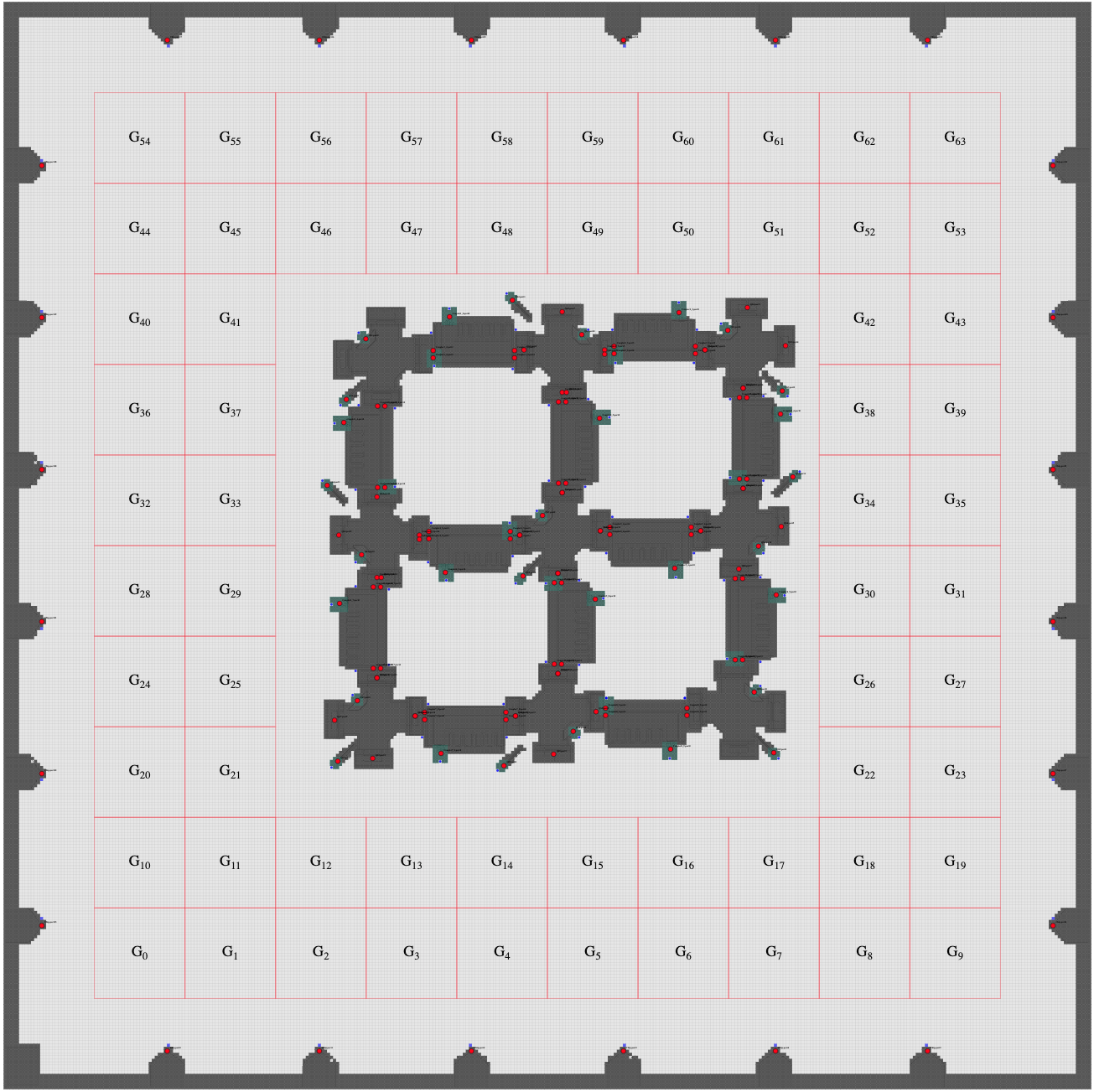}}}%
        }
   \begin{minipage}[b]{0.23\textwidth}
        \subfigure[Layout Rasterization.]{%
            \label{fig:rasterization}
            \scalebox{1.0}{\input{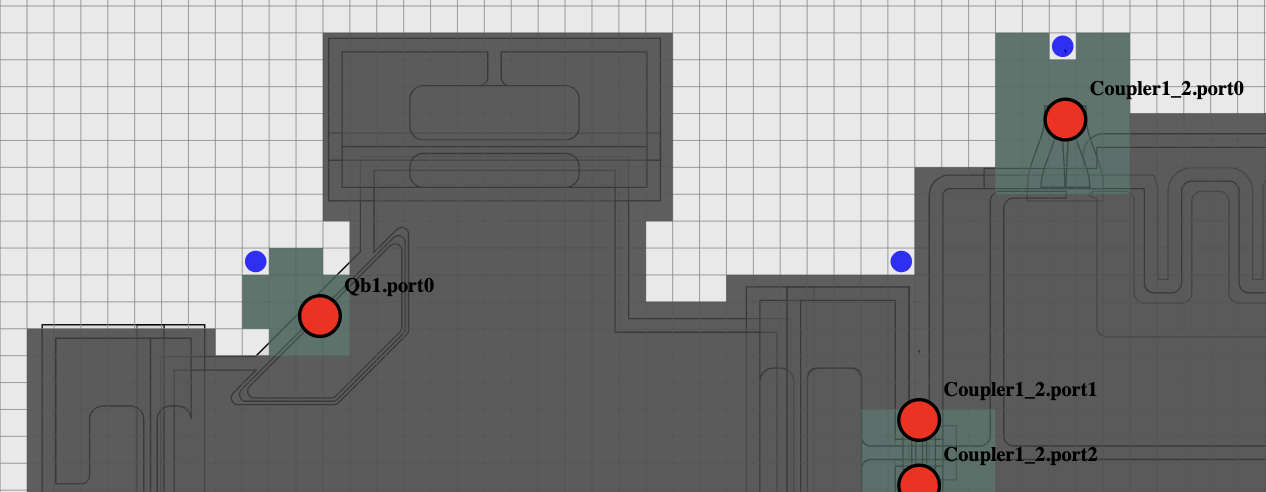}}
            
        }\\[0.5ex]
        \subfigure[Bottleneck Detection.]{%
            \label{fig:bottleneck_detection}
            \scalebox{1.0}{\input{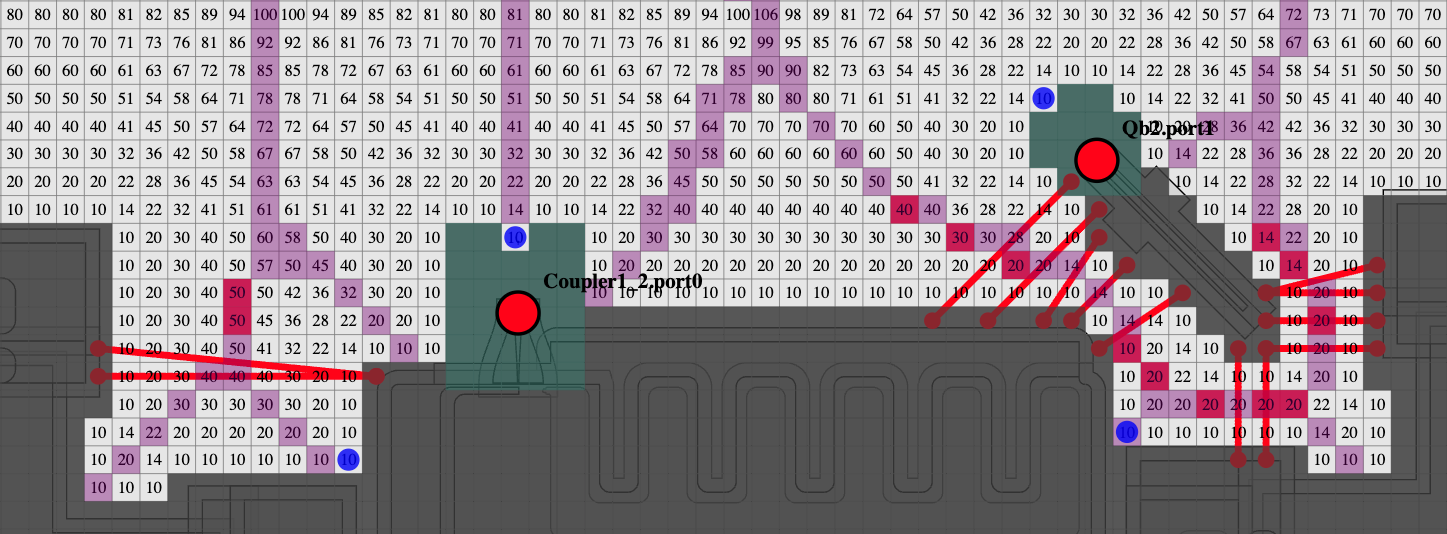}}
            
        }
    \end{minipage}
            \subfigure[Voronoi Graph.]{%
                \label{fig:voronoi_graph}
                \includegraphics[width=0.23\textwidth]{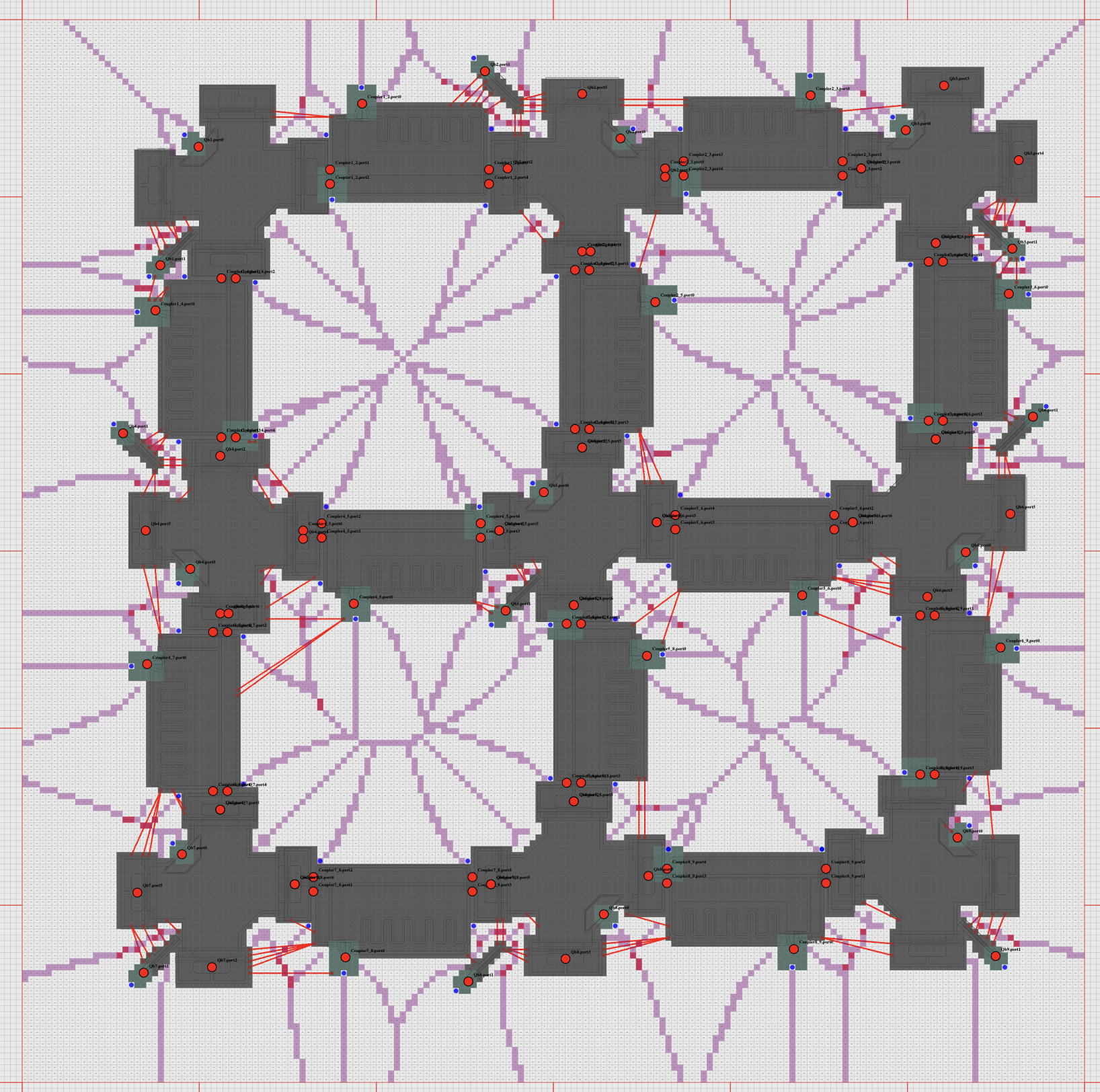}
            }
            \hspace{-3.5mm}
            \begin{minipage}[b]{0.23\textwidth}
            \centering
             
              \subfigure[Cavity Tree.]{%
              \label{fig:cavity_tree}
                \raisebox{0.3\height}{\scalebox{0.4}{\begin{tikzpicture}[
    >=stealth,
    node/.style={
        circle,
        draw=black,
        thick,
        minimum size=7mm,
        inner sep=0pt,
        font=\scriptsize
    },
    terminal/.style={
        rectangle,
        draw=black,
        thick,
        minimum width=8mm,
        minimum height=5mm,
        inner sep=1pt,
        font=\scriptsize
    },
    edge/.style={
        ->,
        thick
    },
    prunednode/.style={
        circle,
        draw=gray!60,
        thick,
        fill=gray!10,
        minimum size=7mm,
        inner sep=0pt,
        font=\scriptsize,
        text=gray!70
    },
    title/.style={
        font=\scriptsize
    }
]

\node[node] (L)  at (0,0) {$L$};
\node[node] (b0) at (-2,-1.2) {$b_0$};

\node[node] (t1) at (-1,-1.2) {$t_1$};
\node[node] (b2) at (0,-1.2) {$b_2$};
\node[node] (t2) at (1,-1.2) {$t_2$};

\node[node] (b1) at (-2,-2.4) {$b_1$};
\node[node] (t0) at (-2,-3.6) {$t_0$};

\node[node] (b3) at (0,-2.4) {$b_3$};
\node[node] (b4) at (0,-3.6) {$b_4$};
\node[node] (b5) at (0,-4.6) {$b_5$};
\node[node] (b6) at (0,-5.6) {$b_6$};
\node[node] (t3) at (0,-6.6) {$t_3$};

\node[node] (b7) at (2,-1.2) {$b_{7}$};
\node[node] (b8) at (2,-2.4) {$b_{8}$};
\node[node] (b9) at (2,-3.6) {$b_{9}$};
\node[node] (b10) at (2,-4.6) {$b_{10}$};
\node[node] (b11) at (2,-5.6) {$b_{11}$};
\node[node] (b12) at (1,-6.6) {$b_{12}$};

\draw[edge] (L) -- (b0);
\draw[edge] (L) -- (t1);
\draw[edge] (L) -- (b2);
\draw[edge] (L) -- (t2);
\draw[edge] (L) -- (b7);

\draw[edge] (b0) -- (b1);
\draw[edge] (b1) -- (t0);

\draw[edge] (b2) -- (b3);
\draw[edge] (b3) -- (b4);
\draw[edge] (b4) -- (b5);
\draw[edge] (b5) -- (b6);
\draw[edge] (b6) -- (t3);

\draw[edge] (b7) -- (b8);
\draw[edge] (b8) -- (b9);
\draw[edge] (b9) -- (b10);
\draw[edge] (b10) -- (b11);
\draw[edge] (b11) -- (b12);
\draw[edge] (b12) -- (t3);

\node[fill=white,font=\tiny,xshift=6mm,yshift=2.0mm] (b0l) at (b0.north) {$c_0=2$};

\node[fill=white,font=\tiny,xshift=6mm,yshift=2.0mm] (b1l) at (b1.north) {$c_1=2$};

\node[fill=white,font=\tiny,xshift=6mm,yshift=2.0mm] (b2l) at (b2.north) {$c_2=2$};

\node[fill=white,font=\tiny,xshift=6mm,yshift=2.0mm] (b3l) at (b3.north) {$c_3=1$};

\node[fill=white,font=\tiny,xshift=6mm,yshift=2.0mm] (b4l) at (b4.north) {$c_4=0$};

\node[fill=white,font=\tiny,xshift=6mm,yshift=2.0mm] (b5l) at (b5.north) {$c_5=0$};

\node[fill=white,font=\tiny,xshift=6mm,yshift=2.0mm] (b6l) at (b6.north) {$c_6=0$};

\node[fill=white,font=\tiny,xshift=6mm,yshift=2.0mm] (b7l) at (b7.north) {$c_7=0$};

\node[fill=white,font=\tiny,xshift=6mm,yshift=2.0mm] (b8l) at (b8.north) {$c_8=0$};

\node[fill=white,font=\tiny,xshift=6mm,yshift=2.0mm] (b9l) at (b9.north) {$c_9=0$};

\node[fill=white,font=\tiny,xshift=6mm,yshift=2.0mm] (b10l) at (b10.north) {$c_{10}=0$};

\node[fill=white,font=\tiny,xshift=6mm,yshift=2.0mm] (b11l) at (b11.north) {$c_{11}=0$};

\node[fill=white,font=\tiny,xshift=6mm,yshift=2.0mm] (b12l) at (b12.north) {$c_{12}=0$};

\end{tikzpicture}}
                }}%
            \subfigure[Pruned Tree.]{%
                \label{fig:cavity_tree_pruned}
                
                \raisebox{1.2\height}{\scalebox{0.4}{\begin{tikzpicture}[
    >=stealth,
    node/.style={
        circle,
        draw=black,
        thick,
        minimum size=7mm,
        inner sep=0pt,
        font=\scriptsize
    },
    terminal/.style={
        rectangle,
        draw=black,
        thick,
        minimum width=8mm,
        minimum height=5mm,
        inner sep=1pt,
        font=\scriptsize
    },
    edge/.style={
        ->,
        thick
    },
    prunednode/.style={
        circle,
        draw=gray!60,
        thick,
        fill=gray!10,
        minimum size=7mm,
        inner sep=0pt,
        font=\scriptsize,
        text=gray!70
    },
    title/.style={
        font=\scriptsize
    }
]

\node[node] (L)  at (0,0) {$L$};
\node[node] (b1) at (-2,-1.2) {$b_1$};
\node[node] (t1) at (-1,-1.2) {$t_1$};
\node[node] (b5) at (0,-1.2) {$b_4$};
\node[node] (t2) at (1,-1.2) {$t_2$};
\node[node] (b8) at (2,-1.2) {$b_{11}$};

\node[node] (t0) at (-2,-2.4) {$t_0$};

\node[node] (t3) at (0,-2.4) {$t_3$};

\draw[edge] (L) -- (b1);
\draw[edge] (L) -- (t1);
\draw[edge] (L) -- (b5);
\draw[edge] (L) -- (t2);
\draw[edge] (L) -- (b8);

\draw[edge] (b1) -- (t0);

\draw[edge] (b5) -- (t3);

\draw[edge] (b8) -- (t3);

\node[fill=white,font=\tiny,xshift=6mm,yshift=2.0mm] (b1l) at (b1.north) {$c_{1}=2$};
\node[fill=white,font=\tiny,xshift=6mm,yshift=2.0mm] (b4l) at (b5.north) {$c_{1}=0$};
\node[fill=white,font=\tiny,xshift=6mm,yshift=2.0mm] (b11l) at (b8.north) {$c_{11}=0$};

\end{tikzpicture}}}
            }
            
                \end{minipage}
           
            \subfigure[Final Partitions.]{%
                \label{fig:final_partitions}
                \includegraphics[width=0.23\textwidth]{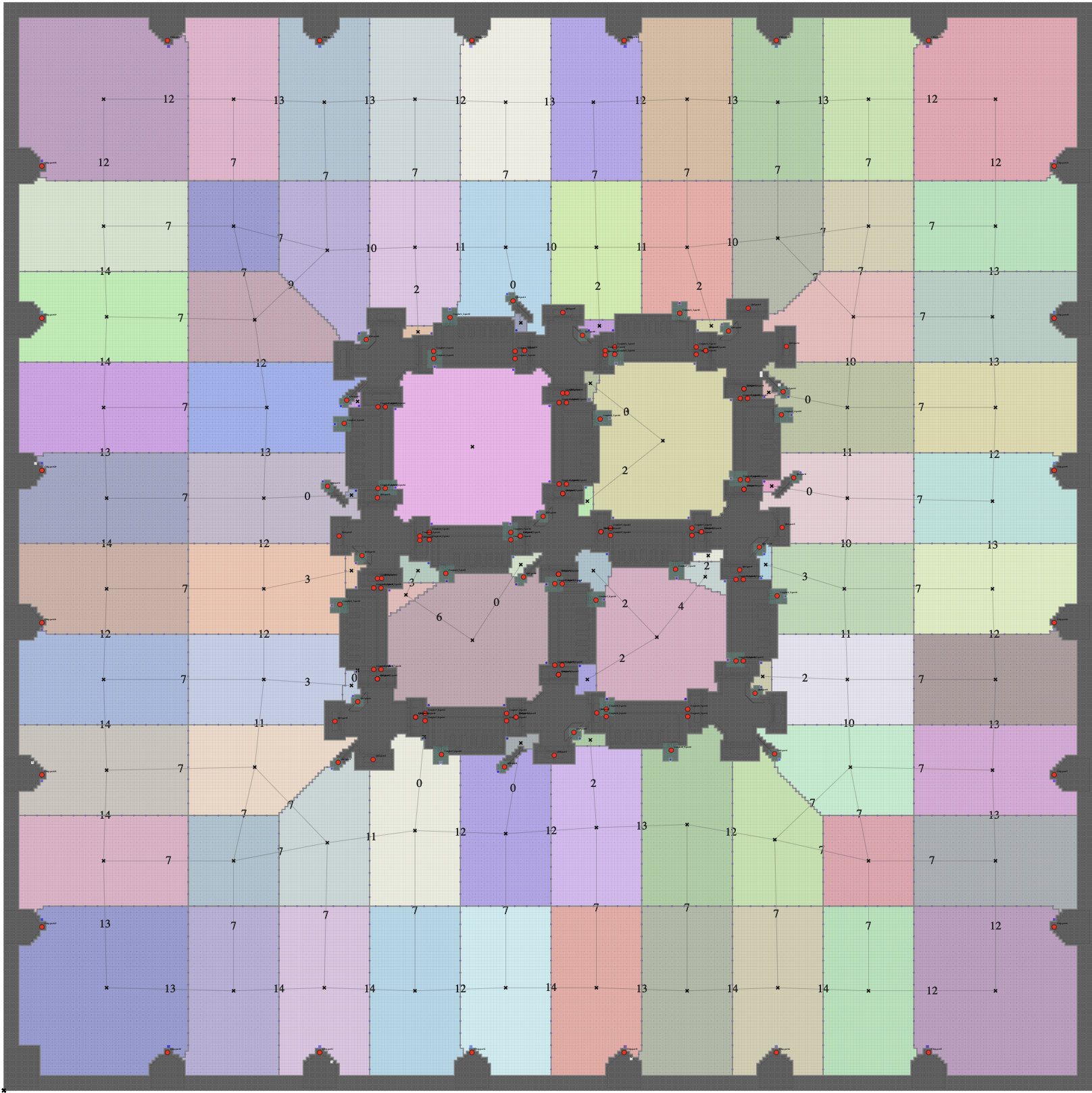}
            }
            \begin{minipage}[b]{0.23\textwidth}
            \centering
              \subfigure[Pruned Bottlenecks.]{%
                \label{fig:pruned_bottlenecks}
                \input{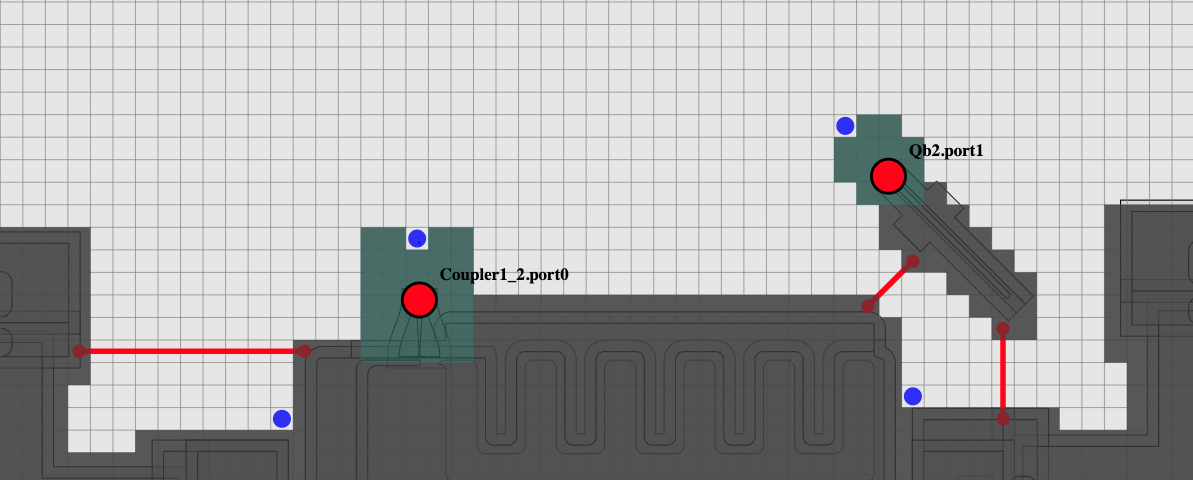}%
            }\\[1ex] %
            \subfigure[Bottleneck Partitioning.]{%
                \label{fig:bottleneck_partitioning}
                \includegraphics[width=\textwidth]{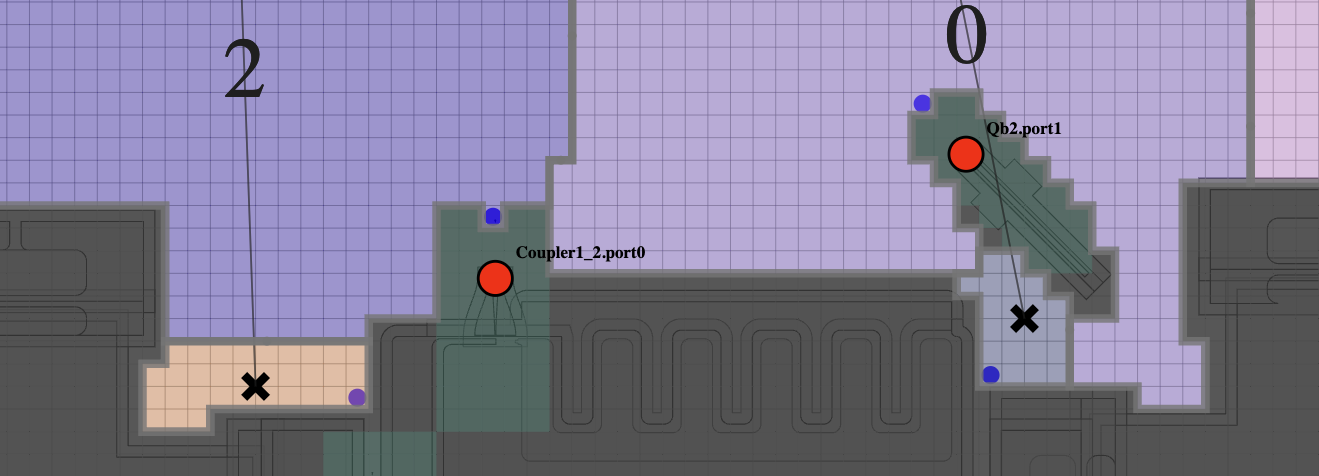}%
            }
            
        \end{minipage}
            \vspace{-2.0mm}
            \caption{Geometry-aware global partitioning: the layout is rasterized and its Voronoi graph analyzed to detect narrow passages, whose topology is captured and pruned in a cavity tree, inducing the final partition boundaries.}
            \label{fig:partitioning}
        \end{figure}
   While \emph{Voronoi Graphs} are already well-established in routing for achieving high clearance (such as in \cite{voronoi}) and in general spatial analysis (such as in \cite{voronoi_analy}), we leverage them here to identify bottlenecks and cavities within our initialized grids and partitions. Specifically, we construct a Voronoi graph of the routing domain (\cref{fig:voronoi_graph}) and rasterize it onto the partitioning grid, yielding a set of paths that trace the free space between obstacles. We then compute the distance transform map $D$ of the partitioning grid, where $D(v)$ assigns each cell $v$ the distance to its nearest obstacle. As an approximating heuristic, bottleneck centers are detected by traversing all non-branching Voronoi nodes $v$ with neighbors $v_0, v_1$ and locating local minima, satisfying $D(v_0) \geq D(v) \land D(v_1) \geq D(v)$, as well as saddle points, satisfying $\big(D(v_0) = D(v) \land D(v_1) \geq D(v)\big) \lor \big(D(v_1) = D(v) \land D(v_0) \geq D(v)\big)$. For each detected center, the physical bottleneck line is reconstructed by first retrieving the nearest obstacle to the Voronoi node and then determining the nearest obstacle on the opposite side of that first obstacle. This yields a complete set of bottleneck lines that describe all narrow passages in the layout.
\subsubsection{Cavity Trees and Geometry-Aware Partitions}\label{subsub:cavity} 
As the bottlenecks alone do not yet reveal which targets they actually 
restrict. We therefore construct a cavity tree as a \emph{Directed Acyclic Graph}~(DAG) rooted at the chip boundary or launcher $L$, containing bottleneck nodes $b_i$ with capacity $c_i = \lfloor \frac{d_{b_i}} {d_\text{clear}} \rfloor$ (derived from the obstacle distance $d_{b_i}$) and target nodes $t_i$. To map the topological dependencies between bottlenecks and target ports, a leaf-to-root \emph{Breadth-First Search}~(BFS) is executed starting from the target nodes (leaves): Initially, all detected bottlenecks are blocked, and the BFS expands outward through unblocked space; encountered bottlenecks are instantiated as parent nodes, while reaching a launcher port terminates the path as a root. This recurses across newly discovered bottlenecks until all branches connect to launcher roots. Finally, sequential bottlenecks along a path are pruned to retain only the most restrictive (smallest) one, and unconnectable dead ends are removed.
\vspace{-1.5mm}
\begin{example}[Voronoi-Based Bottleneck Detection]
    \Cref{fig:voronoi_graph} shows the Voronoi graph of the layout. Since Voronoi edges trace loci of maximal clearance, analyzing the distance transform values along them reveals narrow passages: the detected bottleneck centers are highlighted in red in \cref{fig:bottleneck_detection}, with the reconstructed bottleneck lines indicated by arrows.
    The dedicated cavity for the bottleneck in \cref{fig:bottleneck_detection} is represented as a cavity tree in \cref{fig:cavity_tree}. This tree captures the topological relationships: starting from the outer boundary $L$, the target $t_0$ can be reached only by passing sequentially through the bottlenecks $b_0$ and $b_1$, each with capacity $2$. Conversely, $t_1$ can be reached without traversing any bottlenecks, whereas the two passages leading to $t_3$ are both too narrow for a wire. Furthermore, the cavity tree can be pruned: for sequential bottlenecks, only the most restrictive (smallest) one dominates, and branches without target nodes can be removed. This results in the pruned tree in \cref{fig:cavity_tree_pruned}, which corresponds to the dominating bottlenecks in \cref{fig:pruned_bottlenecks}.
\end{example}
\vspace{-2.0mm}
Once the cavity trees are pruned, the remaining cavities are defined and integrated as additional partitions alongside the initial global cells $G_0, G_1, \dots, G_k$, as shown in \cref{fig:final_partitions} and \cref{fig:bottleneck_partitioning}. Finally, full coverage of the layout is achieved by assigning any remaining unpartitioned grid nodes to the  closest established partition. The resulting partitioning captures all relevant geometric bottlenecks within the layout. On the one hand, these partitions are used during the port assignment stage to enforce routable inner assignments by constraining bridge usage; on the other hand, they serve as the foundational global routing graph for the subsequent routing stage. Based on these partitions, the next step computes the port assignment (\cref{subsec:assignm}).
\vspace{-1.0mm}
\subsection{Port Assignment Computation}
\vspace{-0.5mm}
\label{subsec:assignm}
As the next step in the design automation flow, we introduce a flow that computes a port assignment satisfying $R_\text{assign}$, determining all point-to-point assignments $W_p$ and resonator assignments $W_r$, as well as allocating the feedlines $W_f$, while optimizing the bridge count $O_b$ and the expected wire length $O_l$. As established during partitioning (\cref{subsec:global_part}), some target ports of $W_p$ and $W_r$ are not reachable from the outer chip boundary and must therefore be accessed via dedicated bridges. To handle this, we split the assignment into two steps: First, as detailed in  \cref{subsub:innerass}, the \emph{inner assignment} maps all inner target ports ($t_\text{ctrl}$, $t_\text{read}$, $t_\text{flux}$) to bridges reachable from the outer boundary, modeled as a minimum-cost integer network flow problem. Then, based on these bridge assignments, \cref{subsub:outerass} presents the \emph{outer assignment}, which allocates all nets $W$ by determining source and target ports for $W_p$ and $W_r$, and ordered terminal sequences for $W_f$, for which we introduce an assignment graph representation. Both steps are formulated as Integer Linear Programs (ILPs): As shown in \cref{sec:experiments}, the resulting instances remain small and can be solved exactly within negligible runtime, while $R_\text{assign}$ and the cavity capacity limits are guaranteed to be satisfied, avoiding assignment decisions that would only surface as failures in the later routing stages. Note that while each step is solved exactly, the sequential decomposition renders the overall solution optimal only with respect 
to the fixed inner assignment.
\vspace{-0.5mm}
\subsubsection{Inner Assignment}\label{subsub:innerass}
\vspace{-0.5mm}
For the inner assignment, each inner target port serves as a sink with unit demand, while the dedicated bridges serve as sources whose capacities are constrained by the cavity trees. A feasible flow then corresponds directly to a valid, non-overlapping assignment of targets to bridges, and the minimum-cost solution yields the assignment with the minimum total wire length. To this end, a \emph{Hanan Grid} $H_i$ is initialized for each partition $I_i$ in $I$, considering all contained bridge ports and inner target ports, on which the following ILP is formulated as a minimum-cost flow problem. For each node $i$ in the Hanan grid with adjacent nodes $j \in H_i$, let $w_{ij}$ denote the edge length between $i$ and $j$. Directed integer flow variables $h_{ij}, h_{ji} \in [0, 1]$ are defined for every adjacent node pair. Flow conservation is enforced for all grid nodes: intermediate nodes $i$ satisfy $\sum_{j \in H_i} (h_{ij} - h_{ji}) = 0$, while target nodes $i$ satisfy $\sum_{j \in H_i} h_{ji} = 1$, ensuring a net inflow of one. To further ensure that only physically routable bridge assignments are selected, the cavity trees derived in \cref{subsec:global_part} are integrated into the ILP. For each tree edge between parent $i$ and child $j$, an integer flow variable $e_{ij}$ is introduced, with flow conservation enforced at every bottleneck node $i$ ($\sum_k e_{ki} = \sum_j e_{ij}$). The variables are bounded by $e_{ij} \in [0, c_j]$ if child $j$ is a bottleneck with capacity $c_j$, $e_{ij} \in [0, 1]$ if $j$ is a dedicated bridge terminal, and $e_{ij} = 1$ if $j$ is a mandatory terminal, thereby limiting the number of nets traversing each bottleneck. The cavity tree is then coupled to the Hanan grid: For each bridge terminal $j$ with parent $i$ and corresponding Hanan grid node $k$, the tree inflow must equal the total grid outflow, i.e., $e_{ij} = \sum_{l \in H_k} h_{kl}$. Consequently, a bridge can only carry nets if the cavity tree admits sufficient capacity along its path to the outer boundary.
Finally, the ILP minimizes the total wire length: $\min \sum_{(i,j)} (h_{ij} + h_{ji}) \, w_{ij}$.
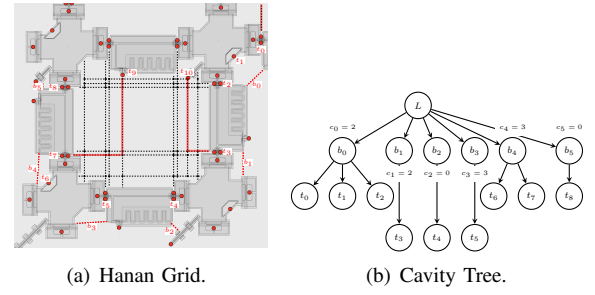
\begin{figure}[!t]  
    \centering
    \subfigure[Hanan Grid.]{%
        \label{fig:hanan_grid}
       \scalebox{0.5}{\hbox{\input{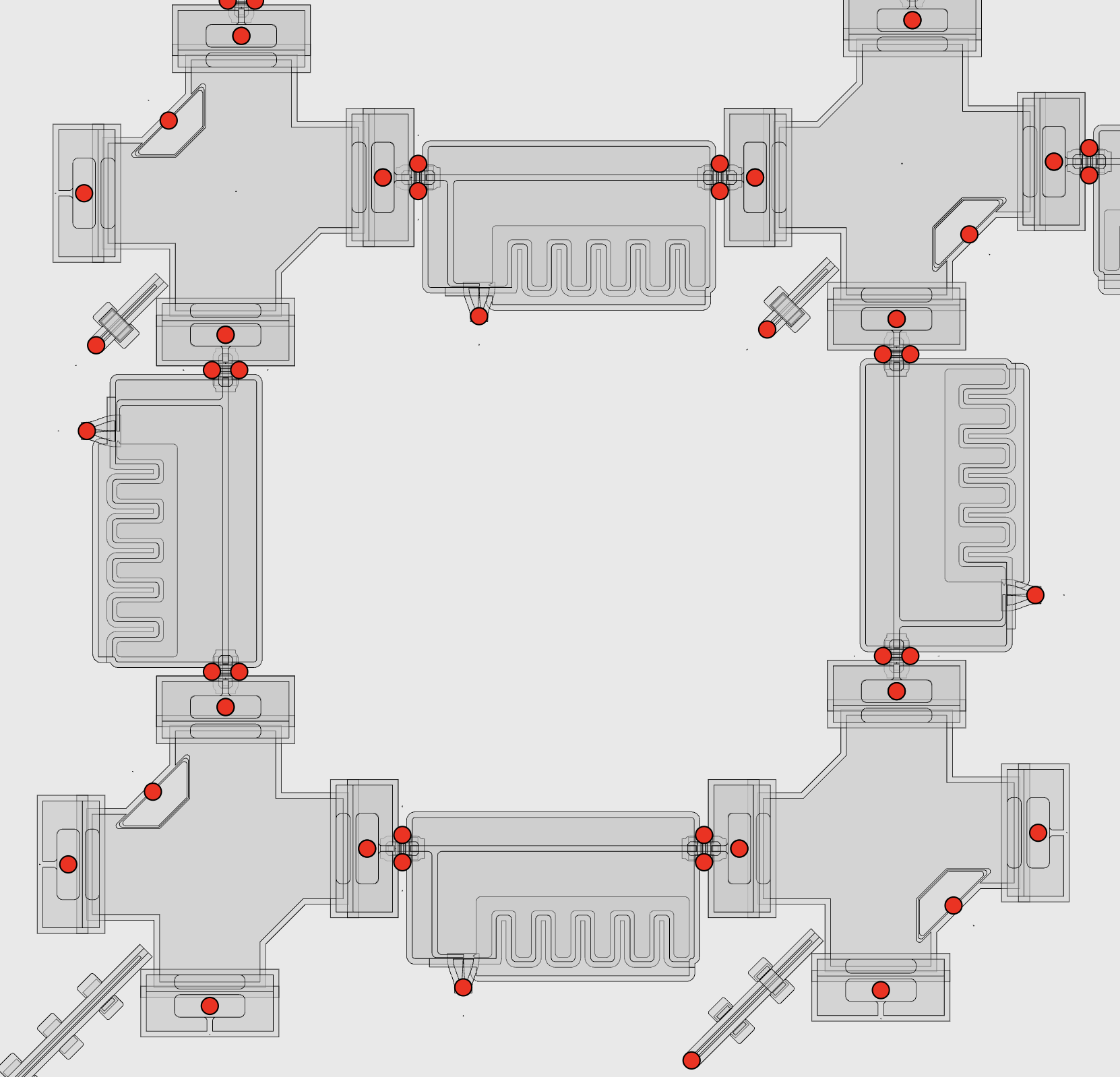}}}%
       }
       \subfigure[Cavity Tree.]{%
       \label{fig:inner_cavity_tree}
       \scalebox{0.5}{\hbox{\begin{tikzpicture}[
    >=stealth,
    node/.style={
        circle,
        draw=black,
        thick,
        minimum size=7mm,
        inner sep=0pt,
        font=\scriptsize
    },
    terminal/.style={
        rectangle,
        draw=black,
        thick,
        minimum width=8mm,
        minimum height=5mm,
        inner sep=1pt,
        font=\scriptsize
    },
    edge/.style={
        ->,
        thick
    },
    prunednode/.style={
        circle,
        draw=gray!60,
        thick,
        fill=gray!10,
        minimum size=7mm,
        inner sep=0pt,
        font=\scriptsize,
        text=gray!70
    },
    title/.style={
        font=\scriptsize
    }
]

\node[node] (L)  at (0,0) {$L$};
\node[node] (b0) at (-2,-1.2) {$b_0$};
\node[node] (b1) at (-0.5,-1.2) {$b_1$};

\node[node] (b2) at (0.5,-1.2) {$b_2$};
\node[node] (b3) at (1.5,-1.2) {$b_3$};
\node[node] (b4) at (2.5,-1.2) {$b_4$};
\node[node] (b5) at (4.0,-1.2) {$b_5$};

\node[node] (t0) at (-3,-2.4) {$t_0$};
\node[node] (t1) at (-2,-2.4) {$t_1$};
\node[node] (t2) at (-1,-2.4) {$t_2$};

\node[node] (t3) at (-0.5,-3.5) {$t_3$};

\node[node] (t4) at (0.5,-3.5) {$t_4$};

\node[node] (t5) at (1.5,-3.5) {$t_5$};

\node[node] (t6) at (2.0,-2.4) {$t_6$};
\node[node] (t7) at (3.0,-2.4) {$t_7$};

\node[node] (t8) at (4.0,-2.4) {$t_8$};

\draw[edge] (L) -- (b0);
\draw[edge] (L) -- (b1);
\draw[edge] (L) -- (b2);
\draw[edge] (L) -- (b3);
\draw[edge] (L) -- (b4);
\draw[edge] (L) -- (b5);

\draw[edge] (b0) -- (t0);
\draw[edge] (b0) -- (t1);
\draw[edge] (b0) -- (t2);

\draw[edge] (b1) -- (t3);

\draw[edge] (b2) -- (t4);

\draw[edge] (b3) -- (t5);

\draw[edge] (b4) -- (t6);
\draw[edge] (b4) -- (t7);

\draw[edge] (b5) -- (t8);

\node[yshift=2.5mm,fill=white,font=\tiny] (b0c) at (b0.north) {$c_0=2$};
\node[yshift=-2.5mm,fill=white,font=\tiny] (b1c) at (b1.south) {$c_1=2$};
\node[yshift=-2.5mm,fill=white,font=\tiny] (b2c) at (b2.south) {$c_2=0$};
\node[yshift=-2.5mm,fill=white,font=\tiny] (b3c) at (b3.south) {$c_3=3$};
\node[yshift=2.5mm,fill=white,font=\tiny] (b4c) at (b4.north) {$c_4=3$};
\node[yshift=2.5mm,fill=white,font=\tiny] (b5c) at (b5.north) {$c_5=0$};

\end{tikzpicture}}}%
       }
       \vspace{-2mm}
    \caption{Inner assignment computation: (a)~Hanan grid of an inner region, showing the detected bottlenecks, the target ports, and the resulting assignment of bridge paths; (b)~the corresponding cavity tree, capturing all target ports and their topological relation to the bottlenecks, which constrains the admissible bridge utilization.}
    \label{fig:inner_assignment}
\end{figure}
\vspace{-1.5mm}
\begin{example}[Constrained Inner Assignment]
    The inner assignment determines which target ports located within inner regions are connected to dedicated bridges, as illustrated by the red highlighted paths in \cref{fig:hanan_grid} (e.g., connecting $t_9$ to $t_7$ and $t_{10}$ to $t_3$). Additionally, the cavity tree constraints shown in \cref{fig:inner_cavity_tree} ensure that only physically routable assignments are made: For instance, assigning $t_8$ is prohibited since bottleneck $b_5$ has a capacity of $c_5=0$. Similarly, of the two bridges with terminals $t_0$ and $t_2$, only one can be utilized, as the bottleneck $b_0$ with capacity $c_0=2$ must also accommodate the mandatory readout net.
\end{example}
\vspace{-1.5mm}
\subsubsection{Outer Assignment}\label{subsub:outerass}
\begin{figure}[!t]  
    \centering
    \subfigure[Assignment Graph]{\ \ \label{fig:assignment_graph}%
       \scalebox{0.5}{\hbox{\begin{tikzpicture}[
    >=stealth,
    node/.style={
        circle,
        draw=black,
        thick,
        minimum size=7mm,
        inner sep=0pt,
        font=\scriptsize
    },
    resonode/.style={
        circle,
        draw=black,
        fill=black!30,
        thick,
        minimum size=7mm,
        inner sep=0pt,
        font=\scriptsize
    },
    terminal/.style={
        rectangle,
        draw=black,
        thick,
        minimum width=8mm,
        minimum height=5mm,
        inner sep=1pt,
        font=\scriptsize
    },
    edge/.style={
        ->,
        thick
    },
    arrowedge/.style={
        ->,
        thick,
        dashed
    },
    prunednode/.style={
        circle,
        draw=gray!60,
        thick,
        fill=gray!10,
        minimum size=7mm,
        inner sep=0pt,
        font=\scriptsize,
        text=gray!70
    },
    title/.style={
        font=\scriptsize
    }
]

\node[node] (t0)  at (0,0) {$t_{\text{ctrl}_1}$};
\node[node] (t1) at (-1,0) {$t_{\text{flux}_0}$};
\node[node] (t2) at (-2,0) {$t_{\text{ctrl}_0}$};

\node[resonode] (t3) at (-2.5,-1) {$t_{\text{read}_0}$};
\node[node] (l0) at (-3.5,-0) {$l_0$};
\node[node] (t4) at (-3,-2) {$t_{\text{flux}_2}$};
\node[node] (t5) at (-3,-3) {$t_{\text{ctrl}_2}$};

\node[resonode] (t6) at (-2,-3.5) {$t_{\text{read}_2}$};
\node[node] (l1) at (-3,-4.5) {$l_1$};
\node[node] (t7) at (-1,-4) {$t_{\text{flux}_1}$};
\node[node] (t8) at (0,-4) {$t_{\text{ctrl}_3}$};

\node[resonode] (t9) at (0.5,-3) {$t_{\text{read}_3}$};
\node[node] (l2) at (1.5,-4) {$l_2$};
\node[node] (t10) at (1,-2) {$t_{\text{flux}_3}$};
\node[resonode] (t11) at (0.5,-1) {$t_{\text{read}_1}$};
\node[node] (l3) at (1.5,0) {$l_3$};

\draw[edge] (t3) -- node[midway, font=\footnotesize,yshift=2.5mm] {$o_{01}$} (t11) node[midway, font=\footnotesize,yshift=-2.5mm] {$m_{01}=3$}(l0);
\draw[edge] (t3) -- node[midway, font=\footnotesize,yshift=-2mm] {$o_{0}$}  (l0) ;
\draw[edge] (t6) -- node[midway, font=\footnotesize,xshift=3mm] {$o_{20}$} node[midway, font=\footnotesize,xshift=5.5mm,yshift=3mm] {$m_{20}=1$} (t3);
\draw[edge] (t9) -- node[midway, font=\footnotesize,yshift=2mm] {$o_{32}$} node[midway, font=\footnotesize,yshift=5mm] {$m_{32}=2$}(t6);
\draw[edge] (t6) -- node[midway, font=\footnotesize,xshift=3mm] {$o_2$}(l1);
\draw[edge] (t11) -- node[midway, font=\footnotesize,xshift=-3mm] {$o_{13}$} node[midway, font=\footnotesize,xshift=-6mm,yshift=2mm] {$m_{13}=1$}(t9);
\draw[edge] (t9) -- node[midway, font=\footnotesize,xshift=3mm] {$o_3$}(l2);
\draw[edge] (t11) -- node[midway, font=\footnotesize,xshift=3mm] {$o_1$}(l3);

\draw[arrowedge] (t3) -- (t2);
\draw[arrowedge] (t2) -- (t1);
\draw[arrowedge] (t1) -- (t0);
\draw[arrowedge] (t0) -- (t11);
\draw[arrowedge] (t11) -- (t10);
\draw[arrowedge] (t10) -- (t9);
\draw[arrowedge] (t9) -- (t8);
\draw[arrowedge] (t8) -- (t7);
\draw[arrowedge] (t7) -- (t6);
\draw[arrowedge] (t6) -- (t5);
\draw[arrowedge] (t5) -- (t4);
\draw[arrowedge] (t4) -- (t3);

\node[yshift=2mm,font=\footnotesize] (l0) at (t3.north) {$f_0$};
\node[yshift=2mm,font=\footnotesize] (l0) at (t2.north) {$f_1$};
\node[yshift=2mm,font=\footnotesize] (l0) at (t1.north) {$f_2$};
\node[yshift=2mm,font=\footnotesize] (l0) at (t0.north) {$f_3$};
\node[yshift=2mm,font=\footnotesize] (l0) at (t11.north) {$f_4$};
\node[yshift=2mm,font=\footnotesize] (l0) at (t10.north) {$f_5$};
\node[xshift=2mm,font=\footnotesize] (l0) at (t9.east) {$f_6$};
\node[yshift=2mm,font=\footnotesize] (l0) at (t8.north) {$f_7$};
\node[yshift=2mm,font=\footnotesize] (l0) at (t7.north) {$f_8$};
\node[yshift=2mm,xshift=2mm,,font=\footnotesize] (l0) at (t6.north) {$f_9$};
\node[xshift=2mm,font=\footnotesize] (l0) at (t5.east) {$f_{10}$};
\node[yshift=2mm,font=\footnotesize] (l0) at (t4.north) {$f_{11}$};

\end{tikzpicture}}}%
    }
    \subfigure[Solved Assignment Graph]{\ \ \label{fig:assignment_graph_solved}%
       \scalebox{0.5}{\hbox{\begin{tikzpicture}[
    >=stealth,
    node/.style={
        circle,
        draw=black,
        thick,
        minimum size=7mm,
        inner sep=0pt,
        font=\scriptsize
    },
    resonode/.style={
        circle,
        draw=black,
        fill=black!30,
        thick,
        minimum size=7mm,
        inner sep=0pt,
        font=\scriptsize
    },
    terminal/.style={
        rectangle,
        draw=black,
        thick,
        minimum width=8mm,
        minimum height=5mm,
        inner sep=1pt,
        font=\scriptsize
    },
    edge/.style={
        ->,
        thick
    },
    arrowedge/.style={
        ->,
        thick,
        dashed
    },
    prunednode/.style={
        circle,
        draw=gray!60,
        thick,
        fill=gray!10,
        minimum size=7mm,
        inner sep=0pt,
        font=\scriptsize,
        text=gray!70
    },
    title/.style={
        font=\scriptsize
    }
]

\node[node] (t0)  at (0,0) {$t_{\text{ctrl}_1}$};
\node[node] (t1) at (-1,0) {$t_{\text{flux}_0}$};
\node[node] (t2) at (-2,0) {$t_{\text{ctrl}_0}$};

\node[resonode] (t3) at (-2.5,-1) {$t_{\text{read}_0}$};
\node[node] (l0) at (-3.5,-0) {$l_0$};
\node[node] (t4) at (-3,-2) {$t_{\text{flux}_2}$};
\node[node] (t5) at (-3,-3) {$t_{\text{ctrl}_2}$};

\node[resonode] (t6) at (-2,-3.5) {$t_{\text{read}_2}$};
\node[node] (l1) at (-3,-4.5) {$l_1$};
\node[node] (t7) at (-1,-4) {$t_{\text{flux}_1}$};
\node[node] (t8) at (0,-4) {$t_{\text{ctrl}_3}$};

\node[resonode] (t9) at (0.5,-3) {$t_{\text{read}_3}$};
\node[node] (l2) at (1.5,-4) {$l_2$};
\node[node] (t10) at (1,-2) {$t_{\text{flux}_3}$};
\node[resonode] (t11) at (0.5,-1) {$t_{\text{read}_1}$};
\node[node] (l3) at (1.5,0) {$l_3$};

\draw[edge] (t3) -- node[midway, font=\footnotesize,yshift=-2mm] {$o_0$}  (l0) ;
\draw[edge] (t6) -- node[midway, font=\footnotesize,xshift=2.5mm] {$o_{20}$} (t3);
\draw[edge] (t9) -- node[midway, font=\footnotesize,yshift=2mm] {$o_{32}$} (t6);

\draw[edge] (t11) -- node[midway, font=\footnotesize,xshift=-2.5mm] {$o_{13}$} (t9);
\draw[edge] (t11) -- node[midway, font=\footnotesize,xshift=2mm,yshift=-1] {$o_1$}(l3);

\draw[arrowedge] (t3) -- (t2);
\draw[arrowedge] (t2) -- (t1);
\draw[arrowedge] (t1) -- (t0);
\draw[arrowedge] (t0) -- (t11);
\draw[arrowedge] (t11) -- (t10);
\draw[arrowedge] (t10) -- (t9);
\draw[arrowedge] (t9) -- (t8);
\draw[arrowedge] (t8) -- (t7);
\draw[arrowedge] (t7) -- (t6);
\draw[arrowedge] (t6) -- (t5);
\draw[arrowedge] (t5) -- (t4);
\draw[arrowedge] (t4) -- (t3);

\node[yshift=-1.0mm,xshift=8.5mm,font=\footnotesize] (l0) at (t3.north) {$f_0=14$};
\node[yshift=2mm,font=\footnotesize] (l0) at (t2.north) {$f_1=1$};
\node[yshift=2mm,font=\footnotesize] (l0) at (t1.north) {$f_2=2$};
\node[yshift=2mm,font=\footnotesize] (l0) at (t0.north) {$f_3=3$};
\node[yshift=-2mm,xshift=-8mm,font=\footnotesize] (l0) at (t11.north) {$f_4=4$};
\node[yshift=2mm,xshift=3mm,font=\footnotesize] (l0) at (t10.north) {$f_5=5$};
\node[xshift=4.5mm,font=\footnotesize] (l0) at (t9.east) {$f_6=7$};
\node[yshift=-9mm,font=\footnotesize] (l0) at (t8.north) {$f_7=8$};
\node[yshift=1.5mm,xshift=2.5mm,font=\footnotesize] (l0) at (t7.north) {$f_8=9$};
\node[yshift=-8mm,xshift=0 mm,,font=\footnotesize] (l0) at (t6.north) {$f_9=10$};
\node[xshift=-10mm,yshift=4.5mm,font=\footnotesize] (l0) at (t5.east) {$f_{10}=11$};
\node[yshift=2mm,xshift=-4mm,font=\footnotesize] (l0) at (t4.north) {$f_{11}=13$};

\end{tikzpicture}}}%
    }
    \vspace{-2mm}
    \caption{Outer assignment computation: (a)~assignment graph, and (b)~solved assignment graph corresponding to the final port assignment.}
    \label{fig:outer_assignment}
\end{figure}
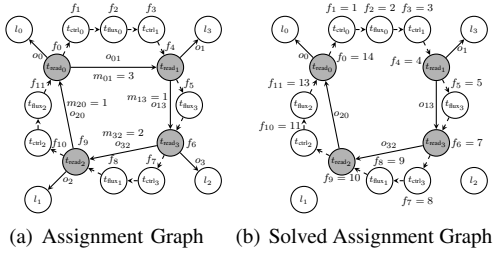

With the inner assignment computed, the remaining \emph{outer assignment} allocates the actual nets $W$ by determining source and target ports for $W_p$ and $W_r$ as well as the ordered terminal sequences for $W_f$. To this end, we propose an assignment graph representation that is converted into an ILP, enforcing the remaining constraints of $R_\text{assign}$ while minimizing bridge usage ($O_b$) and expected wire length ($O_l$). All target and launcher nodes are ordered in \emph{cyclic order}, according to their geometric position along the chip boundary in a clockwise direction. The assignment graph consists of $q$ resonator nodes ($t_{\text{read}_0}, \dots, t_{\text{read}_{q-1}}$) with a dedicated launcher node ($l_0, \dots, l_{q-1}$) assigned to each resonator port, as well as $q+c$ control/flux nodes ($t_{\text{ctrl}_0}, \dots, t_{\text{flux}_0}, \dots$). Each resonator node $i$ is connected in cyclic order via directed edges $o_{ij} \in [0,1]$ to its two adjacent resonator nodes, and via an edge $o_i \in [0,1]$ to its launcher node $l_i$, indicating the start or end of a feedline. As visualized in \cref{fig:assignment_graph}, each edge $o_{ij}$ is annotated with a weight $m_{ij}$ representing the number of bridges required if that edge were assigned to a feedline. To enforce $R_f$ (valid ordered terminal sequences starting and ending at launchers with at most $r_\text{util}$ CPW couplers) and $R_r$ (assigning CPW couplers to resonator ports), we formulate the following ILP constraints: For each resonator node $i$, exactly two adjacent edges are selected, i.e., $\sum_j o_{ij} + o_i = 2$. The number of launcher edges is constrained by $\sum_i o_i = 2 \cdot |W_f|$, ensuring that each feedline starts and ends at a launcher port. Furthermore, each resonator node carries an integer utilization variable $u_i \in [0, r_\text{util}]$, which is set to $1$ if the node starts a feedline ($o_i = 1 \land o_{ki} = 0$) and otherwise increments along the feedline ($u_i = u_{i_\text{prev}} + 1$), thereby bounding each feedline to at most $r_\text{util}$ couplers. To additionally enforce $R_p$, $R_\text{lnc}$, and $R_\text{port}$, i.e. each net $W_p$ connects exactly two terminals, each launcher is used at most once, and every flux and control target node is assigned to a net in $W_p$, we define cyclic assignment variables $f_i \in [0, n_\text{lnc}]$ for all non-launcher nodes, where $n_\text{lnc}$ denotes the number of launchers. Regular ports are constrained to preserve the cyclic ordering, $f_i > f_{i-1}$, preventing crossing net assignments, which would otherwise require additional bridges at each intersection. Furthermore, since $f_i$ is strictly increasing across regular ports, each launcher is assigned at most once ($R_\text{lnc}$), while the bounded domain guarantees that every flux and control node receives a launcher assignment ($R_\text{port}$, $R_p$). Resonator nodes, in contrast, satisfy $f_i \le f_{i-1} + o_i$, indicating that a resonator does not require a separate launcher unless the corresponding launcher edge is explicitly assigned.

 The primary objective minimizes the number of required bridges via the feedline edge weights, $\min \sum_{(i,j)} o_{ij} \cdot m_{ij}$ ($O_b$). For the secondary objective, let $f_i^*$ denote the nearest launcher port for node $i$; the overall expected wire length is then minimized by penalizing the distance of each target to its nearest launcher, $\min \sum_i |f_i - f_i^*|$ ($O_l$). Although ILP is NP-hard in general, our formulation operates on the abstract assignment graph with only $O(q)$ variables and employs symmetry breaking via $f_i > f_{i-1}$ to eliminate equivalent launcher permutations, rendering its solving time negligible compared to the geometric routing stages.
 \vspace{-0.5mm}
\begin{example}[Port Assignment Graph]
    \Cref{fig:assignment_graph} shows the graph representation of the port assignment problem: Each target node from \cref{fig:abstraction_4q} is depicted as a node, arranged in cyclic order. Each resonator node $i$ has three directed edges $o_{ij}$ to the next resonator $j$, $o_{ki}$ from the previous resonator $k$, and $o_i$ to its launcher node, where $o_i$ indicates whether this resonator is the first or last one in a feedline chain. The weights $m_{ij}$ correspond to the number of bridges that need to be instantiated if the edge between $i$ and $j$ were realized. The assignment variables $f_i$ determine which launcher a node is assigned to; for instance, $f_2 = 1$ assigns target $t_{\text{flux}_0}$ to $t_{\text{lnc}_1}$. Furthermore, as seen in \cref{fig:assignment_primary}, the nearest launcher for $t_{\text{ctrl}_1}$ is $t_{\text{lnc}_1}$, yielding $f_1^* = 1$. The solved assignment graph in \cref{fig:assignment_graph_solved} then directly corresponds to the assignment shown in \cref{fig:assignment_primary}.
\end{example}
\vspace{-0.5mm}
The resulting ILP solves the outer port assignment encoded by the variables $f$, satisfying $R_\text{assign}$ while simultaneously minimizing bridge usage ($O_b$) and expected wire length ($O_l$). With the port assignment completed, all nets in $W$ are now mapped to their dedicated ports; the next step is to compute the actual physical routes for these nets (\cref{subsec:routing}).

\vspace{-1mm}
\subsection{Routing}
\vspace{-0.5mm}
\label{subsec:routing}
The previous port assignment stage assigns all  nets in the SCQC layout to dedicated ports; it guarantees a valid assignment as long as sufficient launcher ports are available and the layout cavities admit a feasible assignment; otherwise, the infeasibility is detected and reported by the ILP itself. In this section, we detail the proposed routing flow that takes the assigned nets $W$ as input---at this stage consisting only of source and target ports for $W_p$ and $W_r$, or an ordered terminal sequence for $W_f$---and transforms them into fully routed nets satisfying the minimum straight length ($R_s$), minimum bending radius ($R_r$), minimum clearance ($R_c$), and fixed-length routing ($R_l$) constraints.

The flow builds on two routing algorithms, introduced in 
\cref{subsub:algorithms}: a grid-based A$^*$ operating on discrete cells, and a Dubins A$^*$ that expands arc-move primitives rather than straight grid steps and thereby satisfies $R_r$ and $R_s$ by construction. Since both route a single net at a time, \cref{subsub:conflictres} presents a cyclic rip-up-and-reroute strategy resolving conflicts between nets. Finally, \cref{subsub:pipeline} details how these components form a four-stage hierarchical pipeline, where each stage refines the solution of its predecessor on a finer grid while introducing additional constraints from $R_\text{routing}$. This hierarchy is essential for scalability, as the runtime-critical Dubins router would otherwise have to explore the full layout.
\vspace{-0.5mm}
\subsubsection{Routing Algorithms}
\label{subsub:algorithms}
Both routing algorithms are based on A$^*$ search, expanding the node $n$ with the lowest estimated cost $f(n) = g(n) + h(n)$, where $g(n)$ is the accumulated cost and $h(n)$ is an admissible estimate of the remaining cost. In the plain grid-based variant, nodes are obstacle-free cells of the discretized layout expanded to their eight neighbors, for which the octile distance is an admissible heuristic. A proximity penalty $p_c(n)$, weighted by $\mu_c$, keeps nets spaced apart wherever the layout permits ($O_c$), reducing the blocking situations that rip-up-and-reroute must resolve.

As straight cell-to-cell transitions permit arbitrarily sharp direction changes, this variant cannot express $R_r$ and $R_s$. We therefore additionally employ a Dubins A$^*$ on a finer grid of dimensions $d_\text{width} \times d_\text{height}$, which expands precomputed motion primitives---straight segments and circular arcs respecting the minimum bending radius $r_\text{min}$---with eight discrete orientations in 45-degree increments, as introduced for waveguide routing in \cite{zhou2025lidar}. Since a search state comprises position and orientation, and every expansion is a physically realizable move, $R_r$ and $R_s$ hold by construction. Its cost function adds a bend penalty $p_a(n)$ weighted by $\mu_a$, yielding $f(n) = g(n) + h(n) + \mu_c \cdot p_c(n) + \mu_a \cdot p_a(n)$, which suppresses alternating arc sequences ($O_a$); both penalties are normalized to the grid cell size for resolution-independent behavior. Its enlarged state space, however, makes the Dubins A$^*$ the runtime-critical component of the flow. Finally, the Dubins A$^*$ supports fixed-length routing ($R_l$): if a routed net falls short of $d_{\text{fix}_i}$, point pairs along its path are examined for whether a meander can be inserted without colliding with obstacles, as illustrated by the transition from \cref{fig:dubins_path} to \cref{fig:meander_insertion}.
\vspace{-0.5mm}
\subsubsection{Conflict Resolution} \label{subsub:conflictres}
Since both algorithms route a single net at a time, an already-routed net may block a later one. To resolve such conflicts, all stages employ a cyclic rip-up-and-reroute strategy alternating between forward and backward passes: in a forward pass, the nets are routed in cyclic order and, upon a routing failure, the subsequent nets are iteratively ripped up until the failed net can be routed; in a backward pass, the order is reversed, and the preceding nets are ripped up instead. Since $W_p$ and $W_r$ are, by construction of the assignment, cyclically ordered and mutually crossing-free, a resolvable failure is necessarily caused by the adjacent nets, exactly those covered by the two pass directions.
\vspace{-1.5mm}
\begin{example}[Cyclic Rip-Up-and-Reroute]
    On the example of \cref{fig:global_detailed_routing}, the wires are processed in cyclic 
    order, i.e., $t_{\text{ctrl}_0}$, $t_{\text{flux}_0}$, $t_{\text{ctrl}_1}$, 
    $t_{\text{read}_1}$, $\dots$, $t_{\text{read}_0}$. Upon a routing failure of 
    $t_{\text{ctrl}_0}$ in a forward pass, $t_{\text{flux}_0}$ would be ripped up 
    first, followed by $t_{\text{ctrl}_1}$ if the routing still fails. Conversely, 
    in a backward pass, $t_{\text{read}_0}$ would be ripped up first, followed by 
    $t_{\text{flux}_2}$.
\end{example}
\vspace{-1.0mm}
\subsubsection{Hierarchical Routing Pipeline}
\label{subsub:pipeline}
With the search algorithms and the rip-up-and-reroute routine in place, we can now detail the routing pipeline. The first three stages route the point-to-point and resonator nets ($W_p$, $W_r$), while the final stage generates the feedlines ($W_f$). The former are routed first, since $W_p$ carries the wire-length objective ($O_l$) and would otherwise be forced onto detours around already-routed feedlines.

\begin{figure}[!t]
    \centering
    \subfigure[Global- and Detailed Routing.]{%
       \label{fig:global_detailed_routing}
        \scalebox{1.0}{\input{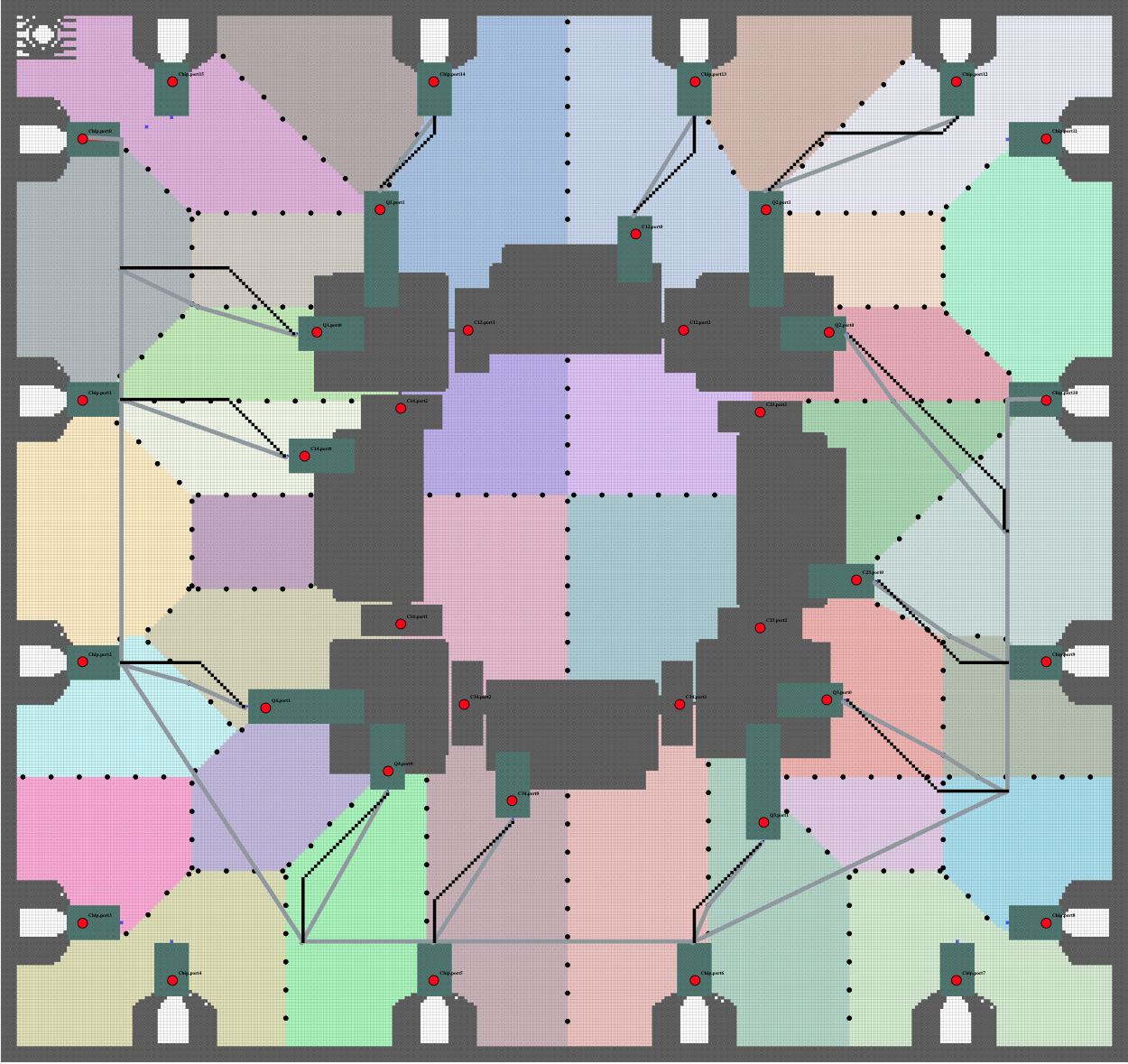}}
    }
    \subfigure[Dubin Routing.]{%
        \label{fig:dubins_routing}
         \scalebox{1.0}{\input{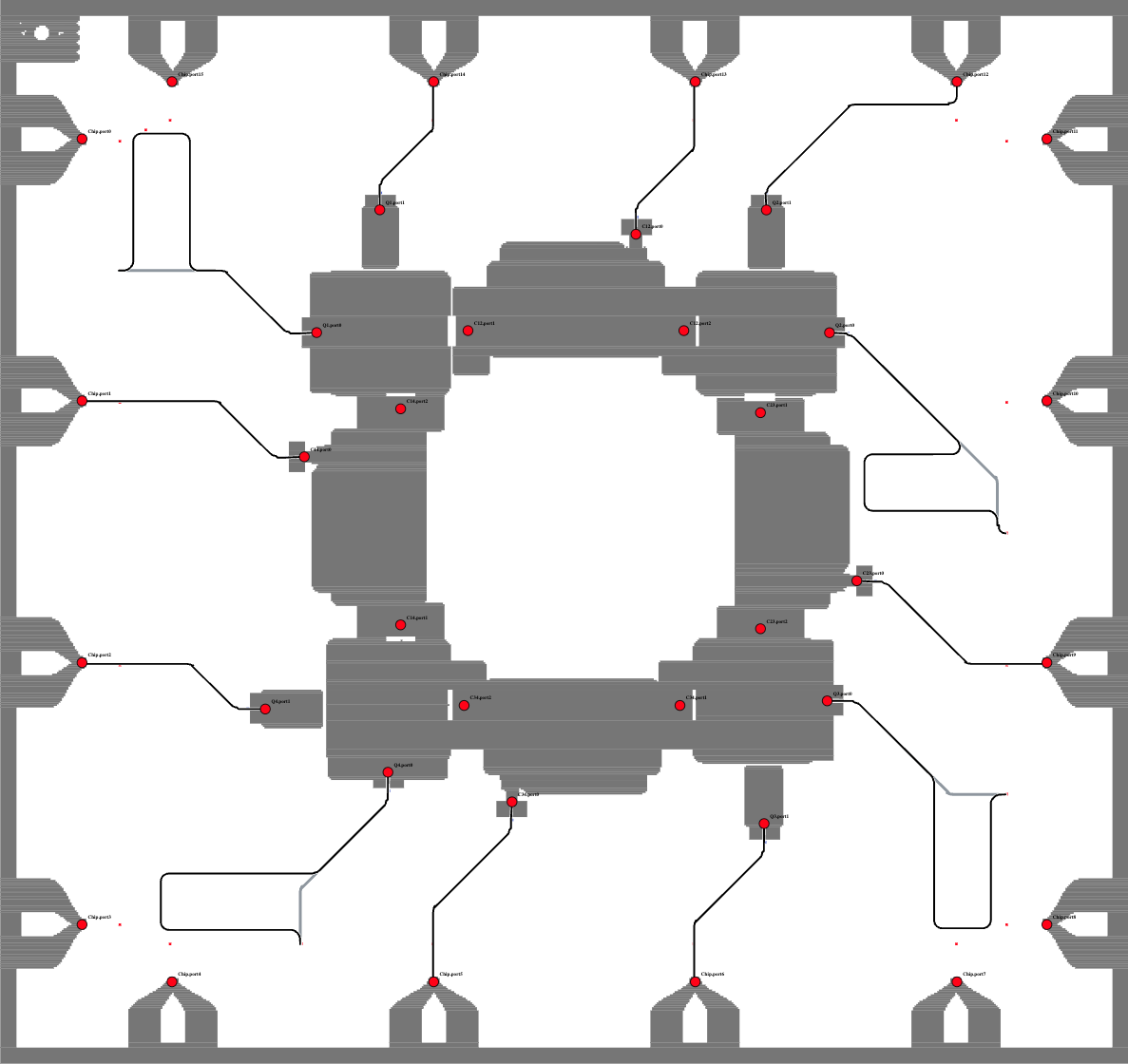}}
    }
    \hspace{-15mm}
    \subfigure[Feedline Routing.]{%
        \label{fig:feedline_routing}    
        \hspace{-15mm}\scalebox{1.0}{\input{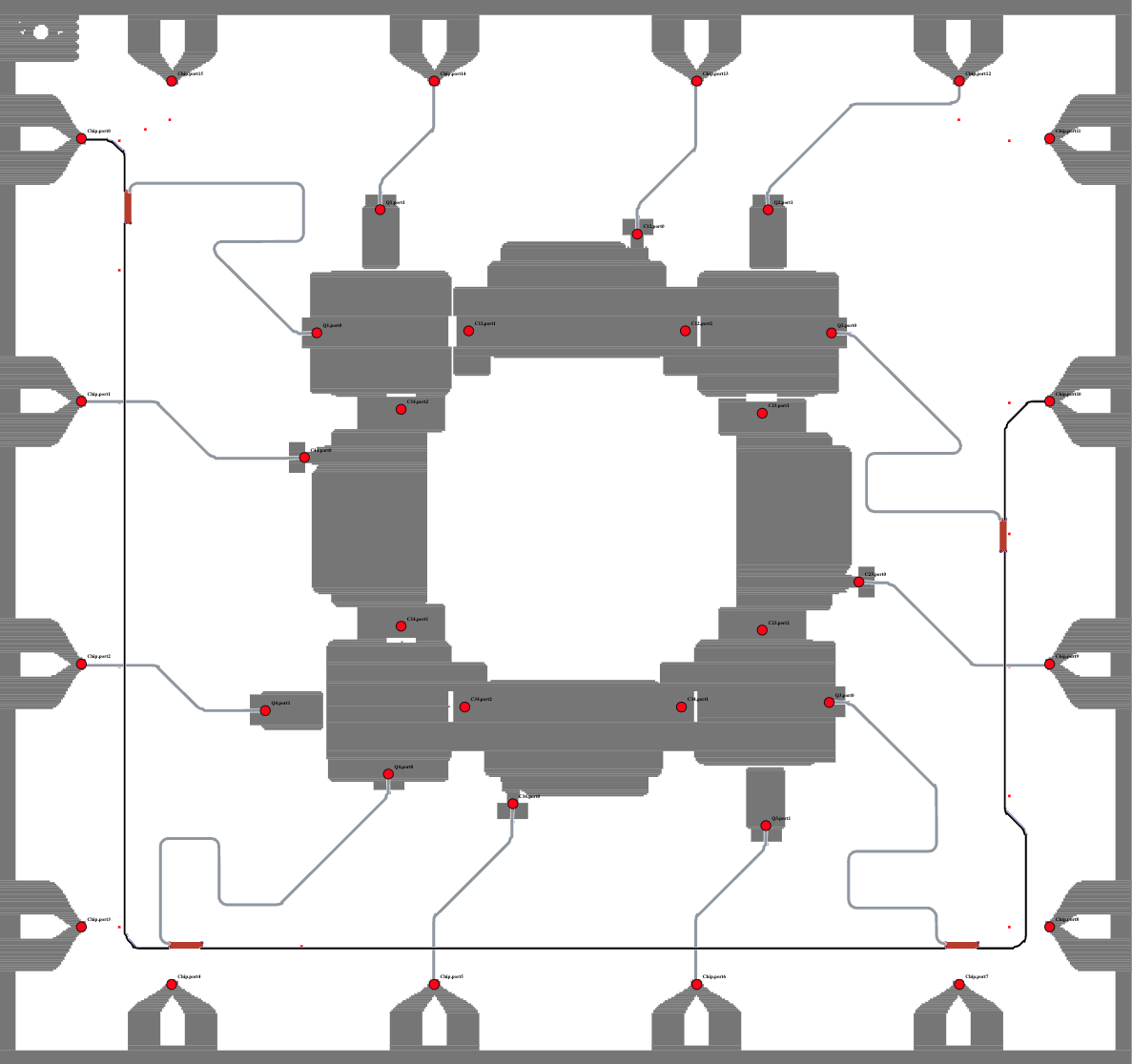}}
    }
    \hspace{2mm}
    \begin{minipage}[b]{0.1\textwidth}
    \centering
    \subfigure[Dubin Path.]{%
        \makebox[1.5\textwidth][c]{%
            \includegraphics[width=0.9\textwidth]{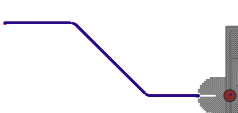}%
        }%
        \label{fig:dubins_path}
    }\\[0.5ex]
    \subfigure[Meander Insertion.]{%
        \makebox[1.5\textwidth][c]{%
            \includegraphics[width=0.9\textwidth]{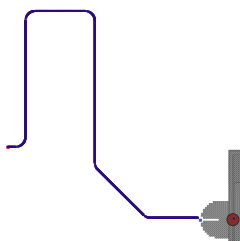}%
        }%
        \label{fig:meander_insertion}
    }
    \end{minipage}
    \caption{Routing pipeline: (a)~global and detailed routing, and (b)~Dubins refinement with bending-radius and meander insertion, followed by (c)~feedline routing; (d-e)~example Dubins path and corresponding meander insertion.}
\label{fig:routing_pipeline}
\end{figure}

The pipeline starts with a global routing phase on the partitions $G$, where each pair of adjacent gcells $i$ and $j$ is assigned a maximum edge capacity $g_{ij}$. On each partition border, exactly $g_{ij}$ ports are initialized at evenly spaced distances as shown in \cref{fig:global_detailed_routing} as grey paths; these border ports form the routing grid for the A$^*$ router, which routes all nets in $W_p$ and $W_r$ while minimizing the wire length ($O_l$). This coarse solution is essential for scalability, as the fine-grained routers of the subsequent stages would otherwise have to explore the full layout.

Then, the detailed router refines the coarse gcell paths on the partitioning grid. For each net, its two direct cyclic neighbors are inserted as obstacles inflated by $d_\text{clear}$, such that 
any collision-free path inherently satisfies $R_c$. Clearance conflicts are deliberately resolved here, as the plain A$^*$ is significantly faster than the runtime-critical Dubins router. The result of the second pipeline stage is illustrated in \cref{fig:global_detailed_routing} as black paths.

Next, as shown in \cref{fig:dubins_routing}, the paths are rerouted with the Dubins A$^*$ on the finer-resolution Dubins grid while retaining the clearance obstacles, thereby ensuring that the minimum straight length ($R_s$) and bending radius ($R_r$) constraints are satisfied. Additionally, each resonator net in $W_r$ must match a fixed length $d_{\text{fix}_i}$ ($R_l$), achieved via meander insertion.

Finally, as shown in \cref{fig:feedline_routing}, the feedlines $W_f$ are generated with the same Dubins A$^*$. If a routed resonator net instead exceeds $d_{\text{fix}_i}$, the corresponding CPW coupler is shifted along the net towards the resonator until the effective length matches, and the CPW couplers are instantiated at the resulting positions. The feedline segments are then sequentially routed along each ordered terminal sequence from the starting launcher through the assigned CPW coupler positions to the terminating launchers. During this stage, $W_p$ and $W_r$ are also selectively rerouted wherever they conflict with the newly introduced feedline segments and bridge crossings, resolving any remaining routing conflicts between the two stages.

\vspace{-2.0mm}
\begin{example}[Routing Pipeline]
    The routing pipeline is visualized in \cref{fig:routing_pipeline}: Global routing first determines coarse paths across the global partitions (\cref{fig:global_detailed_routing}). Subsequently, detailed routing refines these paths while ensuring the minimum clearance $R_c$. In the Dubins routing phase (\cref{fig:dubins_routing}), the minimum bending radius is enforced using discretized angle arcs, and meander structures are inserted for resonators falling short of their mandatory length, e.g., visible at $t_{\text{read}_0}$. Finally, the feedlines are routed using the same Dubins A$^*$ approach in \cref{fig:feedline_routing}.
\end{example}
\vspace{-0.5mm}
The proposed four-stage pipeline transforms the abstract net assignments into a fully routed SCQC layout that satisfies all design rules $R_\text{routing}$ while optimizing wire length ($O_l$), clearance ($O_c$), and bends ($O_a$). Since routing under these constraints is computationally hard, the heuristic pipeline cannot guarantee a valid routing for every instance: a solution may not exist for overly congested layouts, or may be missed despite rip-up-and-reroute. The geometry-aware partitioning and assignment stages, however, are specifically designed to avoid such infeasible configurations upfront. Together, they complete the proposed end-to-end design automation flow, evaluated in the following section on SCQC layouts of varying scales with respect to routability, quality, and runtime.

\vspace{-0.5mm}
\section{Experimental Evaluation}\label{sec:experiments}
\vspace{-0.5mm}
All methods proposed above have been implemented as the open-source tool \texttt{mqt-scpd},\footnote{\href{https://github.com/munich-quantum-toolkit/scpd}{github.com/munich-quantum-toolkit/scpd}} released as part of the \emph{Munich Quantum Toolkit}~(MQT,~\cite{Burgholzer2026}). It implements the three-stage design automation flow described in \cref{sec:desau}, building upon the abstraction introduced in \cref{sec:prelim}, and \emph{automatically} generates complete physical layouts of planar SCQCs, ready for fabrication. To assess the performance and quality of results generated by this flow, evaluations were conducted by our interdisciplinary team of design automation researchers and superconducting hardware experts. In this section, we describe the setup, summarize the results obtained, and discuss the conclusions drawn.
\vspace{-0.5mm}
\subsection{Experimental Setup}
\vspace{-0.5mm}
All experiments were conducted on a macOS 15.7.7 machine equipped with an Apple Silicon M4 Pro CPU and \SI{24}{\giga\byte} of unified RAM.  We implemented all proposed methods and the corresponding workflow in C++20, using \emph{Gurobi}~\cite{gurobi} as the ILP solver due to its outstanding performance on our instance formulation.

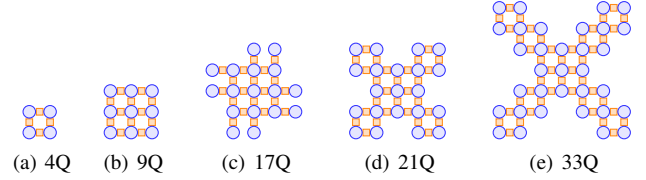
\begin{figure}[!t]
    \centering
    \subfigure[4Q]{%
        \label{fig:benchmark_4q}
        \scalebox{0.50}{\begin{tikzpicture}[qc layout]
  \qcgridlines{2}{2}{1}{1}
  \qcgridnodes{2}{2}{1}{1}
\end{tikzpicture}}
    }
    \hfil
    \subfigure[9Q]{%
        \label{fig:benchmark_9q}
        \scalebox{0.50}{\begin{tikzpicture}[qc layout]
  \qcgridlines{3}{3}{1}{1}
  \qcgridnodes{3}{3}{1}{1}
\end{tikzpicture}}
    }
    \hfil
    \subfigure[17Q]{%
        \label{fig:benchmark_17q}
        \scalebox{0.50}{\begin{tikzpicture}[qc layout]
  \qcgridlines{3}{3}{1}{1}
  \foreach \ax/\ay/\bx/\by in {2/3/2/4, 3/3/3/4,   %
                               3/2/4/2, 3/1/4/1,   %
                               1/1/1/0, 2/1/2/0,   %
                               1/3/0/3, 1/2/0/2}   %
    {\qclink{\ax}{\ay}{\bx}{\by}}
  \qcgridnodes{3}{3}{1}{1}
  \foreach \x/\y in {2/4, 3/4, 4/2, 4/1, 1/0, 2/0, 0/3, 0/2}
    {\node[qubit] at (\x,\y) {};}
\end{tikzpicture}}
    }
    \hfil
    \subfigure[21Q]{%
        \label{fig:benchmark_21q}
        \scalebox{0.50}{\begin{tikzpicture}[qc layout]
  \qcxlayout{1}
\end{tikzpicture}}
    }
    \hfil
    \subfigure[33Q]{%
        \label{fig:benchmark_33q}
        \scalebox{0.50}{\begin{tikzpicture}[qc layout]
  \qcxlayout{2}
\end{tikzpicture}}
    }
    \vspace{-2.0mm}
    \caption{Connectivity of the benchmark layouts: qubits are blue circles,
        couplers are orange squares. (a)~$2\times2$ grid, (b)~$3\times3$ grid,
        (c)~$3\times3$ core with eight boundary qubits,(d)~X-shaped topology with one 3-qubit block added at each corner; (e)~X-shaped topology with two 3-qubit blocks added at each corner.}
    \label{fig:benchmark_layouts}
\end{figure}

To demonstrate its capabilities, we evaluate the proposed flow on a benchmark suite of planar SCQC layouts ranging from four qubits (4Q) to sixty-nine qubits (69Q). For each instance, the flow fully automatically executes global partitioning, port assignment, and routing, as described in \cref{sec:desau}, to generate a physically fabricable layout. We evaluate performance by measuring total and stage-wise runtime, as well as various quality of results metrics. The evaluated chip architectures are illustrated in \cref{fig:benchmark_layouts} for 4Q--33Q. For 45Q--69Q, the layout follows the same X-shaped scheme as 33Q, adding three qubits and four couplers at each corner per step, extending the pattern outward. Note that all evaluated architectures are purely planar; using more than 48 launcher ports is physically impractical due to boundary bottlenecks, beyond which multilayer architectures (e.g., flip-chip) would be required in practice. Consequently, these larger instances primarily serve to benchmark our flow's scalability and algorithmic limits.

Across all experiments, we define the physical design constraints to a minimum clearance of $d_\text{clear}=\SI{185}{\micro\metre}$, a minimum straight length of $d_\text{straight}=\SI{100}{\micro\metre}$, and a minimum bending radius of $r_\text{min}=\SI{50}{\micro\metre}$. The target resonator length~$d_{\text{fix}}$ and maximum feedline capacity $r_\text{util}$ are selected based on the layout size and available launcher ports, with inner-region resonators routed to an adjusted length of $d_{\text{fix}}+\SI{3000}{\micro\metre}$. For each layout instance, we record key structural parameters---including the physical layout dimensions, number of launcher ports, feedline utilization capacity $r_\text{util}$, target resonator length $d_\text{fix}$, and total point-to-point connections broken down by net type ($|W_p|$, $|W_r|$, and $|W_f|$). To quantify design quality, we report the total length of $W_p$ wires ($O_l$), the mean clearance between adjacent $W_p$ and $W_r$ wires ($O_c$), the total count of angle transitions across all wire types ($O_a$), and the number of instantiated bridges $|C_b|$ ($O_b$). Finally, to evaluate how design quality scales with increasing layout complexity, we analyze normalized metrics, including the angle ratio $\left(\frac{O_a}{\text{total p2p}}\right)$, the bridge ratio $\left(\frac{|C_b|}{(|W_p| + |W_r|)}\right)$, and the clearance scaling factor $\left(\frac{O_c \cdot \text{p2p}}{\sqrt{\text{Area}}}\right)$, which normalizes the mean clearance by the connection load relative to the chip size. Having detailed the experimental setup and evaluation methodology, we present the resulting empirical findings in the following section.

\subsection{Experimental Results}
\begin{figure}[tbp]
    \centering
    \includegraphics[width=0.35\textwidth]%
        {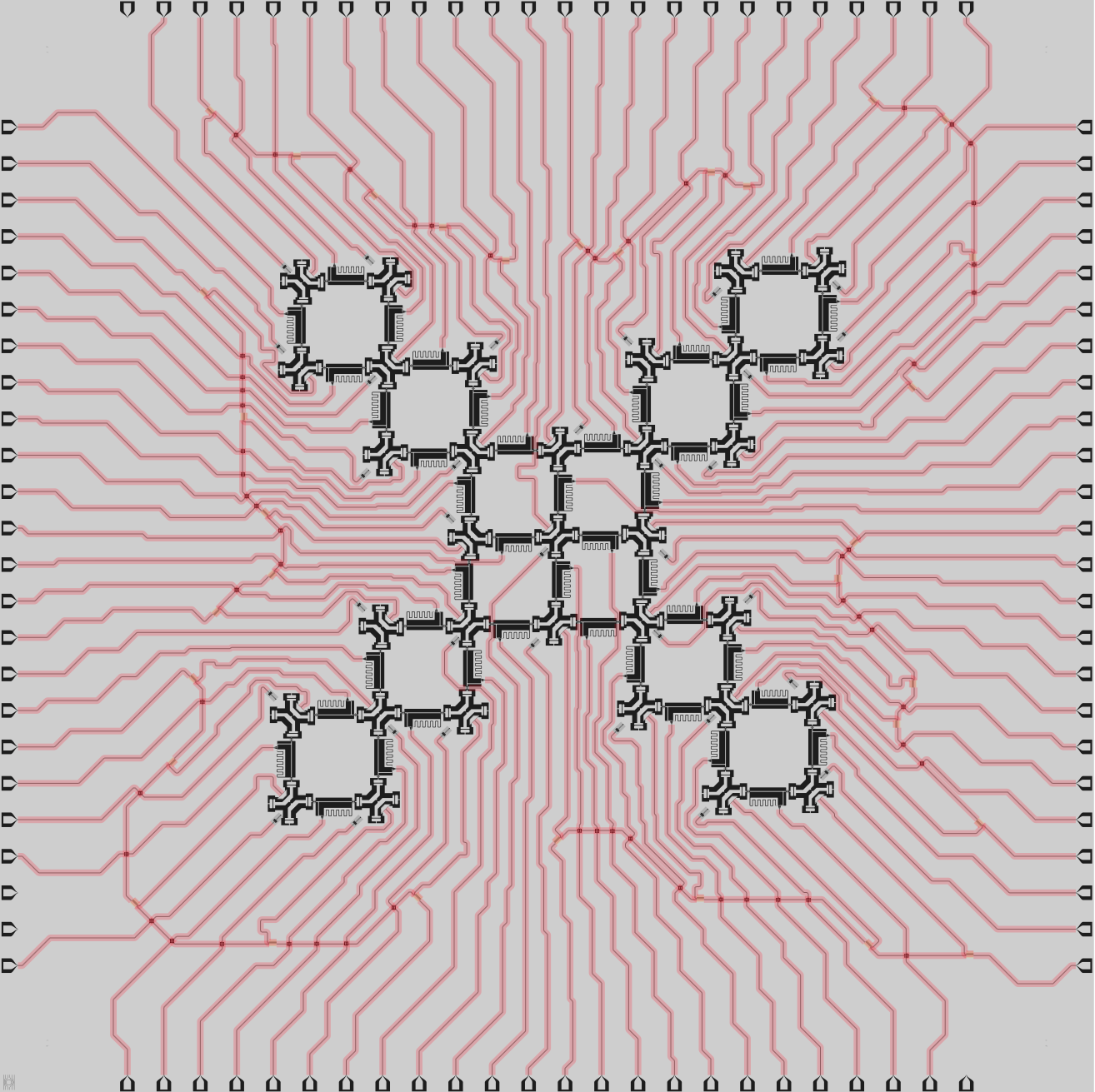}
    \vspace{-2.0mm}
    \caption{Fully routed 33-qubit chip generated by the proposed flow.}
    \label{fig:layout_33q_routed}
\end{figure}

\Cref{tab:benchmark} presents detailed experimental results for all benchmark layouts. As a representative, \cref{fig:layout_33q_routed} shows a fully routed 33Q layout generated by the proposed flow. The detailed designs for \emph{all} instances are provided in the supplementary data repository\footnote{\href{https://github.com/cda-tum/planar-superconducting-pd}{github.com/cda-tum/planar-superconducting-pd}}. For every evaluated instance, the synthesized layout strictly satisfies all physical constraints discussed in \cref{sec:prelim}.

As the current literature lacks other automated approaches to serve as a baseline for this problem formulation, direct quantitative benchmarking of the experimental outcomes is not feasible. Execution times remain highly practical even for large instances, taking seconds to a few minutes across all instances, with the largest runtime of \SI{385}{\second} for the 69-qubit layout. Furthermore, the evaluated metrics exhibit clear structural trends: the angle ratio ranges from $4.71$ to $9.47$ and increases monotonically, indicating that each point-to-point connection requires more angle transitions as the design scales. The bridge ratio remains tightly bounded between $0.42$ and $0.55$ across all instances, confirming that bridge demand grows proportionally with the number of routed nets. The clearance scaling factor remains within a narrow band of $0.62$ to $1.08$, indicating that the router maintains consistent wire spacing quality even as the connection load and chip dimensions increase.
\vspace{-1.5mm}
\begin{table}[!t]
    \centering
    \caption{Architecture Evaluation of 4Q to 69Q layouts.}
    \resizebox{\linewidth}{!}{%
        \begin{tabular}{lcccccccc}
        \toprule
        \textbf{Metric / Layout} & \textbf{4Q} & \textbf{9Q} & \textbf{17Q} & \textbf{21Q} & \textbf{33Q} & \textbf{45Q} & \textbf{57Q} & \textbf{69Q} \\
        \midrule
        Layout size (W $\times$ H) [\si{\square\milli\metre}]  & 6.6x7 & 14.3$\times$14.3 & 14.3$\times$14.3 & 24$\times$24 & 30$\times$30 & 35$\times$35 & 40$\times$40 & 54$\times$54\\
        Launcher ports                                 & 16 & 24 & 48 & 72 & 96 & 120 & 152 & 192 \\
        Feedline max.\ capacity $r_{\text{util}}$     & 4 & 5 & 5 & 5 & 5 & 6 & 7 & 6\\
        Resonator length $d_{\text{fix}}$ [\si{\micro\metre}] & 2500 & 3000 & 3000 & 4000 & 6000 & 6000 & 6000 & 6000\\
        \midrule
        $|W_p|$ & 8 & 21 & 41 & 49 & 78 & 105 & 134 & 163 \\
        $|W_r|$ & 4 & 9 & 17 & 21 & 32 & 45 & 56 & 69 \\
        $|W_f|$ & 1 & 2 & 4 & 5 & 6 & 8 & 9 & 12 \\
        p2p connections in $W_f$ & 5 & 10 & 20 & 26 & 39 & 52 & 65 & 81 \\
        Total p2p connections & 17 & 40 & 78 & 96 & 149 & 202 & 255 & 313 \\
        \midrule
        Total wirelength in $W_p$ ($O_l$) [\si{\milli\metre}] & 10.32 & 109.38 & 174.65 & 433.95 & 867.30 & 1330.11 & 1947.72 & 3346.49 \\
        Mean clearance in $W_p$, $W_r$ ($O_c$) [\si{\micro\metre}] & 454.11 & 295.62 & 198.88 & 286.08 & 263.04 & 251.18 & 228.26 & 227.70 \\
        Total angles ($O_a$) [$45^\circ$] & 80 & 240 & 518 & 677 & 1178 & 1704 & 2287 & 2964 \\
        Bridges $|C_b|$ ($O_b$) & 5 & 15 & 27 & 33 & 57 & 82 & 101 & 121 \\
        \midrule
        \textbf{Ratios \& Scaling Analysis} &  &  &  &  &  &  &  &  \\
        \midrule
        Angle ratio  & 4.71 & 6.00 & 6.64 & 7.05 & 7.91 & 8.44 & 8.97 & 9.47 \\
        Bridge ratio  & 0.42 & 0.50 & 0.47 & 0.47 & 0.52 & 0.55 & 0.53 & 0.52 \\
        Clearance scaling  & 0.80 & 0.62 & 0.81 & 0.83 & 0.96 & 1.08 & 1.08 & 0.98 \\
        \midrule
        Total runtime [s]                              & 3.22 & 10.76 & 13.73 & 57.08 & 60.41 & 102.96 & 162.95& 385.00\\
        Global partitioning [s]                        & 0.08 & 0.25 & 0.05 & 1.18 & 3.64 & 5.54 & 11.63 & 25.43\\
        Port assignment [s]                            & 0.03 & 0.41 & 0.60 & 0.39 & 0.69 & 1.86 & 3.01 & 4.12\\
        Routing [s]                                      & 3.11 & 10.11 & 13.08 & 55.51 & 56.08 & 95.56 & 148.30& 355.45\\
        \bottomrule
        \end{tabular}%
    }
    \label{tab:benchmark}
\end{table}

\subsection{Discussion}
While a direct comparison with existing approaches reveals fundamental differences in problem formulation (as discussed in \cref{subsec:short}), the proposed flow addresses a substantially more rigorous routing challenge. For instance, our 33Q and 69Q variants manages significantly more point-to-point connections than the 32Q and 64Q variants from Yang \etal \cite{yang2025scalable}, respectively (33Q vs.\ 32Q: $150$ vs.\ $40$, \SI{275}{\%} increase; 69Q vs.\ 64Q: $313$ vs.\ $80$, \SI{291}{\%} increase). Furthermore, both layouts simultaneously enforce a far stricter minimum wire clearance (\SI{185}{\micro\metre} compared to \SI{50}{\micro\metre}, representing a \SI{270}{\%} increase) and accommodate additional constraints, such as fixed-length routing $R_l$ and non-uniform chip architectures.

Despite navigating these tighter physical constraints, the proposed flow exhibits exceptional performance and scalability, as outlined in the following comparison. Yang \etal \cite{yang2025scalable} report a runtime of approximately \SI{1500}{\second} for a 32Q layout, whereas the flow proposed here solves the comparable 33Q instance $25\times$ faster. On larger scales, our 69Q layout is still over $5\times$ faster than the 64Q layout. Even compared to Yu \etal \cite{consaware}---which achieves a $2\times$ speedup over the Yang \etal framework---the proposed flow remains over $12.5\times$ faster for the 32Q variant and $2.5\times$ faster for the 69Q layout.

These findings highlight the high efficiency, robust scalability, and performance of the proposed flow in handling complex routing challenges. By successfully generating fully routed layouts for up to 69 qubits while strictly adhering to all physical design rules and keeping quality-of-results metrics ($O_c$, $O_a$, $O_b$) consistently aligned with net counts and layout area, the proposed flow provides a valuable foundational dataset for future benchmarking in this domain. Crucially, while the proposed flow does not provide universal theoretical guarantees of completeness for every conceivable edge case, it consistently and reliably resolves all evaluated problem instances.

Ultimately, the proposed flow successfully automates chip designs at scales and complexities where manual layout drafting becomes impractical, delivering manufacturing-ready, design-rule-compliant layouts within seconds to a few minutes. Although layout is only one part of a design cycle that spans days to weeks including validation and simulation, it recurs with every architectural change and therefore drives a substantial share of the overall turnaround time. By replacing tedious human design with rapid, automated optimization, this flow establishes a vital cornerstone toward realizing next-generation, highly scaled quantum architectures.
\vspace{-1.1mm}
\section{Conclusion and Future Work}\label{sec:conclusion}
\vspace{-1.4mm}
Scaling superconducting quantum chips beyond 20--30 qubits is severely limited by manual inefficiencies, which consume days or weeks of expert effort per iteration. Existing automation approaches sidestep this bottleneck by simplifying the underlying problem as well as relaxing physical constraints, and therefore fall short of fabrication-ready results. To overcome this scalability wall without such compromises, we introduced a comprehensive automated design flow for superconducting quantum chips. As a key methodological contribution, we established a formal abstraction translating complex device physics into a rigorous geometric problem formulation, effectively bridging physical realization and design automation. 

Building upon this foundation, the proposed three-stage flow fully automates planar SCQC layouts, generating complete, design-rule-compliant results at the push of a button and up to 
25× faster than the state of the art. Rather than remaining a proof of concept, all methods are released as the fully open-source tool \texttt{mqt-scpd}, ready for adoption and extension by researchers and hardware developers alike. 

Ultimately, this framework eliminates the manual layout bottleneck within a design cycle that, including validation and simulation, spans days to weeks, reducing the layout step itself to seconds or a few minutes while providing a reproducible dataset for future benchmarking. With this foundation, the proposed flow can be readily extended to incorporate different kinds of superconducting circuit architectures \cite{schirk2026high,huber2025parametric}. This is the natural next step, as purely planar designs eventually hit fundamental routing and congestion limits, whereas distributing signal paths across stacked planes expands the routing footprint and enables functional layer separation. Importantly, the abstraction developed in this work is not tied to the planar case: it carries over directly to the multi-layer setting, where the same formulation extends across stacked planes. Together with the open-source implementation in \texttt{mqt-scpd}, this work therefore contributes not only an automated flow for planar SCQC design, but a foundation on which the community can  build the design automation required for ultra-scalable quantum processors.
\vspace{-2.5mm}
\balance
\section*{Acknowledgments}
While preparing the manuscript, Claude Sonnet 5.5, Claude Opus 5.5, and Gemini 3.1 Pro were used to improve readability, spelling, grammar, and clarity throughout the manuscript. Each LLM output was reviewed by the authors and edited manually as needed. The authors take full responsibility for the final content.

\bibliographystyle{IEEEtran}
\bibliography{bib/Bibliography_ls}

\par
\ifdim\prevdepth>0pt
  \vskip-\ifdim\prevdepth>\maxdepth\maxdepth\else\prevdepth\fi\relax
  \prevdepth=0pt
\fi
\end{document}